\documentclass[fleqn,usenatbib]{mnras}

\usepackage{newtxtext,newtxmath}
\usepackage[T1]{fontenc}

\DeclareRobustCommand{\VAN}[3]{#2}
\let\VANthebibliography\thebibliography
\def\thebibliography{\DeclareRobustCommand{\VAN}[3]{##3}\VANthebibliography}

\usepackage{graphicx}
\usepackage{amsmath}
\usepackage{xcolor}
\usepackage{threeparttable}
\usepackage{pdfpages}
\usepackage{verbatim}
\usepackage{morefloats} 
\usepackage{longtable, multirow, pdflscape}
\usepackage{arydshln}
\usepackage{hyperref}
\hypersetup{colorlinks=true, citecolor=blue, linkcolor=blue, urlcolor=blue, linktocpage}
\usepackage{soul}
\usepackage{orcidlink}

\bibpunct{(}{)}{;}{a}{}{,}
\newcommand{\hi}{\mbox{H\,{\sc i}}}

\newcommand{\kms}{km\,s$^{-1}$}

\newcommand{\degree}{\ensuremath{^\circ}}

\definecolor{green}{rgb}{0,0.4,0}

\newcommand{\lapp}{\mbox{\raisebox{-0.3em}{$\stackrel{\textstyle <}{\sim}$}}}

\newcommand{\be}{\begin{equation}}
\newcommand{\en}{\end{equation}}

\title[Galactic OH in MALS]{The MeerKAT absorption line survey: Galactic OH absorption at 18\,cm\thanks{The MALS images and spectra presented here are part of MALS data release 3 and are publicly available at \href{https://mals.iucaa.in}{https://mals.iucaa.in}.}}

\author[Balashev et al.]{
S. A. Balashev,$^1$\thanks{E-mail: \href{mailto:s.balashev@gmail.com}{s.balashev@gmail.com}}
N. Gupta,\orcidlink{0000-0001-7547-4241}$^{2}$\thanks{E-mail: \href{mailto:ngupta@iucaa.in}{ngupta@iucaa.in}}
H.-W. Chen,$^3$
F. Combes,$^4$
J. Kerp,$^5$
H.-R. Kl\"ockner,$^6$
\and
D. Kosenko,$^1$
J.-K. Krogager,$^{7,8}$
P. Noterdaeme$^9$
\\
$^{1}$ Ioffe Institute, 26 Politeknicheskaya st., St. Petersburg, 194021, Russia \\ 
$^{2}$ Inter-University Centre for Astronomy and Astrophysics, Post Bag 4, Ganeshkhind, Pune 411 007, India \\
$^{3}$ Department of Astronomy and Astrophysics, The University of Chicago, 5640 S. Ellis Avenue, Chicago, IL 60637, USA \\
$^{4}$ Observatoire de Paris, Coll\`ege de France, PSL University, Sorbonne University, CNRS, LUX, Paris, France  \\
$^{5}$ Argelander-Institut für Astronomie, Auf dem Hügel 71, D-53121 Bonn, Germany \\
$^{6}$ Max-Planck-Institut f\"ur Radioastronomie, Auf dem H\"ugel 69, D-53121 Bonn, Germany \\
$^{7}$ Universit\'e Lyon, ENS de Lyon, CNRS, Centre de Recherche Astrophysique de Lyon UMR5574, F-69230 Saint-Genis-Laval, France \\
$^{8}$ French-Chilean Laboratory for Astronomy, IRL 3386, CNRS and Universidad de Chile, Santiago, Chile \\
$^{9}$ Institut d'Astrophysique de Paris, UMR 7095, CNRS-SU, 98bis bd Arago, 75014  Paris, France
}

\date{Accepted XXX. Received YYY; in original form ZZZ}

\pubyear{2026}

\begin{document}
\label{firstpage}
\pagerange{\pageref{firstpage}--\pageref{lastpage}}
\maketitle

% Abstract of the paper
\begin{abstract}
We present a blind search for Galactic OH 18-cm absorption lines towards 19,130 radio sources brighter than 1\,mJy at 1.4\,GHz, using 390 pointings of the MeerKAT Absorption Line Survey (MALS). We apply an automated line search followed by Bayesian nested sampling to identify OH 1667\,MHz and 1612\,MHz absorption features, using \hi\ emission and absorption profiles as physically motivated priors. We identify 15 confident detections of OH 1667\,MHz absorption. The majority of detections are associated with a visual extinction ($A_{\rm V}$) excess and, within the same MALS pointings, with detected CO emission and a high incidence rate of \hi\ absorption, consistent with OH tracing the H$_2$-bearing medium. We also identify 9 candidates of OH 1612\,MHz satellite line absorption, two of which coincide with OH main line detections and show integrated optical depth ratios well above the LTE expectation of 1:9. As in previous studies, $N(\rm OH)$ shows no dependence on $N$(\hi). Accounting for the varying sensitivity across sight lines, we estimate the covering fraction of OH-bearing gas at $\log N(\rm OH) > 12$ to be $C = 0.051 \pm 0.008$, providing an independent and a more precise estimate than previous measurements of the diffuse molecular gas. We also find qualitative evidence for a decrease of the covering fraction with Galactic latitude. Stacking reveals a clear OH absorption signal that becomes increasingly prominent when sight lines are weighted by $A_{\rm V}$ excess, independently confirming the association of OH with diffuse molecular gas. Modeling of the physical conditions yields hydrogen number densities of 10--100\,cm$^{-3}$, typical H$_2$ column densities $\log N(\rm H_2) > 21$, and UV-flux to number density ratios of 0.01--0.3\,Mathis\,cm$^{3}$, typical of the CNM. 
\end{abstract}


%%%%%%%%%%%%%%%%%%%%%%%%%%%%%%%%%%%%%%%%%%%%%%%%%%

%%%%%%%%%%%%%%%%% BODY OF PAPER %%%%%%%%%%%%%%%%%%

\section{Introduction} 
\label{sec:intro}  
%%%%%

%%%%%%
In the Galactic interstellar medium (ISM), the OH radical is a widespread constituent of diffuse and dense gas clouds.  It is commonly observed through its 18-cm lines: the main lines at 1665.402 and 1667.359\,MHz, and the two satellite lines at 1612.231 and 1720.530\,MHz. Under local thermodynamic equilibrium (LTE), the relative strengths of these lines are 1612:1665:1667:1720\,MHz = 1:5:9:1 \citep[e.g.,][]{Draine2011}. However OH lines, especially satellite lines, often exhibit departures from LTE ratios \citep[e.g.,][]{Elitzur1976, Frayer1998}.  
OH is known to exhibit a relatively constant molecular abundance ratio, X$_{\rm OH}$ = OH/H$_2\sim 10^{-7}$ across a wide range of physical conditions \citep[e.g.,][]{Wannier93}. Considering that H$_2$ is not easily observable in emission \citep[however, see][]{Nigou2026} and its proxy CO may not be present at low extinction or relatively low density environments, OH is an attractive alternative to trace molecular gas in the diffuse ISM and CO-dark molecular clouds \citep[][]{Liszt99}. The fraction of such molecular gas is estimated to be up to 50 percent in our Galaxy \citep{Grenier2005,Pineda2013}, and is most directly accessible through absorption studies along sparse sight lines \citep[][]{Wakker2006, Gillmon06}, via permitted electronic transitions of H$_2$ at UV wavelengths ($\lesssim 1100$\,\AA). 

%%%%%%%
The bottleneck of absorption line experiments is usually the low statistics that directly follow from the product of the gas covering fraction and the available lines of sight. This is practically noticeable in UV range, where only the decommissioned space telescopes, mostly Copernicus and FUSE, provided quantitative samples of H$_2$ absorption in the Local group. Recent detections of diffuse OH emission \citep[e.g.,][]{Busch2021, Dawson2022} further motivate systematic absorption surveys as a complementary approach. 
OH absorption surveys offer several important advantages: they operate at radio wavelengths %that are 
accessible from the ground, %and radio sources 
are unaffected by %not subject to 
dust extinction, and radio sources are far more numerous than bright UV stars. The latter is crucial since unlike H$_2$ absorption lines that are very sensitive in UV and can probe H$_2$ column densities\footnote{Hereafter, we will use $N$ to denote column densities in cm$^{-2}$.} down to $\log N(\rm H_2)\sim 14$, i.e. corresponding to the residual molecular fraction in the atomic CNM, OH lines in radio ($\sim$1.6\,GHz) probe the high end of the CNM column density distribution. Typical detection limits for the OH main lines are $\log N(\rm OH) \gtrsim 12-13$ corresponding roughly to $\log N(\rm H_2) \gtrsim 19-20$. Such gas has a very low incidence rate, due to decreasing power law shape of the column density distribution, and therefore its detection and the compilation of statistical samples require very large radio surveys.

The extragalactic radio background sources used at $\sim$1.6\,GHz are uniformly distributed across the sky, though the presence of the Galactic plane introduces a natural trade-off between sky coverage and detection rate.
While in the Galactic disk the detection rate is relatively high \citep[][]{Rugel2025}, extraplanar surveys typically require very high sensitivity to provide meaningful samples of OH absorptions \citep[][]{Li18}. Interestingly, while extragalactic detections of OH absorption at cosmological distances have begun to emerge \citep[][]{Gupta18oh}, the handful of known detections mostly represent gravitationally lensed systems, where the lensing not only increases the background source flux, but also selects lines of sight crossing the central, molecular-dominated, parts of the lensing galaxies, ensuring strong OH absorption \citep[e.g.,][]{Combes21}.

%%%%%
In this paper, we present a Galactic OH absorption line search based on the MeerKAT Absorption Line Survey \citep[MALS;][]{Gupta17mals}, a recently completed key program at the MeerKAT telescope \citep[][]{Jonas16}. The primary objective of the survey is to carry out a sensitive search of \hi\ and OH absorption lines at $0<z<2$ using the L- and UHF-bands of the telescope covering  900--1670\,MHz and 580--1015\,MHz, respectively. Each MALS pointing is centered at a radio source brighter than 200\,mJy at $\sim$1\,GHz. The radiation from the {\it central} bright source as well as numerous other AGN, hereafter referred to as {\it off-axis} sources, within the MeerKAT field of view, naturally, passes through the Galactic ISM.  391 pointings observed at L-band are suitable to %allows us to 
search for \hi\ 21-cm,  OH 18-cm main lines and the 1612\,MHz satellite lines associated with the Milky Way.  
This paper is structured as follows.  In Section~\ref{sec:obscat} we present details of MALS observations, data analysis and automated line search, including a Bayesian nested sampling approach for robust detection of OH absorption features \citep[see][for details of Galactic \hi\ 21-cm absorption lines]{Gupta2025galhi}.  In Section~\ref{sec:res}, we present  the detected OH main and satellite line samples, the covering fraction of OH-bearing gas, stacking analysis, and constraints on OH abundance and physical conditions in the diffuse ISM using a semi-analytical model. Section~\ref{sec:summ} summarizes our results and presents future prospects.

%%%%%%%%%%%%%%%%%%%%%%%%%%%%%%%%%%%%%
\section{Observations, data analysis and catalog}      
\label{sec:obscat}   
%%%%%%%%%%%%%%%%%%%%%%%%%%%%%%%%%%%%%
%

%%%%%%%%%%%%%%%%%%%%%%%%%%%%%%%%%%%%%
\subsection{Observations and data analysis}      
\label{sec:obs}   
%%%%%%%%%%%%%%%%%%%%%%%%%%%%%%%%%%%%%

The sky coverage of the Galactic component of MALS is based on 391 pointings observed in the L-band between April $1^{st}$, 2020 and January $18^{th}$, 2021.  \citet[][]{Gupta21salt} described the overall target selection process \citep[see also][]{Krogager2018}, and the $L$-band survey footprint of 391 pointings at $\delta~ \lapp~  +20^\circ$ along with the continuum catalogs are presented in \citet[][]{Deka24dr1} and \citet[][]{Wagenveld2024}.  We exclude from this footprint the pointing J213418.19-533514.7 with poor quality spectra.
The remaining 390 pointings, each centered at a radio source brighter than $\sim$200\,mJy at 1\,GHz and typically observed for 56\,mins, are considered for the Galactic OH absorption line search presented here.
The total 856 MHz bandwidth for these observations was centered at 1283.9869\,MHz and split into 32768 frequency channels.

%%%%%%
The details of calibration and imaging of $L$-band datasets using the Automated Radio Telescope Imaging Pipeline (ARTIP) are presented in \citet[][]{Gupta21}. In short, after flux density scale, delay and bandpass calibrations of XX and YY polarization products, the total bandwidth was partitioned into 15 spectral windows (SPWs) labelled as SPW0 to SPW14.  For each SPW, the Stokes $I$ radio continuum images, after three rounds of phase-only and one round of amplitude-and-phase self-calibration, have 6k\,$\times$\,6k pixels, with a pixel size of $2\arcsec$. 
For cube imaging, the self-calibration solutions obtained from the continuum imaging were applied to the line datasets and continuum subtraction was performed using the model based on the corresponding continuum images.

%%%%%
Two OH main and the 1612\,MHz satellite lines fall in SPW14 (1618.4 -- 1671.9\,MHz) and SPW13 (1564.9 -- 1625.0\,MHz), respectively.
For these two SPWs, continuum subtracted visibilities for 400 frequency channels centered at 1665.8 and 1612.3\,MHz, respectively, were Fourier inverted with {\tt robust = 0} weighting to obtain spectral line cubes in the kinematic local standard of rest (LSR) system.
These image cubes have a frequency resolution of 26.123\,kHz, that corresponds to a velocity resolution of  4.7 and 4.9\,\kms, respectively. The spatial extent is 3k\,$\times$\,3k pixels, with a pixel size of $2\arcsec$.
We extracted spectra from OH cubes for 19130 radio sources brighter than 1\,mJy\,beam$^{-1}$ at 1.4\,GHz used to define the Galactic \hi\ absorption line catalog presented in \citet[][]{Gupta2025galhi}.  
All the radio sources are within $48\farcm5$ of the pointing center.  We note that the radio continuum and spectral-line properties presented here are corrected for the attenuation of the primary beam pattern using the {\tt katbeam} (version\,0.1) model \citep[][]{Mauch20}.

%%%%%%%%%%%%%%%%%%%%%%%%%%%%%%%%%%%%%
\subsection{Automated line search and catalog}      
\label{sec:cat}   
%%%%%%%%%%%%%%%%%%%%%%%%%%%%%%%%%%%%%

%%%%%%%%%%%%%%%%%%%%%%%%%%%%%%%%%%%%%%%%%%%
\begin{figure*}
    \centering
    \vbox{
    \includegraphics[trim = {0cm -0.5cm 0cm 0cm}, width=1.0\textwidth,angle=0]{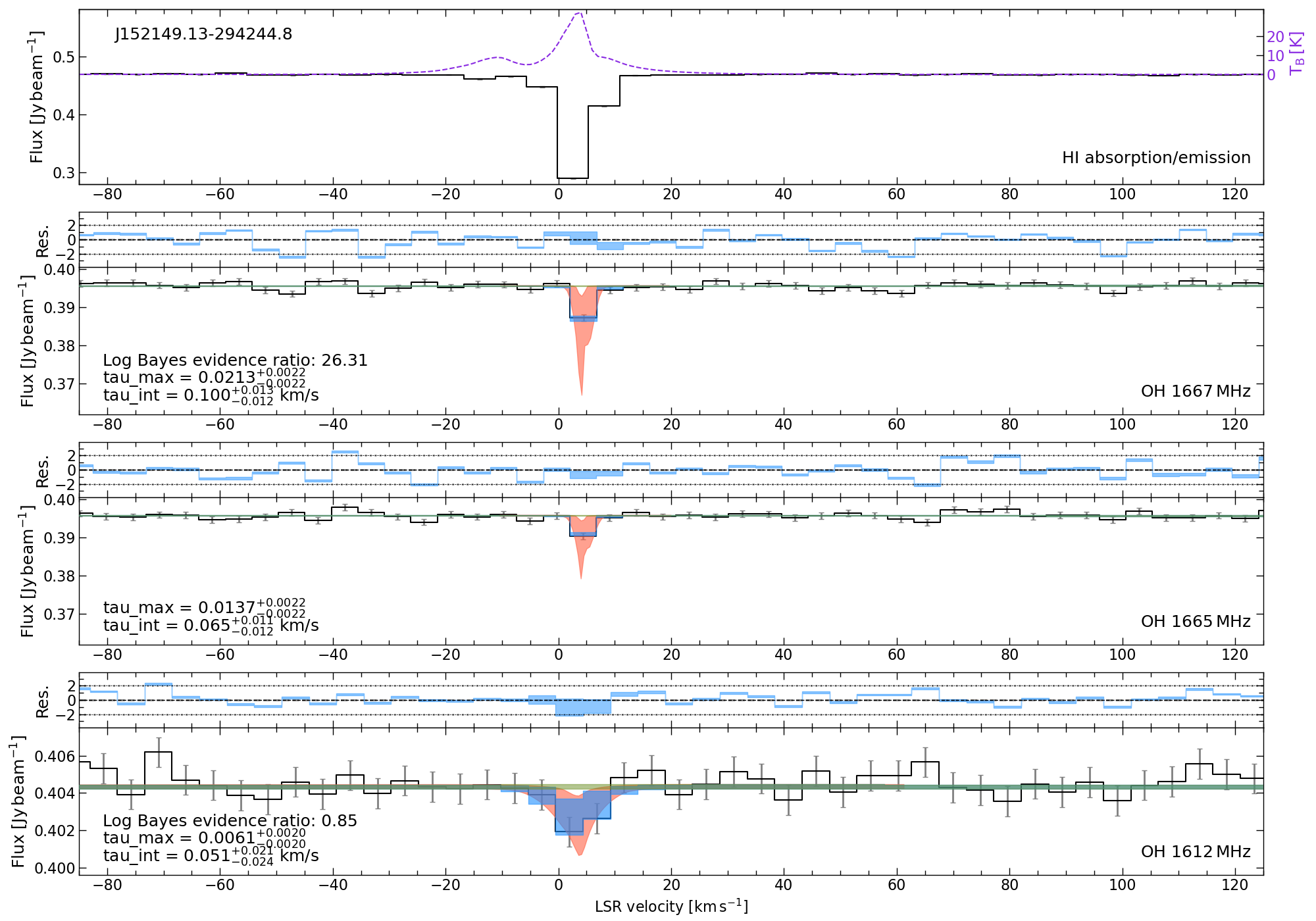} \\
    }
    \caption{The spectra of the sight line toward J152149.13$-$294244.8 showing clear detection of OH absorption. The top panel shows \hi\ absorption (black line) and emission (violet) line profiles. The remaining panels show the observed OH main lines (1667 and 1665\,MHz) and the 1612\,MHz satellite line, each shown as a black curve with pixel uncertainties. The green, red and blue stripes indicate the posterior of the unbinned continuum, absorption and binned absorption model, respectively. The posteriors correspond to 0.683 confidence interval of the posterior distribution of the model parameters derived along the nested sampling. All spectra are plotted in the LSR frame. The values of the log of the Bayesian evidences ratio, and estimates of the maximal and integrated optical depths are presented in the bottom left corner. The fits are obtained using the nested sampling approach described in Section~\ref{sec:cat}. Spectra for all confident detections are presented in Appendix~\ref{sec:ohmain-confid-spec}.}
    \label{fig:ohmain-hi}
\end{figure*}
%%%%%%%%%%%%%%%%%%%%%%%%%%%%%%%%%%%%%%%%%%%

%%%%
We first applied the automatic procedure presented by \citet[][]{Gupta2025galhi} to blindly search for OH 1667\,MHz and 1612\,MHz absorption lines in a 500\,km\,s$^{-1}$-wide spectral window around $v_{\rm LSR}$ = 0\,\kms. For this we adopted a model consisting of the unabsorbed continuum represented by a third order Chebyshev polynomial plus Gaussian absorption components, labeled as $i$ = 1, 2, .., $n$, representing the optical depth of the individual absorption components, defined by three parameters: the amplitude ({\tt comp\_i\_amp}),  central position ({\tt comp\_i\_pos}) and dispersion ({\tt comp\_i\_sigma}), as
\begin{equation}
\tau_i(v) = \frac{\rm comp\_i\_amp}{\sqrt{2\pi} \times {\rm comp\_i\_sigma}} {\rm exp}\left(- \cfrac{(v - {\rm comp\_i\_pos})^2}{2\times{\rm comp\_i\_sigma}^2}\right).
\end{equation}
The resultant absorption line catalogs, following the column definitions presented in \citet[][their Section~2 and Table~A1]{Gupta2025galhi}, are provided in Tables~\ref{tab:ohmaincat} and ~\ref{tab:ohsat1cat}.  In short, the radio continuum properties of the background AGN are provided in columns\,1 - 15, whereas columns\,16 onwards provide spectral line properties.  
%

%%%%%
We supplemented this table by a visual inspection ({\tt Vis\_flag} column). This is simply based on the presence of an absorption feature, visually distinct from other such features in the spectrum and within $\pm$5\,\kms\ of the cataloged value. 
A new visual inspection flag, {\tt Vis\_flag\_HI}, is also included in the OH catalogs to mark the presence of a corresponding \hi\ 21-cm absorption line. This flag represents the reasonable expectation that an OH absorption line be accompanied by a corresponding \hi\ 21-cm absorption.
The dependence of absorption features on S/N and visual inspection flags are presented in Appendix~\ref{sec:oh-inspec}.  
In short, for 1667\,MHz and 1612\,MHz absorption features 256 and 323 absorption features are detected but only 14 and 13 of these, respectively, have a visually coincident \hi\ 21-cm absorption (Table~\ref{tab:ohstat}). This suggests that a vast majority of the detected features are false detections.  

%%%%%%%%%%%%%%%%%%%%%%%%%%%%%%%%%%%%%%%%%%%
% Both main OH line search
\renewcommand{\arraystretch}{1.45}
\setlength{\tabcolsep}{0.1cm}
\begin{table*}
\caption{MALS OH main line detections based on the Bayesian inference. Fifteen sight lines with high Bayesian evidence above the dashed line are used for the analysis. Their spectra are presented in Appendix~\ref{sec:ohmain-confid-spec}.  Five discrepant cases, not used in the analysis here, are presented in Table~\ref{tab:ohmain_add}. 
}
\vspace{-0.4cm}
\begin{center}
\begin{tabular}{cccc|ccc|cc|ccc}
\hline
\hline
Sight line & $l$ & $b$ & $\log \mathcal{R}$ & $v_{\rm 1667}$ & $\sigma_{1667}$ & $\tau_{\rm int}$ = $\int\tau_{1667}dv$& $\log N(\rm OH)$ & $\log N($\hi$)$ & S/N & {\tt F1} & {\tt F2} \\
& (degree) & (degree) & & (km\,s$^{-1}$) & (km\,s$^{-1}$) & (km\,s$^{-1}$) & (cm$^{-2}$) & (cm$^{-2}$) & & \\
(1) & (2) & (3) & (4) & (5) & (6) & (7) & (8) & (9) & (10) & (11) & (12) \\
\hline
J152149.13$-$294244.8 &   337.9956  & 22.7355  & 26.3  &  $4.0^{+0.3}_{-1.2}$  &  $0.70^{+0.50}_{-0.60}$  & $0.101^{+0.011}_{-0.011}$  & $14.01^{+0.17}_{-0.28}$ & $20.85^{+0.08}_{-0.10}$ & 10.5  & T & T \\
J152225.48$-$293625.0 &   338.1795  & 22.7416  & 20.8  &  $3.70^{+0.77}_{-0.30}$  &  $1.70^{+0.70}_{-0.50}$  & $0.0700^{+0.0099}_{-0.0118}$  & $13.85^{+0.18}_{-0.28}$ & $20.71^{+0.08}_{-0.10}$ & 8.2  & T & T \\
J155930.97$+$030448.2 &   13.1602   & 39.0776  & 16.7  &  $1.30^{+0.80}_{-1.00}$  &  $0.50^{+0.30}_{-0.50}$  & $0.0641^{+0.0075}_{-0.0077}$  & $13.81^{+0.18}_{-0.28}$ & $20.38^{+0.08}_{-0.10}$ & 8.6 & T & T \\
J183211.11$-$203947.9 &   12.3641   & -5.2229  & 10.3  &  $6.060^{+0.060}_{-0.870}$  &  $1.40^{+0.60}_{-0.80}$  & $0.086^{+0.014}_{-0.022}$  & $13.94^{+0.18}_{-0.29}$ & $21.14^{+0.08}_{-0.09}$ & 7.1  & T & T \\
J200608.43$-$022334.8 &   39.4845   & -17.6843 & 6.5  &  $3.81^{+1.17}_{-0.24}$  &  $0.70^{+0.50}_{-0.70}$  & $0.0173^{+0.0034}_{-0.0033}$  & $13.24^{+0.19}_{-0.28}$ & $21.05^{+0.08}_{-0.10}$ &  7.5 & T & T \\
J032153.09$+$122114.0 &  170.5908   & -36.2375 & 6.2  &  $9.2^{+1.0}_{-1.4}$  &  $1.30^{+0.90}_{-1.00}$  & $0.0158^{+0.0057}_{-0.0032}$  & $13.20^{+0.23}_{-0.28}$  & $21.56^{+0.08}_{-0.10}$ & 5.9 & T  & T \\
J171456.58$-$055820.4 &   15.9067   & 18.3202  & 3.5  &  $0.90^{+1.30}_{-0.50}$  &  $0.80^{+0.50}_{-0.80}$  & $0.073^{+0.017}_{-0.018}$  & $13.87^{+0.19}_{-0.29}$ & $21.31^{+0.08}_{-0.10}$ & 5.0  & T & T \\
J152101.05$-$294731.9 &  337.7932   & 22.7719  & 3.5  &  $3.40^{+0.60}_{-0.40}$  &  $0.50^{+0.30}_{-0.60}$  & $0.42^{+0.11}_{-0.13}$  & $14.62^{+0.21}_{-0.29}$ & $20.88^{+0.09}_{-0.10}$ & -- & -- & -- \\
J183339.97$-$210339.9 &   12.1659   & -5.7119  & 3.5  &  $4.0^{+1.6}_{-2.0}$  &  $5.6^{+1.4}_{-1.8}$  & $0.0088^{+0.0029}_{-0.0028}$  & $12.95^{+0.22}_{-0.30}$ & $21.10^{+0.08}_{-0.09}$ & 3.3 & T & T \\
J152428.84$-$294107.5 &  338.5186   & 22.4132  & 2.5  &  $5.2^{+1.2}_{-2.0}$  &  $4.0^{+1.5}_{-1.0}$  & $0.372^{+0.103}_{-0.099}$  & $14.57^{+0.21}_{-0.29}$ & $20.76^{+0.08}_{-0.10}$ & 3.6 & T & T \\
J005147.17$+$174711.5 &  123.0493  & -45.0849  & 2.4  &  $-5.70^{+0.80}_{-0.50}$  &  $1.70^{+0.90}_{-0.80}$  & $0.0232^{+0.0067}_{-0.0074}$  & $13.37^{+0.21}_{-0.30}$ & $20.55^{+0.08}_{-0.09}$ & 4.1 & T  & T \\
J071115.10$-$383209.9 &  249.8393  & -12.9424  & 1.7  &  $-3.6^{+2.4}_{-1.5}$  &  $5.70^{+1.40}_{-0.90}$  & $1.58^{+0.58}_{-0.58}$  & $15.20^{+0.23}_{-0.31}$ & $20.60^{+0.12}_{-0.14}$ & --  & --  & -- \\
J183206.63$-$204015.9 &   12.3638  &  -5.2224  & 1.7  &  $6.00^{+0.80}_{-1.30}$  &  $2.7^{+1.6}_{-1.6}$  & $0.078^{+0.038}_{-0.037}$  & $13.90^{+0.25}_{-0.35}$ & $21.09^{+0.08}_{-0.10}$ & --  & -- & -- \\
J162115.62$-$361137.7 &  343.4089  &   9.6964  & 1.4  &  $5.0^{+1.0}_{-1.3}$  &  $2.5^{+1.3}_{-1.7}$  & $0.031^{+0.015}_{-0.014}$  & $13.49^{+0.26}_{-0.34}$ & $21.07^{+0.08}_{-0.10}$ & -- & -- & -- \\
J015710.52$+$001124.2 &  155.6544  & -58.4419  & 1.1  &  $-7.80^{+0.90}_{-0.60}$  &  $1.30^{+0.90}_{-0.80}$  & $0.30^{+0.18}_{-0.13}$  & $14.48^{+0.28}_{-0.33}$  & $19.95^{+0.14}_{-0.13}$ & -- & -- & -- \\
%%%
\hdashline
J115601.47$-$280549.3 &  288.3110  &  33.1950  & 0.9  &  $-6.6^{+1.7}_{-1.4}$  &  $2.0^{+1.2}_{-1.0}$  & $0.26^{+0.15}_{-0.12}$  & $14.42^{+0.27}_{-0.34}$ & $19.98^{+0.13}_{-0.15}$ & -- & -- & -- \\
J155943.47$+$030502.3 &   13.2010  &  39.0359  & 0.9  &  $1.10^{+0.80}_{-1.30}$  &  $0.90^{+0.60}_{-0.70}$  & $1.48^{+0.80}_{-0.69}$  &  $15.17^{+0.27}_{-0.34}$ & $20.67^{+0.23}_{-0.40}$ & -- & -- & -- \\
J034208.50$-$043541.9 &  191.6656  & -43.5108  & 0.8  &  $4.3^{+1.8}_{-1.5}$  &  $2.9^{+1.5}_{-1.1}$  & $0.29^{+0.11}_{-0.11}$  & $14.46^{+0.23}_{-0.32}$ & $20.09^{+0.15}_{-0.15}$ & -- & -- & -- \\
J195629.70$-$053326.2 &   35.3839  & -16.9922  & 0.8  &  $1.0^{+1.0}_{-1.6}$  &  $1.00^{+0.70}_{-0.70}$  & $1.69^{+0.64}_{-0.56}$  & $15.23^{+0.23}_{-0.30}$ & $21.00^{+0.13}_{-0.12}$ & -- & -- & -- \\
J023916.20$-$140804.4 &  191.0240  & -61.7170  & 0.7  &  $-8.3^{+2.5}_{-1.5}$  &  $2.1^{+1.5}_{-3.9}$  & $0.029^{+0.032}_{-0.020}$  & $13.47^{+0.38}_{-0.54}$ & $19.49^{+0.11}_{-0.11}$ & -- & -- & -- \\
J215800.90$+$092545.8 &   67.8560  & -34.3096  & 0.7  &  $-6.2^{+3.0}_{-9.1}$  &  $5.5^{+2.5}_{-2.8}$  & $0.172^{+0.096}_{-0.104}$  & $14.24^{+0.27}_{-0.44}$ & $20.17^{+0.13}_{-0.14}$ & -- & -- & -- \\
%%%
\hline
\hline
\end{tabular}
\label{tab:ohmain_det}
\end{center}
\begin{tablenotes}
\item 
The columns are: (1): sight line or {\tt Source\_name} based on the position of radio source in the MALS continuum image. (2-3): the Galactic longitude and latitude. (4):Bayesian evidence ratio, $\frac{{\rm Pr}(D|M_{\rm abs})}{{\rm Pr}(D|M_{\rm cont})}$. (5-7): central position, width and integrated optical depth of the absorption component. (8-9): derived OH  and \hi\ 21-cm absorption column densities using Equation~\ref{eq:NOH} and $N$(\hi) = 1.823$\times10^{18}T_{\rm spin}\int\tau$ dv \,cm$^{-2}$, respectively. For $N$(\hi) optical depths from \citet[][]{Gupta2025galhi} are used.  The details of $T_{\rm exc}$ and $T_{\rm spin}$ used in the analysis are provided in Section~\ref{sec:oh_hi}. (10-12): Peak S/N of absorption and visual inspection flags, {\tt F1} = {\tt Vis\_flag} and {\tt F2} = {\tt Vis\_flag\_HI} (see Appendix~\ref{sec:oh-inspec} for details).
\end{tablenotes}
\end{table*}
%%%%%%%%%%%%%%%%%%%%%%%%%%%%%%%%%%%%%%%%%%%

%%%%%
The visual inspection is tedious and somewhat subjective. Therefore, we used Bayesian inference with dynamical Nested Sampling (NS) by means of \textsc{dynesty} package \citep{Speagle2018} to disentangle between detection and non-detection models for the 1667\,MHz line using Bayes evidence. Moreover, as we show below, NS allows us to use physically motivated information from \hi\ profiles as priors for OH detection. Additionally, we included the 1665\,MHz line in the fit model, assuming a fixed LTE ratio 9:5 between 1667\,MHz and 1665\,MHz lines. While we included the 1665\,MHz line, which is covered in the same SPW14 as the 1667\,MHz line, the resultant detection sample, especially for the most robust candidates, is quite similar to the search based on 1667\,MHz alone.
Since we assumed a fixed ratio of the main lines, in the following for the main lines we provide and discuss absorption parameters only for the 1667\,MHz lines.

The absorption features detected above comprise single Gaussian components and are weak, as expected here for diffuse molecular gas observed at coarse spectral resolution. Therefore, we consider a model consisting of the unabsorbed continuum represented by a fourth order Chebyshev polynomial and the optical depth in the velocity space caused by OH absorption in the form 
\begin{equation}
\tau(v)_{1667} = \frac{\tau_{\rm int}}{\sqrt{2\pi} \sigma_{1667}} {\rm exp} \left(- \cfrac{(v - v_{\rm 1667})^2}{2\sigma_{1667}^2}\right),
\label{eq:ns}
\end{equation}
described by the following three parameters: $v_{\rm 1667}$ is the position of the peak, $\sigma_{1667}$ is the line width, and $\tau_{\rm int}$ is the integrated optical depth $\int\tau(v)_{1667} dv$. The calculation of the line profiles was performed on a denser adaptive grid and then integrated within the selected spectral bins to compare with the data. 
%

%%%%%%%%%%%%%%%%%%%%%%%%%%%%%%%%%%%%%%%%%%%
\begin{figure*}
    \centering
    \includegraphics[trim = {0cm 0cm 0cm 0cm}, width=1.0\textwidth,angle=0]{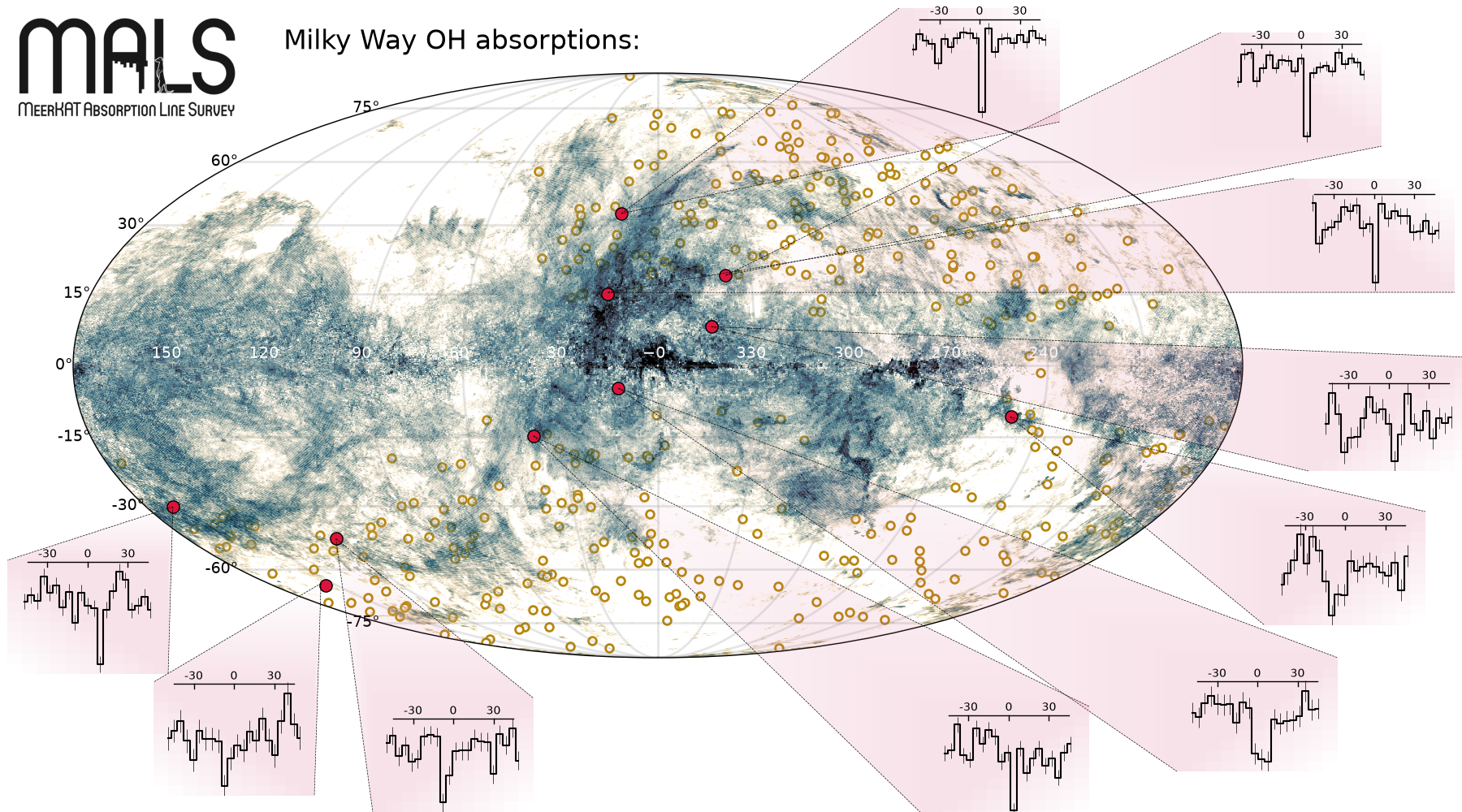}
    \caption{Sky distribution of MALS L-band pointings. Open circles represent individual MALS pointings. Sight lines where OH is confidently detected are coloured by red and their spectra are presented in the zoom-in panels, with the x-axis representing LSR velocity. The underlying color map represents the CO-dark gas column densities \citep[see][]{Kalberla2020} for the velocity range $-8<v_{\rm LSR}<8$\,\kms, characteristic for gas in the local vicinity.}
    \label{fig:ohmap}
\end{figure*}
%%%%%%%%%%%%%%%%%%%%%%%%%%%%%%%%%%%%%%%%%%%

%%%%%%
An important ingredient of Bayesian inference and in particular NS is the prior distribution on the model parameters. It affects the efficiency of the sampling and posterior distribution functions. For the Chebyshev polynomials, we used uniform priors within the relevant ranges defined from 0.997 confidence level ($3\sigma$) obtained from a simple minimization fit to the data and covariance matrix approach. We note that limiting the prior space this way does not duplicate the use of the data. Indeed, the spectra used here over a limited spectral range (500\,\kms) are mostly flat, and in all cases the continuum is robustly defined with Gaussian-like marginalized posteriors with a weak correlation between the coefficients of the polynomial. We checked that the NS results are consistent with the results obtained using priors with several times larger range, and with an Affine invariant MCMC sampler \citep{Goodman2010}.

%%%%%%
For 1667\,MHz absorption, we used priors based on the \hi\ absorption and emission profiles as follows. 
\citet{Gupta2025galhi} noted generally clear consistency between \hi\ absorption detected in MALS spectra and the corresponding emission profiles from the HI4PI survey \citep{HI4PI}.  Therefore, for the position ($v_{1667}$) of the absorption in Equation~\ref{eq:ns} we sampled priors using a probability distribution function based on the shape of \hi\ emission profile within the velocity range, where \hi\ absorption is detected at 2$\sigma$ level over continuum.
Concerning the width ($\sigma_{\rm 1667}$), we used uniform prior between 0 and the interquantile interval corresponding to the central 0.683 ($1\sigma$ for the Gaussian) of the \hi\ emission profile. This limits the OH absorption to be narrower than the \hi\ emission, which is reasonable because OH is expected to represent higher column density clumps along the sight line.
%
%%%%%
For the amplitude parameter ($\tau_{\rm int}$), as a consequence of the relationship for the 1667\,MHz line 
\begin{equation}
\label{eq:NOH}
N({\rm OH}) = 2.24\times10^{14} \frac{T_{\rm exc}}{f} \int\tau(v)_{1667}{\rm d}v \,\,\rm cm^{-2}\,K^{-1}\,(km\,s^{-1})^{-1},
\end{equation}
we used a power-law prior in the range of diffuse gas column densities, $N$(OH) = $10^{13-16}$\,cm$^{-2}$, with a slope of $-1.25$ \citep{Balashev2021}. 
Here, we assumed that an excitation temperature $T_{\rm exc}$ = 4.5$\pm$2.1\,K, based on measurements in the literature \citep[e.g.][]{Hafner2023} and covering factor $f=1$.
An identical approach was adopted to search the 1612\,MHz line.  We discuss the outcomes in the next section.

%%%%%%%%%%%%%%%%%%%%%%%%%%%%%%%%%%%%%
\section{Results}
\label{sec:res}   
%%%%%%%%%%%%%%%%%%%%%%%%%%%%%%%%%%%%%

%%%%%%%%%%%%%%%%%%%%%%%%%%%%%%%%%%%%%%%%%%%
\begin{figure*}
    \centering
    \includegraphics[trim = {0cm 0cm 0cm 0cm}, width=1.0\textwidth,angle=0]{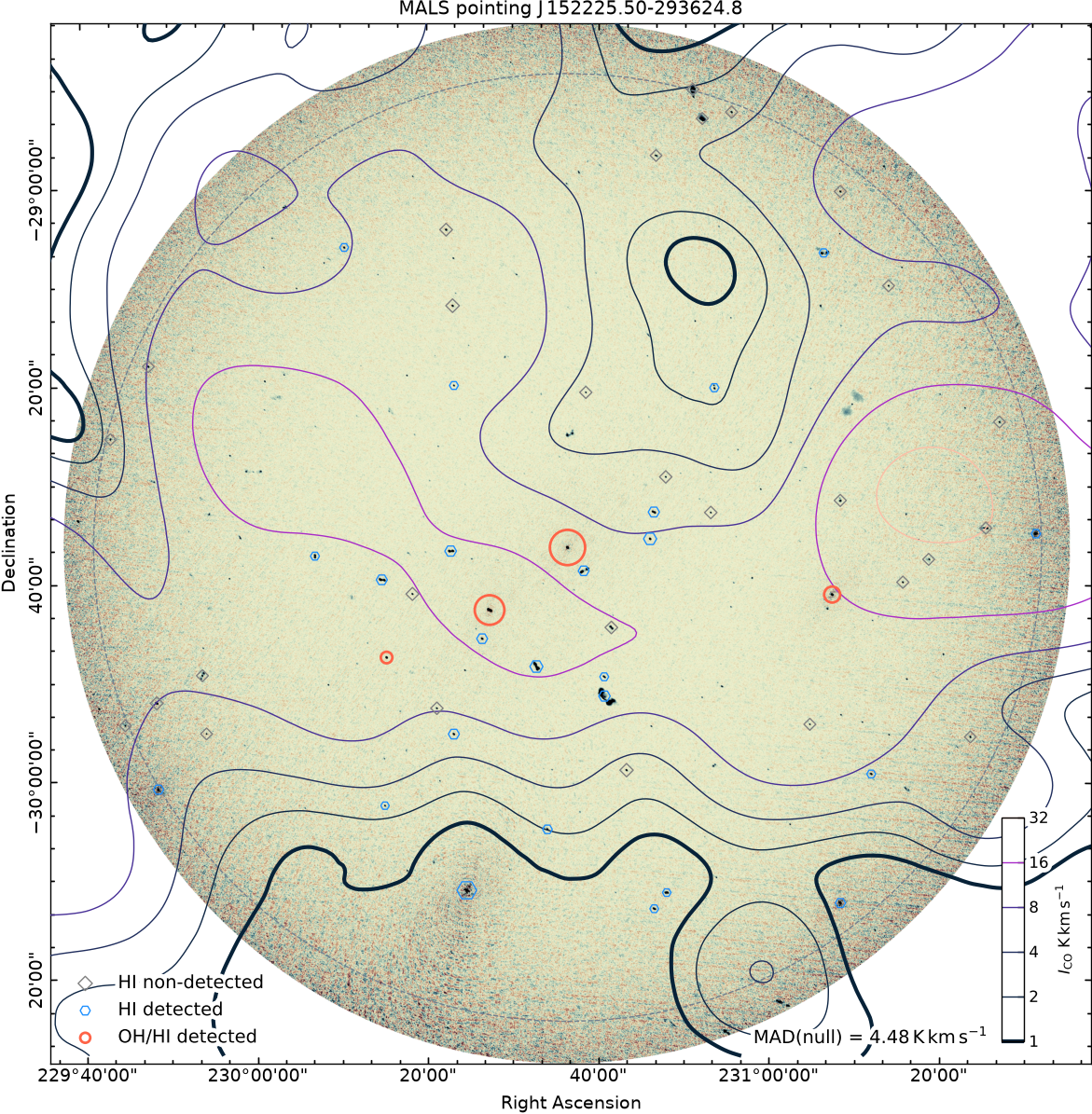}
    \caption{MALS field around J152225.50$-$293624.8. The red circles, blue hexagons and gray rhombi indicate MALS background sources with OH/\hi\ detections, 
    \hi\ detections, and \hi\ non-detections, respectively. The size of the symbols represents the sensitivity (continuum-to-noise ratio) of each sight line. 
    The contours represent the CO $J$ = $1\rightarrow0$ map derived from Planck \citep{Ghosh2024}, with levels defined by the colorbar at the bottom right corner. The local noise, estimated as the median absolute deviation (MAD) of the null map, is given at the bottom of the panel.
    }
    \label{fig:poi}
\end{figure*}
%%%%%%%%%%%%%%%%%%%%%%%%%%%%%%%%%%%%%%%%%%%

%%%%%%%%%%%%%%%%%%%%%%%%%%%%%%%%%%%%%%%%%%%
\renewcommand{\arraystretch}{1.45}
\setlength{\tabcolsep}{0.1cm}
\begin{table*}
\caption{MALS OH 1612\,MHz satellite line detections based on the Bayesian inference.  A discrepant case, not used in the analysis here, is presented in Table~\ref{tab:ohsat_add}.}
\vspace{-0.4cm}
\begin{center}
\begin{tabular}{cccc|ccc|c|ccc}
\hline
\hline
Sight line & $l$ & $b$ & $\log \mathcal{R}$ & $v_{\rm 1612}$ & $\sigma_{1612}$ & $\tau_{\rm int}$ = $\int\tau_{1612}dv$  & OH main & S/N & {\tt F1} & {\tt F2} \\
&  (degree) & (degree) & & (km\,s$^{-1}$) & (km\,s$^{-1}$) & (km\,s$^{-1}$) &  \\
(1) & (2) & (3) & (4) & (5) & (6) & (7) & (8) & (9) & (10) & (11) \\
\hline
J200322.53$-$023347.9 & 39.0024  & -17.1614  & 1.5  &  $1.8^{+1.0}_{-1.1}$  &  $1.10^{+0.80}_{-0.80}$  & $1.58^{+0.59}_{-0.51}$  & -- & 5.3 & T & T      \\
J161615.34$-$774404.6 & 313.0226 & -19.0939 & 1.5  &  $7.20^{+0.80}_{-1.10}$  &  $1.20^{+0.80}_{-1.00}$  & $1.47^{+0.55}_{-0.49}$   & -- & -- & -- & -- \\
J171944.78$+$075951.5 & 29.6657  &  23.9986 & 1.2  &  $-2.3^{+1.4}_{-1.2}$  &  $1.00^{+0.70}_{-0.70}$  & $0.44^{+0.17}_{-0.15}$  & -- & -- & -- & --    \\
J114335.39$-$261102.0 & 284.4687 &  34.2429 & 1.1  &  $1.00^{+1.90}_{-0.60}$  &  $1.20^{+0.80}_{-1.00}$  & $7.5^{+3.4}_{-2.9}$   & -- & 7.4 & T & T \\
J032153.09$+$122114.0 & 170.5908 & -36.2375 & 1.0  &  $10.4^{+2.3}_{-1.0}$  &  $2.9^{+1.6}_{-1.6}$  & $0.0112^{+0.0052}_{-0.0050}$  & Y & -- & -- & --\\
J164922.32$+$063155.9 & 24.3742  &  30.0570 & 0.9  &  $0.5^{+1.1}_{-1.0}$  &  $1.00^{+0.70}_{-0.70}$  & $0.61^{+0.24}_{-0.22}$  &  -- & -- & -- & -- \\
J152149.13$-$294244.8 & 337.9956 &  22.7354 & 0.9  &  $3.30^{+2.20}_{-0.70}$  &  $3.2^{+1.7}_{-2.2}$  & $0.049^{+0.022}_{-0.024}$  & Y  & -- & -- & -- \\
J131207.87$-$202652.7 & 309.4765 &  42.1608 & 0.7  &  $-7.1^{+1.9}_{-1.8}$  &  $1.6^{+1.1}_{-1.2}$  & $0.0163^{+0.0086}_{-0.0071}$  &  -- & -- & -- & --\\
\hline
\end{tabular}
\label{tab:ohsat_det}
\end{center}
\begin{tablenotes}
\item
The columns are: (1): sight line or {\tt Source\_name}. (2-3) the Galactic longitude and latitude. (4): log of the Bayesian evidence ratio, $\frac{{\rm Pr}(D|M_{\rm abs})}{{\rm Pr}(D|M_{\rm cont})}$. (5-7): the central position, width and integrated optical depth of the absorption component. (8) a flag indicating whether the OH main line is detected. (9-11): peak S/N of absorption and visual inspection flags, {\tt F1} = {\tt Vis\_flag} and {\tt F2} = {\tt Vis\_flag\_HI} (see Appendix~\ref{sec:oh-inspec} for details).
\end{tablenotes}
\end{table*}
%%%%%%%%%%%%%%%%%%%%%%%%%%%%%%%%%%%%%%%%%%%

%%%%%%%%%%%%%%%%%%%%%%%%%%%%%%%%%%%%%
\subsection{Detection sample}
\label{sec:sample}
%%%%%%%%%%%%%%%%%%%%%%%%%%%%%%%%%%%%%

We compiled the OH\,1667\,MHz absorption sample by comparing the Bayesian evidences of the continuum-only (non-detection) and one-component (detection) models from the aforementioned NS runs. The actual value of the threshold to be used for the confident detection is debatable and depends on the chosen credibility level. In Table~\ref{tab:ohmain_det} we provide a list of 21 sight lines which have Bayesian evidence ratio, $\mathcal{R}\equiv \frac{{\rm Pr}(D|M_{\rm abs})}{{\rm Pr}(D|M_{\rm cont})}>5$ ($\log\,\mathcal{R} > 0.7$) along with their absorption properties sampled during the NS run. We adopted $\mathcal{R}>10$ ($\log\,\mathcal{R} >$1; dashed line in Table~\ref{tab:ohmain_det}) as the threshold for confident detection, based on visual inspection and the fact that sight lines with $0.7 < \log\,\mathcal{R} < 1$ tend to have systematically higher OH column densities than those with $\log\,\mathcal{R} >1$. The further confirmation of the credibility of chosen threshold is provided in Section~\ref{sec:global}.  
We validated this threshold by re-running the Bayesian analysis with the velocity prior shifted by $\pm$100\,\kms, where no real OH absorption is expected. The resulting false-detection sample shows a few sight lines with $1 < \log\,\mathcal{R} \lesssim 3$, suggesting that the confident detection sample may contain a few false positives (see Appendix~\ref{sec:oh-check} for details). 
That leaves 15 sight lines in the confident detection sample. Hereafter, we use these for the analysis of OH main lines. An example of the OH detection and comparison with \hi\ emission and absorption spectra is shown in Fig.~\ref{fig:ohmain-hi}. The spectra for all confident detections are shown in Appendix~\ref{sec:ohmain-confid-spec}.

For completeness, the peak S/N and visual inspection flags based on the initial search are presented in the last three columns of Table~\ref{tab:ohmain_det}. Ten out of fourteen systems with {\tt Vis\_flag\_HI} = {\tt True} overlap with the confident detections based on the Bayesian evidence ratio.  The overlap is also consistent with the high completeness fraction ($>$90\%) expected for absorbers with peak S/N $>$6 in the MALS spectra \citep[][their Fig.~7]{Gupta2025galhi}.  
The remaining four sight lines along with another discrepant case (J040519.73+180610.9) are presented in  Table~\ref{tab:ohmain_add}. These are not relevant for the results presented here and therefore discussed only in Appendix~\ref{sec:oh-inspec}.

%%%%%%%%%%%%%%%%%%%
\subsection{Global distribution of OH detections and CO association}
\label{sec:global}
%%%%%%%%%%%%%%%%%%%

The sky distribution of MALS pointings and OH detections is shown in Fig.~\ref{fig:ohmap}. More than half of the fields with OH detections are located towards the Galactic center/anticenter. Although the MALS footprint avoids the Galactic plane by design, the OH detections preferentially lie at the lower latitudes it covers. At lower latitudes, sight lines traverse longer paths through the Galactic disk and hence relatively larger total columns of the molecule-bearing gas, increasing the probability of intercepting the patchy molecular component. 
As an indirect tracer of the molecular gas, we use the CO-dark column density map of   \citet{Kalberla2020}, derived from HI4PI Gaussian decomposition of \hi\ emission combined with dust measurements at $16\farcm2$ resolution. 
Indeed, the majority of confident OH detections $(\log\,\mathcal{R}> 1)$ are found toward sight lines with elevated CO-dark gas column densities, consistent with OH tracing the H$_2$-bearing medium.
We discuss this dependence statistically in Sections~\ref{sec:av} and \ref{sec:cf}. 

%%%%%%%%%%%%%%%%%%%%%%%%
\begin{figure}
    \centering
    \includegraphics[trim = {0cm 0cm 0cm 0cm}, width=1.0\columnwidth,angle=0]{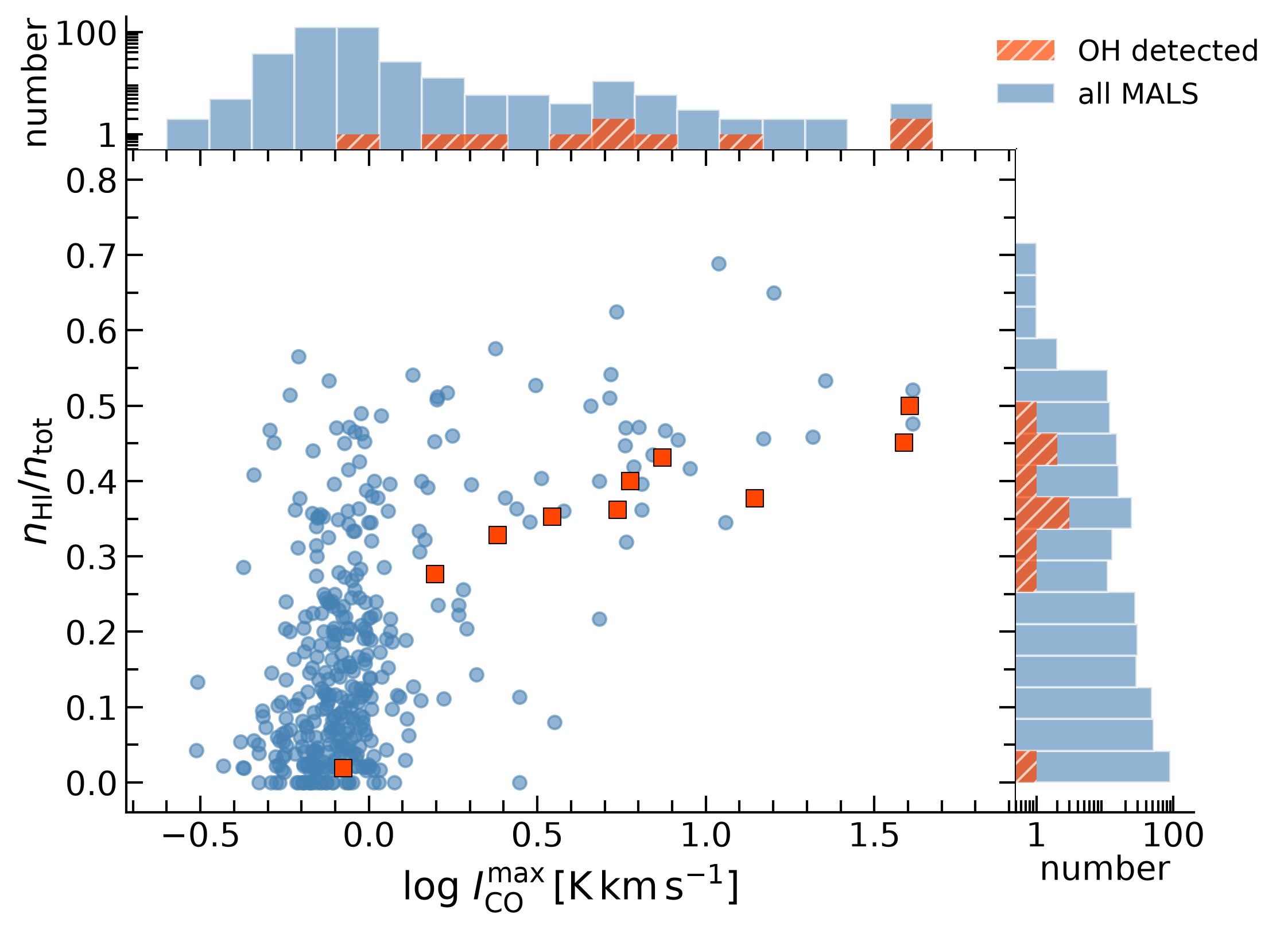}
    \caption{
    The distributions of the maximal CO $J$ = $1\rightarrow0$ integrated intensity \citep[from][]{Ghosh2024} and fraction of sight lines with \hi\ detections within each pointing for the whole sample (blue; Table~\ref{tab:ohmaincat}) and for the pointings with confident OH detections (red; Table~\ref{tab:ohmain_det}). Pointings with more than one OH detection are counted once. The \textit{top} and \textit{right} panels show the marginalized distributions in the log scale. 
    }
    \label{fig:HI_CO_OH}
\end{figure}
%%%%%%%%%%%%%%%%%%%%%%%%

%%%%%
Here, we explore the association between OH absorption detections and a direct tracer of the molecular gas i.e., CO emission, using the full-sky CO $J$ = $1\rightarrow0$ map at 10$\arcmin$ resolution from \citet[][]{Ghosh2024}. Figure~\ref{fig:poi} shows the field around J152225.50$-$293624.8, the richest field in terms of OH detections with four confident absorbers, overlaid with CO emission. The MALS sight lines are indicated by red circles (OH/\hi\ detections), blue hexagons (\hi-only detections), and gray rhombi (\hi\ non-detections). The symbol sizes illustrate the wide range of optical depth sensitivities across the field. The spatial correspondence between the CO emission and the OH detections further supports the association of OH with molecular gas. We present similar CO overlays for the pointings containing all 15 confident OH detections listed above the dashed line in Table~\ref{tab:ohmain_det} in Appendix~\ref{sec:co-overlays}. In general, we notice that the more confident detections at the top of Table~\ref{tab:ohmain_det} tend to show better spatial proximity to the CO emission, while less confident detections toward the bottom show increasing offsets between the OH absorption and the CO emission peak.

%%%%%
Figure~\ref{fig:HI_CO_OH} extends the OH absorption and CO emission comparison to the full sample, showing for each pointing the maximum CO $J=1\rightarrow0$ intensity within the field and the fraction of sight lines with \hi\ detections, estimated using the sample of \citet{Gupta2025galhi}.
Pointings with confident OH detections systematically show $I_{\rm CO}^{\rm max}\gtrsim1$\,K\,km\,s$^{-1}$ and higher \hi\ detection fractions. All but one have $n_{\rm HI}/n_{\rm tot}\gtrsim0.3$, roughly the top quartile of MALS pointings. In contrast, the bulk of MALS pointings have $I_{\rm CO}^{\rm max}\lesssim1$\,K\,km\,s$^{-1}$ (the typical MAD of the null map is 0.16\,K\,km\,s$^{-1}$), consistent with the noise maximum expected within a pointing, i.e. no detected CO. The only exception, J015710.52$+$001124.2, lies just above the detection threshold ($\log\mathcal{R}=1.1$) and is likely one of the false positives expected from Appendix~\ref{sec:oh-check}.

%%%%%%
At the level of individual sight lines, however, the correspondence is not one-to-one. For example, J005147.17$+$174711.5 shows a confident OH detection despite CO emission only at the edge of the pointing, highlighting the ability of OH to trace molecular gas in CO-faint environments.
In contrast, in Fig.~\ref{fig:poi_J0445} we also show the pointing corresponding to the sight line J045541.91$+$185010.9, which has CO contours passing over the bright central source, yet no OH absorption is detected down to $\log N(\rm OH) \le 13.6$ at 0.99 confidence (estimated using formalism presented in Section~\ref{sec:cf}). Interestingly, there are two OH candidates near the boundary of the pointing and CO emission with $\log\mathcal{R}\approx0.5$ (below the chosen threshold $\log\mathcal{R}=1$). The non-detection toward the central source likely reflects the patchy nature of the molecular ISM on scales below the $10\arcmin$ resolution of the CO map. Hence, the presence of CO emission does not guarantee OH absorption.
%

%%%%%%%
A detailed quantitative analysis of the CO-OH spatial correspondence is limited by the angular resolution and sensitivity. The pointing-level statistics in Fig.~\ref{fig:HI_CO_OH} are therefore more robust. Both CO emission and elevated \hi\ absorption incidence rate within pointing indicates the presence of cold H$_2$-bearing gas, including CO-dark envelopes, which may contain sufficient column densities of OH. Indeed, CO-bearing clumps are usually embedded in diffuse cold and molecular gas envelope, which can be traced by \hi\ and OH absorption, respectively. In line with this, the OH-bearing pointings from the detection sample correspond to the high end of both CO emission and \hi\ incidence rate distributions. This further supports the reliability of the confident OH detections presented in Table~\ref{tab:ohmain_det}.

%%%%%%%%%%%%%%%%%%%%%%%%%%%%%%%%%%%%%%%%%%%
\subsection{Satellite 1612\,MHz sample}
\label{sec:oh_sat}
%%%%%%%%%%%%%%%%%%%%%%%%%%%%%%%%%%%%%%%%%%%

The results of the Bayesian inference search of 1612\,MHz line are presented in Table~\ref{tab:ohsat_det}. As expected from the weakness of the line, we found far fewer candidates, with Bayesian evidence ratio (between absorption and non-absorption models) not exceeding $\log \mathcal{R} = 1.5$. Two (J152149.13$-$294244.8 and J032153.09$+$122114.0) out of nine candidates coincide with detections found based on the OH main lines, while most others do not exhibit OH main line absorption. Their spectra are presented in Figs.~\ref{fig:ohmain-hi} and \ref{fig:J032153.09+122114.0}, respectively. Interestingly, even for these candidates the ratio of the integrated optical depths are $\sim0.5 - 0.7$, which is well above the LTE expectation of 1:9 i.e., 0.11.  This may indicate that the OH satellite line is detached from LTE as in the case of 1612\,MHz emission line detection associated with an AGB star (J171749.17$-$622903.4; see Appendix~\ref{sec:oh-inspec}). 

%%%%%
Notably, in the extragalactic absorber at $z = 0.247$ towards PKS\,1413+135, \hi\ 21-cm and OH satellite lines are detected but among the main lines, 1667\,MHz absorption is barely detected \citep[][]{Combes23}. 
In MALS Galactic sample, J114335.39$-$261102.0 and J200322.53$-$023347.9 present similar properties, where \hi\ 21-cm and OH 1612\,MHz lines are clearly detected, but there is no statistical detection of main lines (see Figs.~\ref{fig:J114335.39-261102.0} and \ref{fig:J200322.53-023347.9}). Although in the former case, a weak 1667\,MHz may be present.

%%%%%%
For main lines we used $\log \mathcal{R}\approx 1$ for identifying credible detections.   Considering this, we cannot rule out a significant number of false detections in Table~\ref{tab:ohsat_det}.  Nevertheless, the detection of a few cases, all exhibiting non-LTE conditions, is a direct consequence of the fact that MALS spectra are not sensitive enough to detect satellite lines. These are expected to be weaker by an order of magnitude with respect to the 1667\,MHz line, so we detected only unusual cases. 
This signifies that forthcoming surveys require a wider dynamical range for detecting both main and satellite OH lines to uncover the full range of OH in different ISM conditions.

%%%%%%%%%%%%%%%%%%%%%%%%%%%%%%%%%%%%%
\subsection{$N$(OH) and $N$(\hi)}
\label{sec:oh_hi}
%%%%%%%%%%%%%%%%%%%%%%%%%%%%%%%%%%%%%

\begin{figure}
    \centering
    \includegraphics[trim = {0cm 0cm 0cm 0cm}, width=1.0\columnwidth,angle=0]{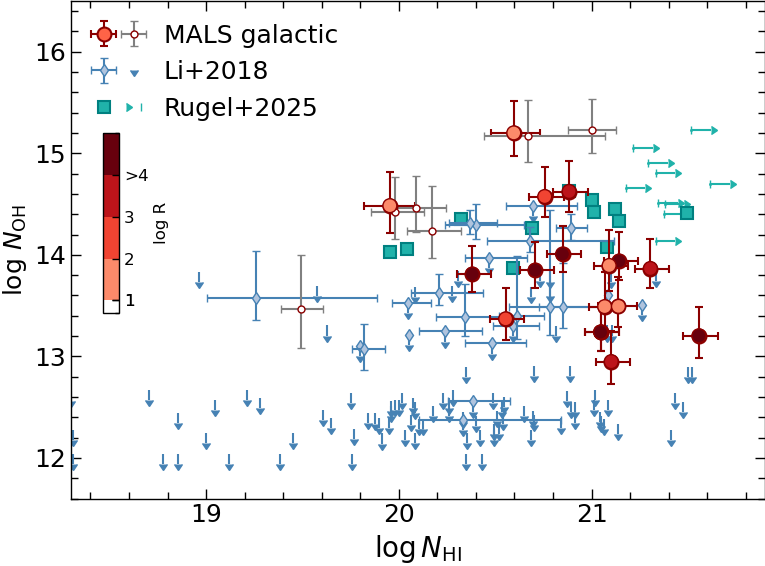}
    \caption{
    OH vs \hi\ column density. The red circles, cyan squares and blue diamonds correspond to the measurements from this work, \citet{Rugel2025} and \citet{Li18}, respectively. The color saturation of the red circles depicts the Bayesian evidence ratio. Arrows represent the upper and lower limits from literature data. 
    }
    \label{fig:ohhi}
\end{figure}

%%%
We estimated the OH column density using Eq.~\ref{eq:NOH}, assuming an excitation temperature $T_{\rm exc} = 4.5\pm2.1\,\rm K$ based on the recent observations by \citet[][]{Hafner2023} and covering factor $f=1$. 
The \hi\ column densities are derived from the optical depths measured by \citet{Gupta2025galhi} assuming $T_{\rm spin}=100\pm20$\,K and covering factor $f=1$.
The derived column densities are presented in Table~\ref{tab:ohmain_det}. 
In Fig.~\ref{fig:ohhi} we show the derived OH versus \hi\ column densities from the MALS sample and compare with previous measurements in the diffuse ISM, obtained by \citet{Li18} and \citet{Rugel2025}. The color of the points from our sample provided in Table~\ref{tab:ohmain_det} corresponds to the Bayesian evidence ratio. One can see that the most confident candidates have OH and \hi\ column densities consistent with previous observations from literature, while the less confident candidates preferentially tend to have slightly lower $N(\text{\hi})$ and/or higher $N(\rm OH)$. These include the strongest absorptions $\log N(\rm OH) \gtrsim 15$, which already tend to correspond to the values measured in the dense molecular clouds, along with most of the sight lines with $5<\mathcal{R}<10$.  Given the coarse velocity resolution and relatively low sensitivity for some off-axis targets, we note that these values should be considered with caution (see also Section~\ref{sec:oh_sat}).

%%%%
As in previous measurements, we do not find any significant relation between $N(\rm OH)$ and $N($\hi$)$, consistent with expectations from  OH chemistry modeling \citep[see e.g.,][]{Balashev2021}. This is because OH predominantly traces the H$_2$-bearing medium, forming only deeper inside the cloud than the \hi/H$_2$ transition. In contrast, \hi\ signal may come from the outer layer of the CNM clouds as well as from more diffuse surrounding WNM, diluting the relation between OH and \hi. Therefore, we examine dependencies of OH with tracers that more directly probe the H$_2$-bearing medium.
We note that assuming $f=1$ provides a lower limit on the true OH column density, since for 
$f<1$ the inferred column density scales as $N$(OH) $\propto$ $f^{-1}$. Moreover, the covering factors of OH and H\,{\sc i}\ absorbing gas need not be equal because OH traces denser, more compact clumps than the surrounding diffuse \hi, which could introduce a systematic bias in the $N$(OH)/$N$(\hi) relationship, though a detailed investigation of this effect is beyond the scope of this paper.
%

%%%%%%%%%%%%%%%%%%%%%%%%
%%%%%%%%%%%%%%%%%%%%%%%%

%%%%%%%%%%%%%%%%%%%%%%%%%%%%%%%%%%%%%
\subsection{$N$(OH) and $A_V$}
\label{sec:av}
\begin{figure}
    \centering
    \includegraphics[trim = {0cm 0cm 0cm 0cm}, width=1.0\columnwidth,angle=0]{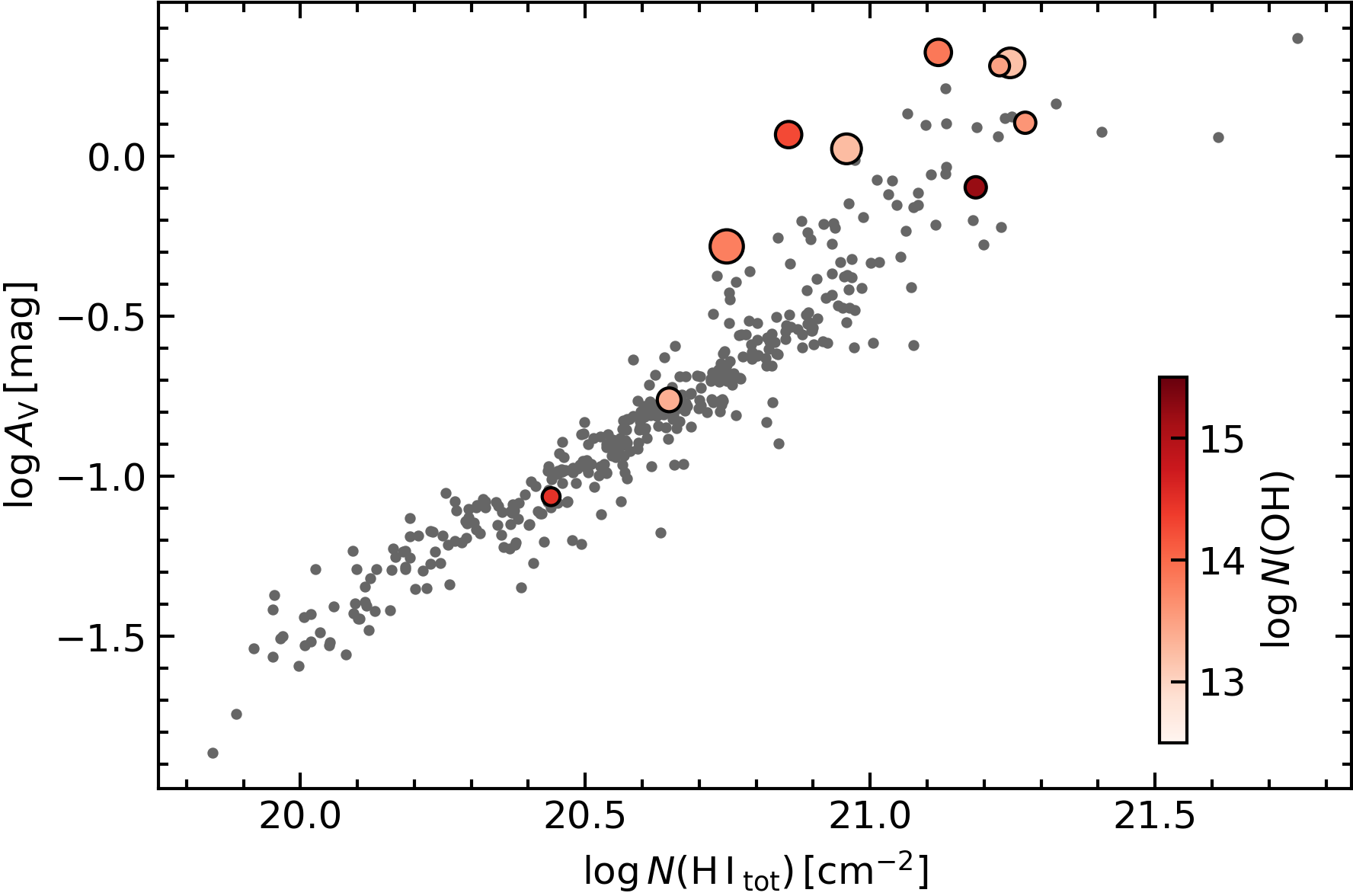}
    \caption{
    \hi\ column densities versus $A_{\rm V}$ for the central sight lines. The red and gray symbols correspond to the sight lines toward which OH has been detected and non-detected, respectively. The color-scale and size of the red points correspond to the OH column density and the probability of OH detection provided in Table~\ref{tab:ohmain_det}, respectively. 
    }
    \label{fig:ohav}
\end{figure}

%%%%%%%%%%%%%%%%%%%%%%%%%%%%%%%%%%%%%
One of the possible and easily available tracers of the H$_2$-bearing medium is the A$_{\rm V}$ excess -- estimated from the sight lines at high \hi\ column densities having A$_{\rm V}$ higher than expected from the linear relation between A$_{\rm V}$ and $N($\hi$)$ from the lower column density end. In Fig.~\ref{fig:ohav} we plot A$_{\rm V}$ -- $N$(\hi$_{\rm tot}$) measurements from \citet{Gupta2025galhi} for central sight lines. 
The majority of the OH detections and the most reliable ones are attributed to high A$_{\rm V}$, with a clear excess. We note that by construction, $N($\hi$)$ and A$_{\rm V}$ correspond to averaged values, since they both were obtained using data with coarse spatial resolution, while OH absorption is measured along pencil-beam sight lines. Therefore, one-to-one comparison between the presence of OH and the A$_{\rm V}$ excess can be done only in a statistical sense, taking into account the covering fraction of the OH-bearing gas.

%%%%%%%%%%%%%%%%%%%%%%%%%%%%
%%%%%%%%%%%%%%%%%%%%%%%%%%%%%

%%%%%%%%%%%%%%%%%%%%%%%%%%%%%%%%%%%%%%%%%%%
\subsection{OH covering fraction}
\label{sec:cf}
%%%%%%%%%%%%%%%%%%%%%%%%%%%%%%%%%%%%%%%%%%%

To accurately estimate the covering fraction of the OH-bearing gas, we need to account for two factors: (i) the presence of OH towards a given sight line is accidental, and can be conveniently described by the column density distribution function, which is a straightforward quantity in absorption analysis, and (ii) different sight lines within the MALS survey, or a particular MALS pointing, have different sensitivities, i.e., their ability to detect OH absorption line depends on the column density threshold corresponding to the noise level in the spectra. Fortunately, the extensive statistics from MALS allow us to account for both. 

%%%
A naive first estimate of the covering fraction can be obtained by dividing the number of detections by the number of sight lines i.e., $15/19130 \approx 10^{-3}$. However, this is a conservative lower limit because it assumes that all the sight lines have equal sensitivity. In reality, the detections are biased toward the brightest background sources which, due to their lower spectral rms, allow detection of lower OH column densities. Since the column density distribution is a steeply declining function, higher sensitivity significantly increases the probability of intersecting detectable amounts of OH-bearing gas. Therefore, many of the 19130 non-detections simply reflect insufficient sensitivity rather than the absence of OH, and the true covering fraction is higher than $10^{-3}$. Lowering the Bayesian evidence threshold would recover more real `weaker' detections but simultaneously increase false positives. Moreover, the Bayesian evidence ratios depend on the chosen priors, which already embed information about the OH column density distribution, making it difficult to characterize the false positive rate correctly. Therefore, a robust estimate of the covering fraction requires a proper account of the varying sensitivity across sight lines, as we describe below.

%%%%%%%%
In many cases, the column density distribution of the species can be described by a power law
\begin{equation}
\label{eq:cddf}
f(N) = C\frac{(\beta - 1)}{N_0} \left(\frac{N}{N_0}\right)^{-\beta},
\end{equation}
where we introduce the normalization column density $N_0$. For the power law index, we use $\beta\approx1.3$ based on the H$_2$ statistics and OH/H$_2$ chemical connection \citep{Balashev2021}. In some cases, the column density distribution is described as a gamma-function -- power-law with exponential cutoff -- but this cutoff is usually located at higher column densities, and since $\beta>1$ the change in the shape at large column densities does not significantly impact the detection statistics. The normalization constant, $C$ is chosen to correspond to covering fraction at $N_0$, i.e., the probability to intersect the gas with OH column density higher than $N>N_0$ is,
\begin{equation}
\label{eq:cf}
\mathbb{P}(N\ge N_0) = \int\limits_{N_0}^{\infty} f(N) dN = \int\limits_{N_0}^{\infty}C\frac{(\beta - 1)}{N_0} \left(\frac{N}{N_0}\right)^{-\beta}dN  = C.
\end{equation}
Given the MALS sensitivity function we take $N_0 = 10^{12}$\,cm$^{-2}$. Obviously, with a given slope $\beta$, and measured $C$ at $N_0$ one can easily estimate $C$ at any $N$ using Eq.~\ref{eq:cf} as long as the power law approximation is valid. We note that this approximation is expected to break down below OH column densities of $\sim10^{11}$\,cm$^{-2}$. Under typical ISM conditions with OH/H$_2$ abundance ratio of $\sim 10^{-7}$, this corresponds to $\log N ({\rm H_2})\sim 18$, below which the H$_2$ column density distribution abruptly drops \citep[see e.g.,][]{Balashev18} due to \hi/H$_2$ transition.

%%%%
We then estimated the covering factor $C$ using the MALS data as follows. First, for each sight line, we estimated the limiting OH column density based on its continuum-to-noise ratio. To do this, we assumed that most OH profiles span one spectral pixel of the MALS spectra (i.e., $\Delta v = 4.7$\,\kms), which is well motivated by the association of OH with CNM and the fact that the profiles of the most confidently detected candidates are narrow (see Fig.~\ref{fig:ohmain-hi}). In general, we do not know the position of the OH line, but it should be associated with \hi\ gas. We therefore used \hi\ emission profiles to estimate the likely velocity range of OH absorption, rather than \hi\ absorption, since we want to constrain the global covering factor of OH-bearing gas, without introducing a prior on its detection. The \hi\ emission profiles typically cover several pixels of the MALS spectrum (see Fig.~\ref{fig:ohmain-hi}). 

%%%%%%
For each sight line $l$, we determined the limiting OH column density, $\tilde N_l$, using extreme value statistics. Using the continuum constructed from a wide spectral region, we defined the residuals as
\begin{equation}
r_i \equiv \frac{(1 - \tilde f_i)}{\tilde \sigma_i},
\end{equation}
where $\tilde f_i$ and $\tilde \sigma_i$ are source flux and pixel uncertainty at velocity pixel $v_i$, respectively, both normalized by the continuum. For the chosen $n_{l}$ spectral pixels, in the case of no absorption and Gaussian statistic (i.e., $r_i$ follows Gaussian distribution with zero mean and unit dispersion), the value of the lowest residual at a chosen significance level $\alpha$ is 
\begin{equation}
\Delta_{\alpha}(n_l) = \Phi^{-1}\left((1 -\alpha)^{\frac1{n_l}}\right),
\end{equation}
where $\Phi^{-1}$ is the quantile function of standard normal distribution. 
The limiting integrated optical depth is then estimated as $\tilde \tau_{\rm int} \approx \Delta_{\alpha}(n_l) \Delta v \langle\tilde\sigma_i\rangle_{n_l}$, where $\langle\tilde\sigma_i\rangle_{n_l}$ is the mean relative rms within the considered $n_l$ spectral pixels of sight line $l$. 
This definition works only in the optically thin regime, which is appropriate for the typical OH column densities expected here. We therefore considered only sight lines for which $\Delta_{\alpha}(n_l) \langle\tilde \sigma_i\rangle_{n_l} < 1$. Using $\tilde \tau_{\rm int}$ and Eq.~\ref{eq:NOH} (assuming $T_{\rm exc} =4.5$\,K and $f=1$) we determined the limiting column density for each $l$ sight line, $\tilde N_{l}$, satisfying the optically thin regime. 
Finally, using $\Delta_{\alpha}(n_l)$ we determined the detection sample, $S_{\rm det}$, as sight lines where at least one pixel within the selected velocity range satisfies $\tilde f_i < 1 - \Delta_{\alpha}(n_l) \langle\tilde \sigma_i\rangle_{n_l}$. The non-detection sample, $S_{\rm non}$, comprised all remaining sight lines in the subset. We note that these samples are defined at a chosen significance level $\alpha$, which can be roughly considered as the false positive detection rate.

%%%%%
Given the significance level $\alpha$ and limiting column densities, we constructed the likelihood function for joint detection and non-detection sample using Bernoulli distribution, which is appropriate for binary detection/non-detection outcomes, as follows
\begin{equation}
\label{eq:like}
%\mathcal{L} = \prod\limits_{l} (p_l + (1-p_l) \alpha)^{X_l} ((1-p_l)(1-\alpha))^{1-X_l},
\mathcal{L} = \prod\limits_{l} (p_l + \alpha)^{X_l} (1-p_l-\alpha)^{1-X_l},
\end{equation}
where $X_l \in\{0,1\}$ correspond to detection or non-detection statistics and 
\begin{equation}
p_l = C \left(\frac{\tilde N_l}{N_0}\right)^{1-\beta}
\end{equation}
is the probability to have column density higher than limiting $\tilde N_l$ that was used as the threshold for classifying each sight line as a detection or non-detection. 

%%%%%
The $\alpha$ plays an important role in Eq.~\ref{eq:like}  since it describes the expected false positive detection rate, which can be larger than the true detection rate. Within this formalism, the likelihood depends only on the covering factor, $C$, and using the detection and non detection sample we can write
\begin{equation}
\ln \mathcal{L}(C) = \sum\limits_{l \in S_{\rm det}} \ln (p_l + \alpha) + \sum\limits_{l \in S_{\rm non}} \ln (1-p_l - \alpha),
\end{equation}
We used maximum likelihood estimate, by solving $\partial \ln \mathcal{L} /\partial C = 0$, which converts to
\begin{equation}
\sum\limits_{l \in S_{\rm det}}\frac{\left(\frac{\tilde N_l}{N_0}\right)^{1-\beta}}{p_l + \alpha} = \sum\limits_{l \in S_{\rm non}} \frac{\left(\frac{\tilde N_l}{N_0}\right)^{1-\beta}}{1 - p_l - \alpha}.
\end{equation}
We solved this equation numerically and found the maximum likelihood estimate of the covering factor (at $N_0$) $\hat C=0.051$ and formal statistical uncertainty defined from the covariance to be $0.001$. We bootstrapped the original sample and estimated the systematic uncertainty to be $0.008$ (1$\sigma$), i.e., the error budget is dominated by this. 

As an aside, if one imagines that we have uniform sensitivity, with $\tilde N_l \equiv N_0$ we can get a simple asymptotic estimate of the covering factor as
\begin{equation}
\label{eq:asymptotics}
C = \frac{n_{\rm det}}{n} - \alpha,
\end{equation}
where $n_{\rm det}$ and $n$ are the number of detections and total sight lines within the optically thin limit, respectively. The first term in Eq.~\ref{eq:asymptotics} describes the simple algebraic rate from the detection sample, while the second term approximates the false detection rate in the detection sample. For the significance level of $\alpha = 0.01$ we formally get 86 detections and 4737 non-detections. Therefore, if we do not take into account sensitivity, we significantly underestimate the covering factor using the asymptotic Eq.~\ref{eq:asymptotics}.

\begin{figure}
    \centering
    \includegraphics[trim = {0cm 0cm 0cm 0cm}, width=1.0\columnwidth,angle=0]{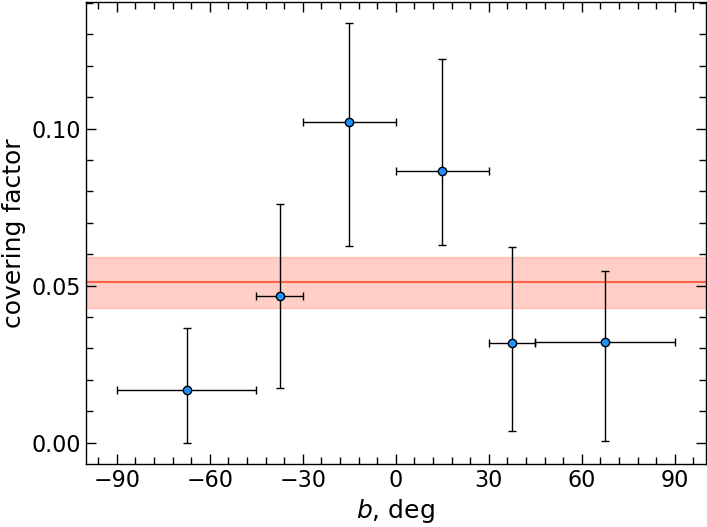}
    \caption{
    The covering fraction of OH absorption in MALS as a function of the Galactic latitude. The covering factors are obtained for $\log N (\rm OH) > 12$. The blue error bars and red stripe indicate the estimates in the relevant latitude bins and the whole MALS footprint, respectively.
    }
    \label{fig:cf_latitude}
\end{figure}

In Fig.~\ref{fig:ohmap} one can notice that the detections tend to have low Galactic latitude consistent with naive expectations from Galactic disk model. Nevertheless, since MALS does not have uniform sky coverage in the Galactic coordinates, we explored the dependence of the covering factor on the Galactic latitude by splitting the full sample into six latitude bins containing similar numbers of probed sight lines, and performed a covering factor estimation for each bin. The results of this binning are presented in Fig.~\ref{fig:cf_latitude}. There is an evident decrease of the covering factor with increasing latitude. Given the limited statistics and sparse MALS coverage at $|b| < 15{\degree}$, this provides qualitative evidence of the latitudinal dependence of the OH covering fraction.

%%%%%%%%%%%%%%%%%%%%%%%%%%%%%
%%%%%%%%%%%%%%%%%%%%%%%%%%%%%

\begin{figure}
    \centering
    \includegraphics[trim = {0cm 0cm 0cm 0cm}, width=1.0\columnwidth,angle=0]{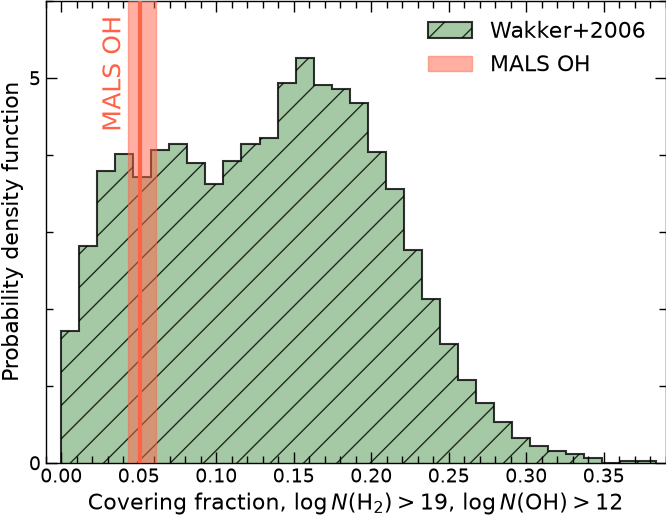}
    \caption{
    Comparison of the covering fraction of the cold diffuse gas estimated from OH absorption (red lines and shaded stripe) and direct H$_2$ ultraviolet absorption (green histogram). The OH estimate is obtained using MALS statistics for $\log N (\rm OH) > 12$, and the H$_2$ estimate is obtained by bootstrapping of the FUSE sample of high Galactic latitude sight lines from \citet[][]{Wakker2006}.  
    }
    \label{fig:h2comparison}
\end{figure}

Returning to the formal analysis, the sky-averaged covering factor $C=0.051\pm0.008$ represents the fraction of the sky that has OH column densities $\log N(\rm OH) > 12$. This corresponds to  $\log N(\rm H_2)\gtrsim 19$, which are the typical values where \hi/H$_2$ transition occurs. However, the exact OH/H$_2$ conversion factor depends on the geometry of the cloud and the local physical conditions, and can easily be a factor of two higher \citep{Balashev2021}. We can compare our estimate of the covering fraction with the statistics based on the H$_2$ absorption survey toward high Galactic latitude obtained by FUSE \citep{Wakker2006}. Out of the 73 studied sight lines, they detected 12 and 4 sight lines with H$_2$ column densities $\log\,N({\rm H_2}) > $ 19 and 19.3, respectively, illustrating the sensitivity of the sample size to the adopted threshold. To compare with our estimate from OH, we bootstrap the H$_2$ sample assuming $N({\rm OH}) = 10^{-7 + G(\mu=0, \sigma=0.3)} N({\rm H_2})$, i.e., a 0.3\,dex Gaussian dispersion on the OH/H$_2$ column density scaling relation. We found that both estimates are formally consistent with each other (see Fig.~\ref{fig:h2comparison}), providing independent validation of our result. It is worth mentioning that due to the large number of probed sight lines, MALS allows us to obtain an estimate with significantly smaller uncertainty than the H$_2$ survey, which contains only 73 sight lines. 
However, a direct one-to-one comparison of the derived covering factor values is complicated by systematic uncertainties arising from the OH/H$_2$ conversion factor, the different sky regions probed by the two surveys, and the fact that our estimates are mostly driven by higher column densities $\log N(\rm OH )>13$, corresponding to $\log N(\rm H_2) \gtrsim 20$ (see also Section~\ref{sec:phys}), meaning the two surveys probe different regions of the H$_2$ column density distribution.

%%%%%%%%%%%%%%%%%%%%%%%%%%%%%
%%%%%%%%%%%%%%%%%%%%%%%%%%%%%

%%%%%%%%%%%%%%%%%%%%%%%%%%%%%%%%%%%%%%%%%
\subsection{OH detection in stack}
\label{sec:stack}
%%%%%%%%%%%%%%%%%%%%%%%%%%%%%%%%%%%%%%%%

Stacking has proven to be a powerful tool to detect weak average signals buried in the noise. We tried different stacks of the dataset, which mainly differ by weighting of individual spectra. 
For all versions of the stack, each sight line was first shifted and rebinned to a common velocity grid, with the shift set to the pixel of maximum \hi\ optical depth, except for the stack of all sight lines where no shift was applied.
We then calculated the stack as weighted average:
\begin{equation}
    F(v) = \cfrac{\sum\limits_{i} \tilde f_i(v) \cfrac{w_i}{\tilde\sigma^2_i(v)}}{\sum\limits_{i} \cfrac{w_i}{\tilde\sigma^2_i(v)}}, 
\end{equation}
where $\tilde f_i(v)$ and $\tilde\sigma_i$ are the flux and uncertainty of each sight line, respectively, both normalized by the continuum of the source, and $w_i$ are weights described below. The continuum was obtained by fitting a fifth-order polynomial to the source flux in a $500\,\rm km\,s^{-1}$-wide window around the OH line,  using sigma clipping to iteratively reject pixels corresponding to possible absorption features. The weight, $w_i$, described below can be chosen differently to maximize the signal. 
\begin{equation*}
\begin{aligned}
    \begin{cases}
        & w_i = 1  \quad\quad\quad\quad\quad\quad\quad\quad\,\,\,\, \text{\rm uniform} \\
        & w_i(v) = (1 - \tilde f_{i, \rm HI}(v)) + \tilde\sigma_i \quad\, \text{\rm regularized \hi\ flux weighted} \\
        & w_i(v) = (1 - m_{i,\rm HI}(v)) + \tilde\sigma_i \quad \text{\rm regularized \hi\ model weighted} \\
        & w_i = A_V \quad\quad\quad\quad\quad\quad\quad\quad \text{\rm $A_V$ weighted} \\
        & w_i = A_V - A^{\rm trend}_V(N({\rm \hi}))  \quad\,\,\,\, \text{\rm $A_V$ excess weighted},
    \end{cases}
\end{aligned}
\end{equation*}
where $\tilde f_{i, \rm HI}(v)$ and $m_{i,\rm HI}(v)$ are the continuum normalized spectrum of \hi\ absorption, rebinned to the same velocity grid, and the model of the \hi\ profile \citep[see][]{Gupta2025galhi}, respectively.

Among the choice of weights, the simplest is the uniform weights, where each sight line contributes equally. However, since OH absorptions are aligned with \hi\ absorptions, both representing the cold phase of the neutral ISM, we also used \hi\ profile weights based either on the observed \hi\ spectrum i.e., $\tilde f_{i, \rm HI}(v)$, or the best-fit model to the profile i.e., $m_{i,\rm HI}(v)$. This weighting was regularized by an additional $\sigma_i$ term to suppress noise in the velocity bins  where absorption is weak or absent. Effectively, when \hi\ absorption is detected, the term $\sigma_i$ is smaller than the \hi\ line profile, ensuring that the weighting is dominated by the \hi\ absorption signal. Given that the presence of OH is tightly linked with the presence of $\rm H_2$, we additionally used weighting by either $A_V$ or the excess of $A_V$ over the linearly derived trend $A_V-N({\rm \hi})$. The latter is a better indicator of the amount of $\rm H_2$ since $A_V$ scales with dust abundance, and this excess should be proportional to the fraction of the hydrogen in molecular rather than atomic form.

%%%%%%%%%%%%%%%%%%%%%%%%%%%%
\begin{figure}
    \centering
    \includegraphics[trim = {0cm 0cm 0cm 0cm}, width=1.0\columnwidth,angle=0]{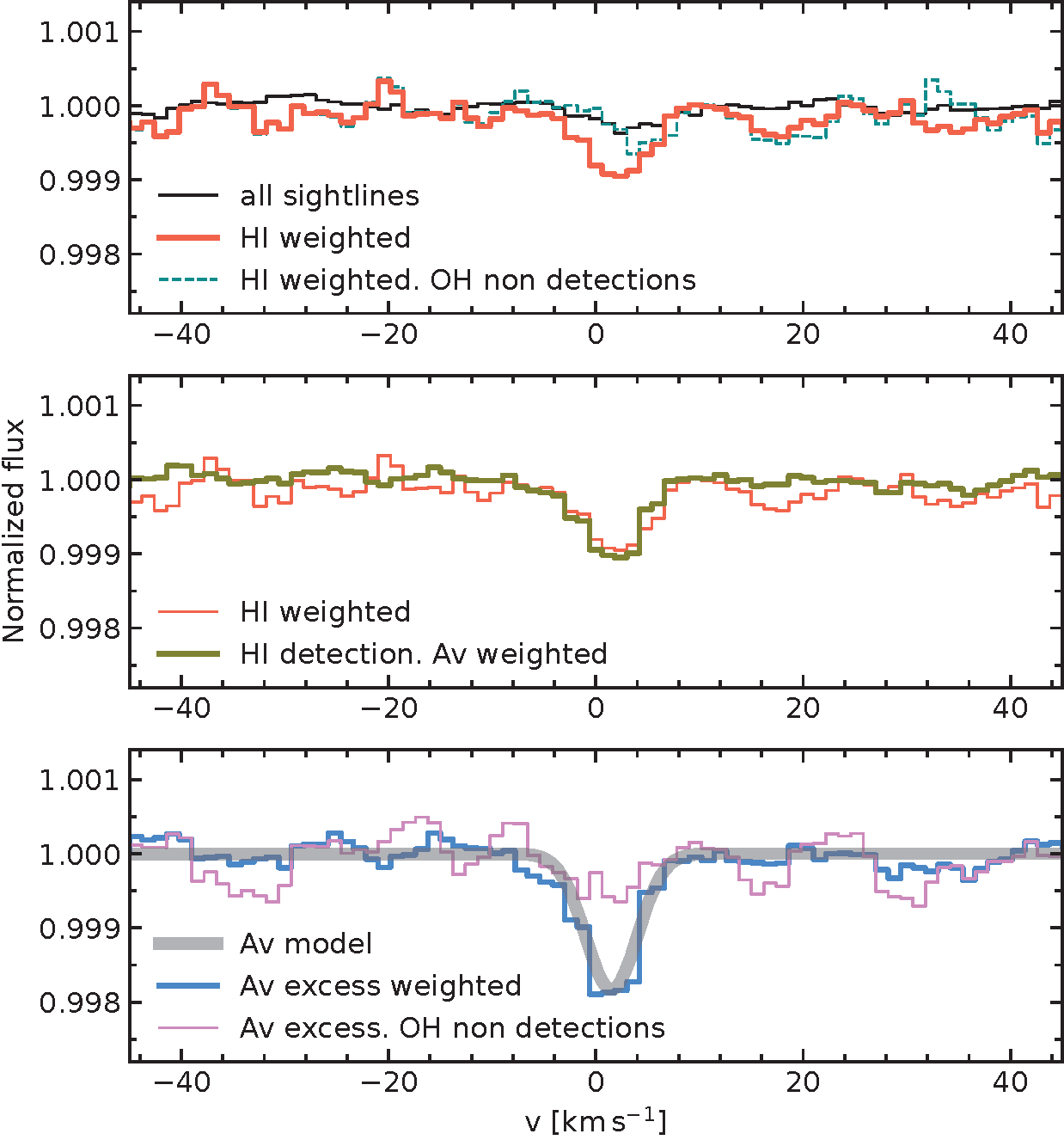}
    \caption{
    The different versions of the OH stack. {\sl Top panel:} the black line shows the stack of all sight lines with uniform weighting, while the red line shows the \hi-flux weighted stack obtained after shifting the individual spectra. 
    The blue dashed line shows a version of the \hi-flux weighted stack obtained after excluding the OH detections listed in Table~\ref{tab:ohmain_det}. {\sl Middle panel:} the red and olive lines show the stack of sight lines, where \hi\ absorption is detected, weighted by the \hi\ absorption profile (same as in the top panel) and $A_V$ value, respectively. {\sl Bottom panel:} the blue and pink lines represent the stack of sight lines weighted by $A_V$ excess, based on the whole sample and where OH is not individually detected, respectively. The thick gray line corresponds to the modeled OH stack profile using $A_V -\rm H_2 - OH$ association and a covering factor of 0.05 (see Section~\ref{sec:stack}).
    }
    \label{fig:ohstack}
\end{figure}
%%%%%%%%%%%%%%%%%%%%%%%%%%%%

In Fig.~\ref{fig:ohstack} we plot different versions of the OH stack. The stack with uniform weighting and no velocity shift applied shows only a very tentative OH signal (black line, top panel). In contrast, the regularized \hi\ flux weighted stack, obtained by shifting spectra to the position of the maximum optical depth of \hi\ profiles, clearly features OH signal (red line, top panel). We checked that this signal is predominantly driven by the OH detections listed in Table~\ref{tab:ohmain_det}, since once we removed these the stacked signal becomes roughly three times smaller (blue dashed line, top panel). 
We also checked that the \hi-model and \hi-flux weighting produce essentially the same stacked signal. In the middle panel we compare the \hi-weighted stack with $A_V$-weighting (with the velocity shift), indicating that while the OH stack does not change dramatically, it becomes more prominent relative to the noise for $A_V$-weighting. This is expected due to the choice of weighting and the fact that the strength of \hi\ absorption (column density or spin temperature) and $A_V$ are correlated \citep{Gupta2025galhi}. The most significant increase of the OH signal is seen in the $A_V$ excess weighted stack (bottom panel), for both the whole sample and the sample without individual OH detections, implying the close association of the presence of OH with $\rm H_2$. 

%%%%%%
To interpret this latter signal we modeled it using the $A_V$--OH--H$_2$ association. First for each sight line we estimated the H$_2$ column density by subtracting the measured \hi\ column density in emission from the total hydrogen column density inferred from the observed $A_V$. We then derived the OH column density using OH/H$_2 = 10^{-7}$, the average value measured in diffuse clouds. We then constructed the stack signal using the same Gaussian profile for all sight lines, with a velocity dispersion $\sigma=2.5$\,\kms\ and a mean velocity $+1.5$\,\kms, and adjusted the covering factor to be 0.05 to visually match the observed stack signal. This small positive velocity shift of $\approx+1.5$\,\kms\ could be due to \hi\ self-absorption (HISA) at the molecular gas velocity, or a particular sky distribution of the detections within the MALS footprint. CO and \hi\ peaks are known to be systematically shifted in other galaxies, but interpreting this shift here is difficult, given the limited velocity binning and non-uniform sky coverage of the MALS sight lines. We find that a covering factor of $0.05$ (which implies that only 5 per cent of the sight lines with $A_V$ excess are covered on average with OH absorption) that we used, agrees well with the incidence rate derived in the previous section using detection statistics. 

%%%%%%%%%%%%%%%%%%%%%%%%%%%%%%%%%%%%%
\subsection{OH abundance and physical conditions}
\label{sec:phys}
%%%%%%%%%%%%%%%%%%%%%%%%%%%%%%%%%%%%%

To assess the physical conditions we used the modeling of diffuse ISM using a code based on \citet{Balashev2021} with modifications regarding the thermal balance described by \citet{Balashev2024}. Briefly, the code is based on static thermo-chemical calculations of the abundances of the species in the diffuse medium for a fixed hydrogen number density and metallicity, exposed to the cosmic rays and UV field. The code is focused on the cold neutral medium and therefore resolves the \hi/H$_2$ transition using an analytical prescription based on \citet{Sternberg2014, Bialy2016}. Consequently, it assumes a plane-parallel geometry of the cloud, illuminated by UV field from one side. The chemical and thermal balances consider the typical reactions, including ionization balance, cooling and heating terms in the ISM, and are solved jointly using an iterative procedure. The comparison between code calculations for a fixed temperature and with the actual thermal balance solution is shown in Fig.~\ref{fig:thermal}. The OH chemistry is calculated before onset of the CO in the medium, so it is applicable only for the diffuse medium \citep{Balashev2021}. 
The CO emission found in the pointings of the OH detections (Section~\ref{sec:global}) suggests that some sight lines, particularly in the fields with the strongest CO emission (e.g., J152225.50$-$293624.8 and J183211.11$-$203947.9), may approach this limit. However, since $I_{\rm CO}^{\rm max}$ is a field maximum rather than CO emission along the sight line, it does not necessarily imply CO-bearing dense gas toward the background sources.

For the considered modeling the parameters are the following: $n_{\rm tot}$ -- hydrogen number density; $Z$ -- metallicity; $I_{\rm UV}$ -- UV field strength; $\zeta$ -- primary hydrogen cosmic ray ionization rate, and $\rm DTG$ -- dust to gas ratio. The latter is important for the calculation of the ionization balance in the medium, since it determines the abundance of polycyclic aromatic hydrocarbons, whose charge state abundances are solved as part of the chemical calculation.  

%%%%%%
We derived constraints on the physical parameters in the medium using a likelihood function that compares the observed integrated \hi\ optical depth and derived OH column densities with modeled values. Since we solve thermal balance, it is more consistent to use the integrated \hi\ optical depth as it encodes both the \hi\ column density and the kinetic temperature; the latter is assumed to be equal to the spin temperature. For OH the excitation is more complex, and therefore we used OH column densities derived assuming an excitation temperature of $4.5\pm2.1$\,K.  We used Bayesian inference and sampled the posterior distribution function with the Affine-invariant sampler provided by the \textsc{emcee} package. We used priors on  $\log Z$ to be Gaussian, $G(0, 0.2)$ -- with mean and sigma to be 0 (i.e., solar metallicity) and 0.2, respectively. We also tied $\rm DTG$ to scale quadratically with metallicity with 0.5 dispersion in log. Initially, we assumed flat priors on $\log n_{\rm tot}$, $\log I_{\rm UV}$ and  $\log \zeta$ within ranges appropriate for CNM conditions. We additionally used the condition of the CNM formation from \citet{Wolfire03}, which ties the main parameters, to ensure that the modeled medium corresponds to CNM. 

%%%%%%%%%%%%%%%%%%%%%%%%%%%%%%%%
\begin{figure*}
    \centering
    \includegraphics[trim = {0cm 0cm 0cm 0cm}, width=1.0\textwidth,angle=0]{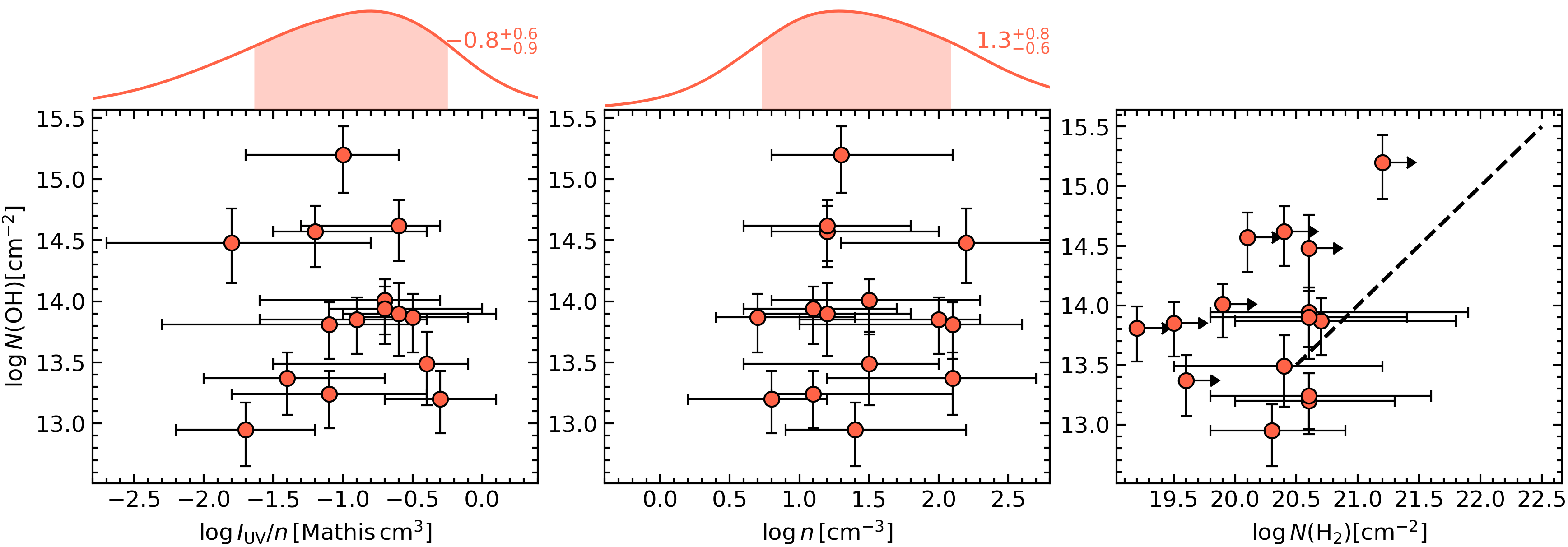}
    \caption{
    OH column densities versus derived physical parameters $I_{\rm UV}/n$ (\textit{left}), $n$ (\textit{middle}) and H$_2$ column density (\textit{right}). The values are provided in Table~\ref{tab:oh_phys}. The upper panels show the kernel density estimate of the marginalized distributions for each parameter. The black dashed line in the right panel indicates the OH/H$_2$ ratio $10^{-7}$.
    }
    \label{fig:oh_phys}
\end{figure*}
%%%%%%%%%%%%%%%%%%%%%%%%%%%%%%%%

%%%%%%
The posteriors for the initial run are shown in Fig.~\ref{fig:J1521_phys}. One can see that, since we have only two observables, the key ISM parameters for which we used less informative priors are significantly degenerate, and we cannot say anything definitive about them independently. However, we noted that the combination of $I_{\rm UV} / n$ is well constrained in most sight lines and found to be $\log I_{\rm UV} / n$ between $\approx -2$ and $-0.5$ (where $I_{\rm UV} / n$ in units Mathis\,cm$^3$), which is consistent with typical values measured in CNM \citep{Bialy2016,Klimenko2024,Kosenko2024}. We further re-ran the sampling assuming more informative priors on the UV field and CRIR to be $\log I_{\rm UV} \propto G(0, 0.3)$ and $\log \zeta \propto G(-16, 1)$, which are reasonable values for the solar vicinity, studied by MALS sight lines. This allows us to break the degeneracy and to get reasonable estimates on the number densities (see an example in Fig.~\ref{fig:1521_phys_uv}). 
Using this, for the detection sample, we infer number densities between 10 and 100 cm$^{-3}$. We also obtained constraints on H$_2$ column densities, typically $\log N(\rm H_2) > 21$. However, these values should be considered with caution, since they are obtained assuming simplistic one-layer plane-parallel medium. Moreover, for the sight lines with $\log N(\rm OH) \gtrsim 14$ -- the exact value strongly depends on the physical conditions -- the diffuse chemistry formalism used here may no longer apply. At these column densities, the gas starts to represent CO-bearing dense gas, where OH can be significantly enhanced at the same H$_2$ column \citep{Balashev2021}. We did not find any significant correlation between the derived parameters except some trivial ones inherited from the priors and the assumed parameter dependencies. The obtained parameters are shown in Table~\ref{tab:oh_phys} and their dependencies on the OH column densities are presented in Fig.~\ref{fig:oh_phys}.

%%%%%%%%%%%%%%%%%%%%%%%%%%%%%%%%
\setlength{\tabcolsep}{0.1cm}
\begin{table}
\caption{Derived physical conditions for the OH detection sample.}
\vspace{0cm}
\begin{center}
\begin{tabular}{cccc}
\hline
\hline
Sight line   &  $\log \left(I_{\rm UV} / n_{\rm tot}\right)$  &  $\log n_{\rm tot}$ $\dag$ & $\log N(\rm H_2)$  $\ddagger$    \\
& $[\rm Mathis\,cm^{3}]$ & $\rm cm^{-3}$ &  $ [\rm cm^{-2}]$ \\
\hline          
J005147.17+174711.5 & $-1.4^{+0.7}_{-0.6}$ & $2.1^{+0.6}_{-0.9}$ & $>19.6$   \\
J015710.52+001124.2 & $-1.8^{+1.0}_{-0.9}$ & $2.2^{+1.0}_{-0.9}$ & $>20.6$   \\
J032153.09+122114.0 & $-0.3^{+0.4}_{-0.4}$ & $0.8^{+0.4}_{-0.6}$ & $20.6^{+0.7}_{-0.6}$   \\
J071115.10-383209.9 & $-1.0^{+0.4}_{-0.7}$ & $1.3^{+0.8}_{-0.5}$ & $>21.2$   \\
J152428.84-294107.5 & $-1.2^{+0.8}_{-0.3}$ & $1.2^{+0.8}_{-0.4}$ & $>20.1$   \\
J152225.48-293625.0 & $-0.9^{+0.5}_{-0.7}$ & $2.0^{+0.3}_{-1.0}$ & $>19.5$   \\
J152149.13-294244.8 & $-0.7^{+0.4}_{-0.9}$ & $1.5^{+0.8}_{-0.7}$ & $>19.9$   \\
J152101.05-294731.9 & $-0.6^{+0.3}_{-0.7}$ & $1.2^{+0.6}_{-0.6}$ & $>20.4$   \\
J155930.97+030448.2 & $-1.1^{+0.6}_{-1.2}$ & $2.1^{+0.5}_{-1.1}$ & $>19.2$   \\
J162115.62-361137.7 & $-0.4^{+0.3}_{-1.1}$ & $1.5^{+0.5}_{-0.9}$ & $20.4^{+0.8}_{-0.9}$   \\
J171456.58-055820.4 & $-0.5^{+0.4}_{-0.4}$ & $0.7^{+0.7}_{-0.3}$ & $20.7^{+1.1}_{-0.7}$   \\
J183339.97-210339.9 & $-1.7^{+0.5}_{-0.5}$ & $1.4^{+0.8}_{-0.5}$ & $20.3^{+0.6}_{-0.5}$   \\
J183211.11-203947.9 & $-0.7^{+0.7}_{-0.4}$ & $1.1^{+0.6}_{-0.5}$ & $20.6^{+1.3}_{-0.8}$   \\
J183206.63-204015.9 & $-0.6^{+0.7}_{-0.4}$ & $1.2^{+0.6}_{-0.5}$ & $20.6^{+0.8}_{-0.8}$   \\
J200608.43-022334.8 & $-1.1^{+0.7}_{-0.7}$ & $1.1^{+1.0}_{-0.3}$ & $20.6^{+1.0}_{-0.8}$   \\
\hline
\hline
\end{tabular}
%%%
\label{tab:oh_phys}
\end{center}
$\dag$, $\ddagger$: The number density and H$_2$ column density estimates, respectively, assuming an informative prior on $I_{\rm UV} \approx 1$ (see Section~\ref{sec:phys} for details).
\end{table}
%%%%%%%%%%%%%%%%%%%%%%%%%%%%%%%%

%%%%%%%%%%%%%%%%%%%%%%%%%%%%%%%%%%%%%%%%%%%%%%%%%%%%%%%%%%%%%%%%%%%%%%%%%%%%%%%%%%
\section{Summary and concluding remarks}    
\label{sec:summ}  
%%%%%%%%%%%%%%%%%%%%%%%%%%%%%%%%%%%%%%%%%%%%%%%%%%%%%%%%%%%%%%%%%%%%%%%%%%%%%%%%%%

We presented a blind search for Galactic OH 18-cm absorption lines toward 19130 radio sources brighter than 1\,mJy at 1.4\,GHz, using 390 L-band pointings of the MeerKAT Absorption Line Survey (MALS). Our main findings are as follows:

%%%%%%%%%%%%%%%%%%%%%%%%%%
\begin{enumerate}

\item We identified 15 confident detections of OH 1667\,MHz absorption ($\log \mathcal{R} > 1$).  The majority of confident detections are located toward sight lines with elevated CO-dark gas column densities, consistent with OH tracing the H$_2$-bearing medium.

\item We also identified 9 candidates of OH 1612\,MHz satellite line absorption ($\log \mathcal{R} > 0.7$), two of which coincide with OH main line detections and show 1612/1667\,MHz integrated optical depth ratios well above the LTE expectation of 1:9, indicating non-LTE excitation. Two further candidates have no OH main line counterpart detected.

\item Consistent with previous measurements, we do not find any significant dependence between $N(\rm OH)$ and $N$(\hi). This is likely because OH predominantly traces the H$_2$-bearing medium, forming only deeper in the cloud than the \hi/H$_2$ transition. 

\item Comparison with the CO $J=1\rightarrow0$ maps of \citet{Ghosh2024} shows that nearly all confident OH detections lie in pointings with detected CO emission ($I_{\rm CO}^{\rm max}\gtrsim1$\,K\,km\,s$^{-1}$) and a high incidence of \hi\ absorption ($n_{\rm HI}/n_{\rm tot}\gtrsim0.3$), whereas most MALS pointings show no detected CO. For individual sight lines the CO–OH correspondence is not one-to-one, reflecting the patchy molecular ISM on scales below the 10$\arcmin$ resolution of the CO maps.

\item The majority of OH 1667\,MHz detections are associated with high $A_{\rm V}$ sight lines with a clear $A_{\rm V}$ excess with respect to that expected from the linear $A_{\rm V}$--$N$(\hi) relation, further supporting the association of OH with H$_2$-bearing gas.

\item Using a maximum likelihood analysis that accounts for the varying sensitivity across sight lines, we estimate the covering fraction of OH-bearing gas at $\log N(\rm OH) > 12$ to be $C = 0.051 \pm 0.008$. This is formally consistent with the covering fraction of H$_2$-bearing gas derived from the FUSE UV absorption survey \citep{Wakker2006}. We note that the MALS result is based on much higher statistics, but probes the high column density end of the H$_2$ distribution function, corresponding to $\log N(\rm H_2) \gtrsim 19$.

\item We also note a qualitative decrease of the covering factor with Galactic latitude, consistent with the expected distribution of molecular gas in the Galactic disk.

\item Stacking of MALS sight lines reveals a clear OH absorption signal, which becomes increasingly prominent when weighted by $A_{\rm V}$ excess. Modeling of the $A_{\rm V}$ excess weighted stack using the $A_{\rm V}$--OH--H$_2$ association yields a covering factor of $\sim$0.05, consistent with the value derived from counting statistics.

\item Using a semi-analytical model of diffuse ISM physical conditions, we constrain the hydrogen number densities to be in the range 10--100\,cm$^{-3}$ for the OH detection sample, with H$_2$ column densities typically $\log N(\rm H_2) > 21$. The combination $I_{\rm UV}/n$ is well constrained between 0.01 and 0.3\,Mathis\,cm$^{-3}$, consistent with typical CNM conditions.

\end{enumerate}
%%%%%%%%%%%%%%%%%%%%%%%%%%

The large number of sight lines probed by MALS, combined with the sensitivity of MeerKAT, demonstrates the power of wide-field radio absorption line surveys for characterizing the diffuse molecular ISM. The methodology developed here for characterizing the diffuse molecular ISM through OH absorption will be directly applicable to the full MALS dataset spanning $0<z<2$ \citep[][]{Gupta17mals}. In the near future, the Deep Synoptic Array (DSA-2000), the Square Kilometre Array (SKA),  and the next-generation Very Large Array (ngVLA) will dramatically increase the number of available background sources, enabling a comprehensive census of OH-bearing gas across a wide range of Galactic environments and cosmic epochs.

%%%%%%%%%%%%%%%%%%%%%%%%%%%%%%%%%%%%%%%%%%%%%%%%%%%%%%%%%%%%%%%%%%%%%%%%%%%%%%%%%%%
\section*{Acknowledgements}
%%%%%%%%%%%%%%%%%%%%%%%%%%%%%%%%%%%%%%%%%%%%%%%%%%%%%%%%%%%%%%%%%%%%%%%%%%%%%%%%%%%

The MeerKAT telescope is operated by the South African Radio Astronomy Observatory, which is a facility of the National Research Foundation, an agency of the Department of Science and Innovation. 
The MeerKAT data were processed using the MALS computing facility at IUCAA (\url{https://mals.iucaa.in/releases}).
NG acknowledges NRAO for generous financial support for the sabbatical visit at Socorro during which a part of this work was done. 
The Common Astronomy Software Applications (CASA) package is developed by an international consortium of scientists based at the National Radio Astronomical Observatory (NRAO), the European Southern Observatory (ESO), the National Astronomical Observatory of Japan (NAOJ), the Academia Sinica Institute of Astronomy and Astrophysics (ASIAA), the CSIRO division for Astronomy and Space Science (CASS), and the Netherlands Institute for Radio Astronomy (ASTRON) under the guidance of NRAO.
The National Radio Astronomy Observatory is a facility of the National Science Foundation operated under cooperative agreement by Associated Universities, Inc.  This research has made use of the NASA/IPAC Extragalactic Database (NED), which is funded by the National Aeronautics and Space Administration and operated by the California Institute of Technology. 

%%%%%%%%%%%%%%%%%%%%%%%%%%%%%%%%%%%%%%%%%%%%%%%%%%%%%%%%%%%%%%%%%%%%%%%%%%%%%
\section*{Data Availability}
%%%%%%%%%%%%%%%%%%%%%%%%%%%%%%%%%%%%%%%%%%%%%%%%%%%%%%%%%%%%%%%%%%%%%%%%%%%%%

The MALS images, spectra and catalogs used in this article are publicly available as part of MALS DR3 at \url{https://mals.iucaa.in}.

%%%%%%%%%%%%%%%%%%%%%%%%%%%%%%%%%%%%%%%
%----------------- Bibliography and bibfile 
%------------------------------------------
\bibliographystyle{mnras}
\bibliography{mybib}

%%%%%%%%%%%%%%%%%%%%%%%%%%A
\appendix
%%%%%%%%%%%%%%%%%%%%%%%%%%%

\onecolumn

%%%%%%%%%%%%%%%%%%%%%%%%%%%
\section{MALS Galactic OH main and satellite absorption line catalog }
\label{sec:listpoint}
%%%%%%%%%%%%%%%%%%%%%%%%%%%

Here we present sample rows of the Galactic OH main (1667\,MHz) and satellite (1612\,MHz) absorption line catalogs in Tables~\ref{tab:ohmaincat} and \ref{tab:ohsat1cat}, respectively. The catalogs follow the same format as the MALS Galactic \hi\ absorption line catalog presented in \citet{Gupta2025galhi}, where detailed descriptions of all columns are provided; the table notes here give only brief descriptions. The only addition is Column\,37, which provides a boolean flag indicating whether an \hi\ absorption feature is detected in the spectrum based on visual inspection. The continuum images of the background radio sources and the absorption line spectra including non-detections are available as part of MALS DR3 at \url{https://mals.iucaa.in/}.

%%%%%%%%%%%
%\clearpage
\onecolumn
\begin{landscape}
\scriptsize
\setlength\LTcapwidth{\linewidth}
\begin{longtable}{lccccccccccccccccccccccccccccccccccc}
\caption{Sample rows from the MALS Galactic OH 1667\,MHz main line spectral line 
catalog, including both detections and non-detections. The complete catalog and the spectra are available as part of MALS DR3 at \protect\url{https://mals.iucaa.in/}.\label{tab:ohmaincat}}\\
\hline
Source\_name & Pointing\_id & Obs\_date\_U & Obs\_date\_L & Obs\_band & SPW\_id  & Maj\_cbeam & Min\_cbeam & PA\_cbeam \\ 
        &      &   &   &   &   &  (arcsec)  & (arcsec)  & (degree)  \\
   ~~~~~~~~(1)  &   (2)  & (3) & (4) & (5) & (6) & (7) & (8) & (9)  \\
    &  Peak\_pos  & Peak\_pos\_l & Peak\_pos\_b & Distance\_ponting & Scale\_factor & Peak\_flux & Spec\_rms &  Vis\_flag & n\_clump & n\_comp \\ 
            &   (J2000) &   (degree) &   (degree) & (arcmin) &   & (mJy\,beam$^{-1}$) & (mJy\,beam$^{-1}$) &  &  \\ 
        &   (10) &   (11) &   (12) & (13) & (14) & (15) & (16) & (17) & (18) & (19) \\  
    & tau\_max\_val & tau\_max\_+err  & tau\_max\_-err & pos\_max & tau\_int\_val & tau\_int\_+err & tau\_int\_-err & delta\_v \\
            &     &    &    & (\kms)  & (\kms) & (\kms) \\ 
        &   (20) &   (21) &   (22) & (23) & (24) & (25) & (26) & (27) \\ 
    & Comp\_1\_amp & Comp\_1\_amp\_+err & Comp\_1\_amp\_-err & Comp\_1\_pos & Comp\_1\_pos\_+err & Comp\_1\_pos\_-err & Comp\_1\_sigma & Comp\_1\_sigma\_+err & Comp\_1\_sigma\_-err & Vis\_flag\_HI \\
            &    &   &    & (\kms) & (\kms) & (\kms) & (\kms) & (\kms) & (\kms)   \\   
        &   (28) &   (29) &   (30) & (31) & (32) & (33) & (34) & (35) & (36) & (37)  \\ 
\hline
J000450.96-153224.1 & J000141.57-154040.6 & \ldots & 2020-07-16T01:40 & L & LSPW\_14G & 7.90 & 6.40 & -6.60 \\
 & 00:04:51.035 -15:32:23.502 & 77.5368 & -74.1461 & 46.35 & 0.112 & 13.17 & 6.69 & \ldots & \ldots & 0 \\
 & \ldots & \ldots & \ldots & \ldots & \ldots & \ldots & \ldots & \ldots \\
 & \ldots & \ldots & \ldots & \ldots & \ldots & \ldots & \ldots & \ldots & \ldots & \ldots \\
J000426.61-152106.7 & J000141.57-154040.6 & \ldots & 2020-07-16T01:40 & L & LSPW\_14G & 7.90 & 6.40 & -6.60 \\
 & 00:04:26.250 -15:21:06.739 & 77.7124 & -73.9390 & 44.31 & 0.143 & 4.54 & 4.98 & \ldots & \ldots & 0 \\
 & \ldots & \ldots & \ldots & \ldots & \ldots & \ldots & \ldots & \ldots \\
 & \ldots & \ldots & \ldots & \ldots & \ldots & \ldots & \ldots & \ldots & \ldots & \ldots \\
J000421.87-153741.2 & J000141.57-154040.6 & \ldots & 2020-07-16T01:40 & L & LSPW\_14G & 7.90 & 6.40 & -6.60 \\
 & 00:04:21.901 -15:37:40.951 & 77.0024 & -74.1350 & 38.71 & 0.242 & 22.57 & 3.23 & \ldots & \ldots & 0 \\
 & \ldots & \ldots & \ldots & \ldots & \ldots & \ldots & \ldots & \ldots \\
 & \ldots & \ldots & \ldots & \ldots & \ldots & \ldots & \ldots & \ldots & \ldots & \ldots \\
J000217.64-050134.4 & J000234.52-052324.5 & \ldots & 2020-08-19T20:40 & L & LSPW\_14G & 9.10 & 8.00 & -37.50 \\
 & 00:02:17.923 -05:01:30.477 & 92.8479 & -64.9864 & 22.23 & 0.626 & 7.78 & 1.09 & T & 1 & 1 \\
 & 0.6550 & 0.3030 & 0.2430 & -248.0 & 5.5033 & 2.4699 & 2.3178 & 23.5 \\
 & 6.1000 & 2.7000 & 2.4000 & -250.4 & 2.8 & 2.3 & 4.0 & 1.8 & 2.9 & F \\
J000617.70+171241.2 & J000647.35+172815.4 & \ldots & 2020-07-16T01:40 & L & LSPW\_14G & 11.70 & 6.40 & -4.30 \\
 & 00:06:20.131 +17:12:41.281 & 107.7963 & -44.3307 & 17.10 & 0.752 & 40.43 & 0.87 & T & 1 & 1 \\
 & 0.0929 & 0.0263 & 0.0258 & -147.7 & 0.5920 & 0.3752 & 0.4335 & 28.2 \\
 & 0.7000 & 0.3000 & 0.3000 & -146.6 & 2.0 & 2.5 & 4.9 & 2.6 & 5.9 & F \\
J001106.36-220824.1 & J001053.65-215703.9 & \ldots & 2020-08-19T20:40 & L & LSPW\_14G & 8.70 & 6.80 & -25.60 \\
 & 00:11:06.317 -22:08:23.871 & 59.3114 & -79.5710 & 11.71 & 0.870 & 4.52 & 0.86 & F & 1 & 1 \\
 & 1.4712 & 1.5913 & 0.6178 & 177.6 & 9.0689 & 7.5958 & 4.3524 & 18.8 \\
 & 6.0000 & 3.0000 & 2.6000 & 178.4 & 1.4 & 1.4 & 2.9 & 1.4 & 5.2 & F \\
\hline
\end{longtable}
\begin{tablenotes}
    \item Column\,1: MALS source name based on its right ascension and declination in SPW\_id image. Column\,2: MALS Pointing ID.  Column\,3-4: The date and time of observation.  Columns\,5-6: The observing band and SPW\_id.  Columns\,7-9: The restoring beam of the cube.  Column\,10: The right ascension and declination (J2000) of the pixel corresponding to flux density peak at which the spectrum is extracted.  Columns\,11-12: The Galactic longitude and latitude of the Peak\_pos.  Columns\,13-14: The distance of source from the pointing center and the scale factor applied for the primary beam correction.  Columns\,15-16: The peak flux density of the source and the unsmoothed spectral rms.  Column\,17: This is a boolean flag indicating whether an absorption feature based on visual inspection is detected in the spectrum.  Columns\,18-19: The number of clumps and the total number of Gaussian components fitted to the absorption features. Columns\,20-22: The peak optical depth and errors.  Column\,23: The position of {\tt tau\_max\_val} in the LSR frame.  Columns\,24-26: The integrated optical depth and errors. Column\,27: The velocity width containing 50\% of {\tt tau\_int\_val}.  Columns\,28-30 provide the amplitude of the $1^{st}$ Gaussian component and errors.  Columns\,31-33 provide the position of the $1^{st}$ Gaussian component relative to {\tt pos\_max} and errors.  Columns\,34-36 provide the width ($\sigma$) of the $1^{st}$ Gaussian component and errors.  For an absorption system, the columns\,28-36 are repeated {\tt n\_comp} times. Column\,37 provides a boolean flag indicating whether an H\,{\sc i}\ absorption 
feature is detected in the spectrum based on visual inspection.
\end{tablenotes}
\end{landscape}
\twocolumn

%\clearpage
%%%%%%%%%%%%

%%%%%%%%%%%
%\clearpage
\onecolumn
\begin{landscape}
\scriptsize
\setlength\LTcapwidth{\linewidth}
\begin{longtable}{lccccccccccccccccccccccccc}
\caption{Sample rows from the MALS Galactic OH 1612\,MHz satellite line spectral line 
catalog, including both detections and non-detections. The complete catalog 
and the spectra are available as part of MALS DR3 at \protect\url{https://mals.iucaa.in/}.\label{tab:ohsat1cat}}\\
\hline
Source\_name & Pointing\_id & Obs\_date\_U & Obs\_date\_L & Obs\_band & SPW\_id  & Maj\_cbeam & Min\_cbeam & PA\_cbeam \\ 
        &      &   &   &   &   &  (arcsec)  & (arcsec)  & (degree)  \\
   ~~~~~~~~(1)  &   (2)  & (3) & (4) & (5) & (6) & (7) & (8) & (9)  \\
    &  Peak\_pos  & Peak\_pos\_l & Peak\_pos\_b & Distance\_ponting & Scale\_factor & Peak\_flux & Spec\_rms &  Vis\_flag & n\_clump & n\_comp \\ 
            &   (J2000) &   (degree) &   (degree) & (arcmin) &   & (mJy\,beam$^{-1}$) & (mJy\,beam$^{-1}$) &  &  \\ 
        &   (10) &   (11) &   (12) & (13) & (14) & (15) & (16) & (17) & (18) & (19) \\  
    & tau\_max\_val & tau\_max\_+err  & tau\_max\_-err & pos\_max & tau\_int\_val & tau\_int\_+err & tau\_int\_-err & delta\_v \\
            &     &    &    & (\kms)  & (\kms) & (\kms) \\ 
        &   (20) &   (21) &   (22) & (23) & (24) & (25) & (26) & (27) \\ 
    & Comp\_1\_amp & Comp\_1\_amp\_+err & Comp\_1\_amp\_-err & Comp\_1\_pos & Comp\_1\_pos\_+err & Comp\_1\_pos\_-err & Comp\_1\_sigma & Comp\_1\_sigma\_+err & Comp\_1\_sigma\_-err & Vis\_flag\_HI \\
            &    &   &    & (\kms) & (\kms) & (\kms) & (\kms) & (\kms) & (\kms)   \\   
        &   (28) &   (29) &   (30) & (31) & (32) & (33) & (34) & (35) & (36) & (37)  \\ 
\hline
J000450.96-153224.1 & J000141.57-154040.6 & \ldots & 2020-07-16T01:40 & L & LSPW\_13G & 8.70 & 6.90 & -14.20 \\
 & 00:04:51.035 -15:32:23.502 & 77.5368 & -74.1461 & 46.35 & 0.118 & 14.14 & 4.60 & \ldots & \ldots & 0 \\
 & \ldots & \ldots & \ldots & \ldots & \ldots & \ldots & \ldots & \ldots \\
 & \ldots & \ldots & \ldots & \ldots & \ldots & \ldots & \ldots & \ldots & \ldots & \ldots \\
J000426.61-152106.7 & J000141.57-154040.6 & \ldots & 2020-07-16T01:40 & L & LSPW\_13G & 8.70 & 6.90 & -14.20 \\
 & 00:04:26.250 -15:21:06.739 & 77.7124 & -73.9390 & 44.31 & 0.149 & 4.73 & 3.93 & \ldots & \ldots & 0 \\
 & \ldots & \ldots & \ldots & \ldots & \ldots & \ldots & \ldots & \ldots \\
 & \ldots & \ldots & \ldots & \ldots & \ldots & \ldots & \ldots & \ldots & \ldots & \ldots \\
J000421.87-153741.2 & J000141.57-154040.6 & \ldots & 2020-07-16T01:40 & L & LSPW\_13G & 8.70 & 6.90 & -14.20 \\
 & 00:04:21.901 -15:37:40.951 & 77.0024 & -74.1350 & 38.71 & 0.249 & 21.68 & 2.90 & \ldots & \ldots & 0 \\
 & \ldots & \ldots & \ldots & \ldots & \ldots & \ldots & \ldots & \ldots \\
 & \ldots & \ldots & \ldots & \ldots & \ldots & \ldots & \ldots & \ldots & \ldots & \ldots \\
J000036.98-153042.6 & J000141.57-154040.6 & \ldots & 2020-07-16T01:40 & L & LSPW\_13G & 8.70 & 6.90 & -14.20 \\
 & 00:00:36.949 -15:30:46.006 & 74.8965 & -73.4400 & 18.47 & 0.725 & 5.39 & 0.90 & T & 2 & 1 \\
 & 0.8489 & 0.4153 & 0.3041 & 157.2 & 5.9550 & 2.9026 & 2.6095 & 19.4 \\
 & 6.4000 & 2.9000 & 2.2000 & 157.8 & 1.9 & 1.7 & 3.2 & 1.5 & 3.5 & F \\
J000338.55-044947.7 & J000234.52-052324.5 & \ldots & 2020-08-19T20:40 & L & LSPW\_13G & 9.50 & 8.60 & -36.30 \\
 & 00:03:38.615 -04:49:48.258 & 93.7619 & -64.9593 & 37.20 & 0.279 & 14.03 & 2.18 & F & 2 & 1 \\
 & 0.7181 & 0.3529 & 0.2715 & 101.0 & 10.5074 & 4.4898 & 4.6018 & 72.9 \\
 & 6.9000 & 3.1000 & 2.1000 & 95.0 & 8.0 & 7.0 & 11.0 & 5.0 & 5.0 & F \\
J001345.75+084938.0 & J001508.49+082803.8 & \ldots & 2020-07-16T01:40 & L & LSPW\_13G & 10.40 & 6.90 & -11.20 \\
 & 00:13:41.996 +08:50:21.216 & 107.3602 & -52.8878 & 29.72 & 0.448 & 45.44 & 1.54 & T & 1 & 1 \\
 & 0.0946 & 0.0320 & 0.0315 & 192.7 & 2.1175 & 0.8854 & 0.8591 & 87.4 \\
 & 2.2000 & 0.9000 & 1.0000 & 196.0 & 6.0 & 6.0 & 14.0 & 3.0 & 3.0 & F \\
\hline
\end{longtable}
\begin{tablenotes}
    \item The column descriptions are the same as in Table~\ref{tab:ohmaincat}. 
\end{tablenotes}
\end{landscape}
\twocolumn

%\clearpage
%%%%%%%%%%%%

\twocolumn
\normalsize

\clearpage

%%%%%%%%%%%%%%%%%%%%%%%%%%%%%%%%%%%%%
\section{Details of visual inspection and discrepant candidates}
\label{sec:oh-inspec}
%%%%%%%%%%%%%%%%%%%%%%%%%%%%%%%%%%%%%
\suppressfloats[t]

\begin{figure}
    \centering
    \vbox{
    \includegraphics[trim = {0cm 0cm 0cm 0cm}, width=0.4\textwidth,angle=0]{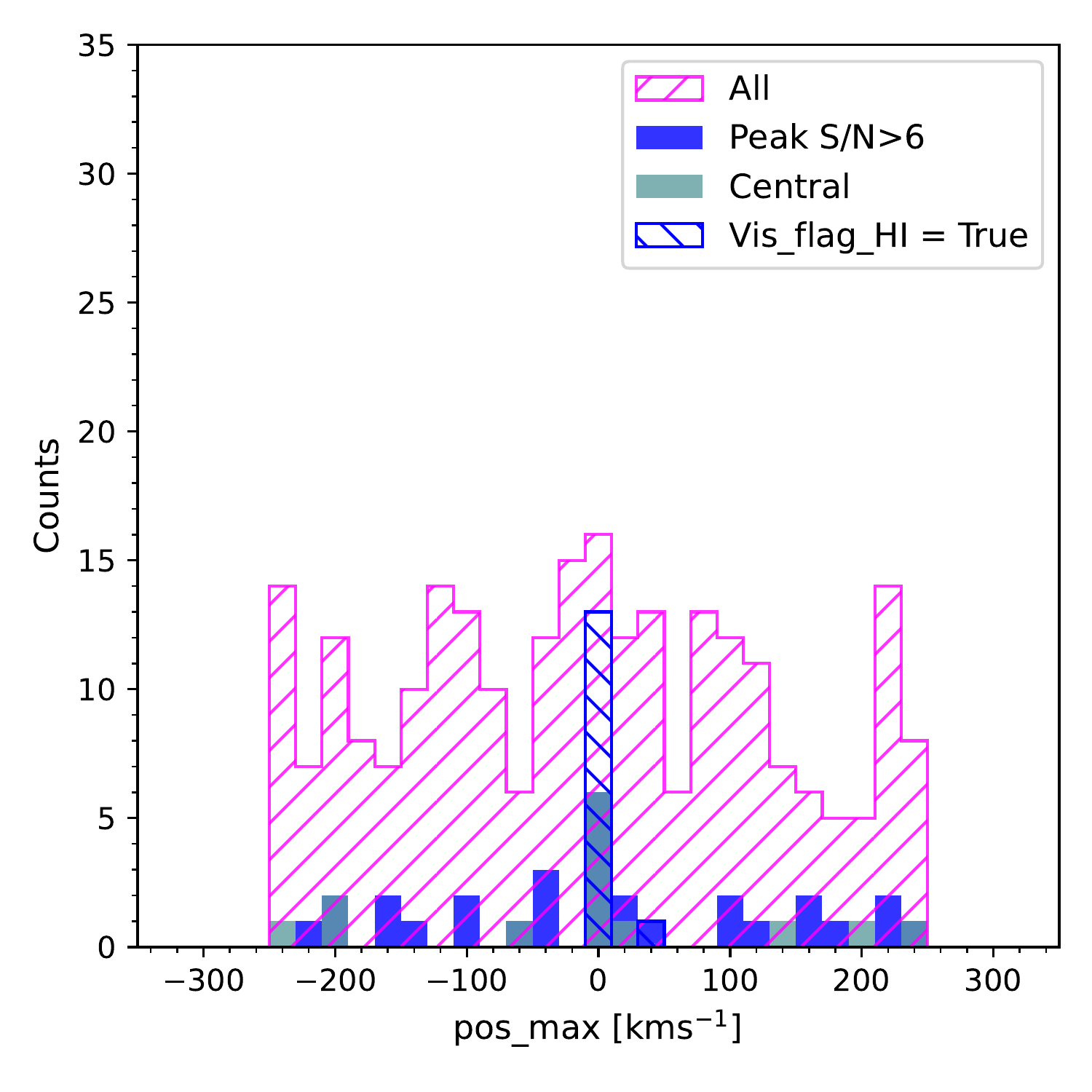}
    \includegraphics[trim = {0cm 0cm 0cm 0cm}, width=0.4\textwidth,angle=0]{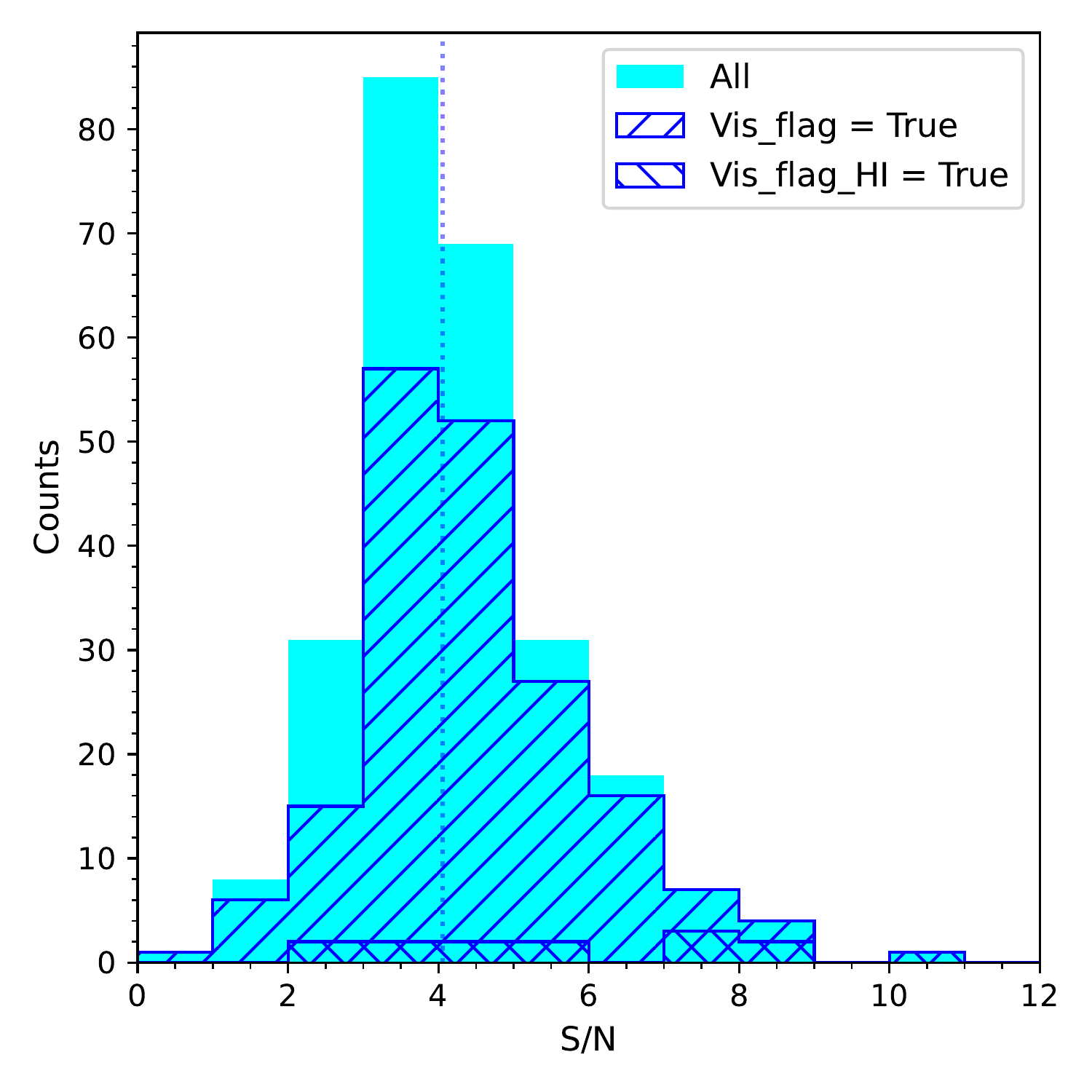}
    }
    \caption{
    Distribution of the positions of absorption peak i.e., {\tt pos\_max} ({\it top}) and peak S/N ({\it bottom}) for 1667\,MHz absorption candidates.  
    }
    \label{fig:ohmaindist}
\end{figure}

%%%%%%%%%%%
Following the automated absorption line search procedure presented in \citet[][]{Gupta2025galhi}, we detected 256 OH main line (1667\,MHz) features from 19130 sight lines.  The distribution of the positions of peak optical depth, provided as {\tt pos\_max} in the catalog (Table~\ref{tab:ohmaincat}),  is shown in the {\it top} panel of Fig.~\ref{fig:ohmaindist}.  The abrupt fall at $\pm$250\,\kms\ corresponds to the edges of the search window.  Otherwise, the positions of absorption features are uniformly distributed over the velocity range searched for absorption, implying that a significant fraction could be false detections due to statistical fluctuations, or even weak radio frequency interference (RFI) at these frequencies in the unprotected frequency range.  
Unlike \hi\ 21-cm absorption line \citep[][their Fig.~4]{Gupta2025galhi}, there is no excess of absorbers at radial velocities of -25$<v_{\rm LSR}$[\kms]$<$+25 apparent in Fig.~\ref{fig:ohmaindist}. 
However, a peak is apparent at $\sim$0\,\kms\ when only 30 absorbers with peak S/N greater than 6, an empirically determined criterion based on Galactic \hi\ absorption line search to distinguish {\it true} detections from statistical fluctuations, are considered.  
%

%%%%%%%%%%%%%%%
%%%%%%%%%%%%%

%%%%%%%%%%%%%%%
\begin{figure*}
    \centering
    \hbox{
    \includegraphics[trim = {0cm 0cm 0cm 0cm}, width=0.4\textwidth,angle=0]{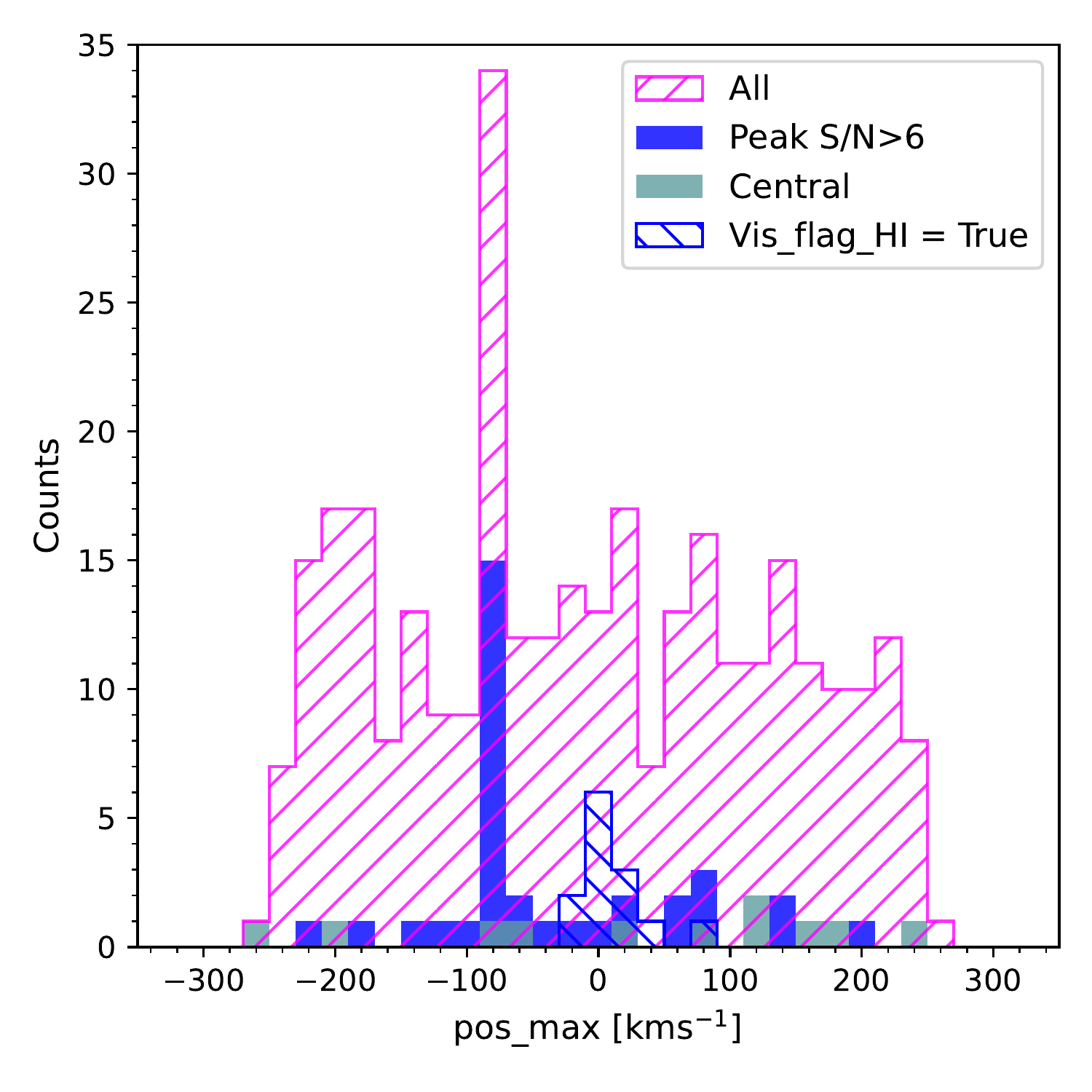}
    \includegraphics[trim = {0cm 0cm 0cm 0cm}, width=0.4\textwidth,angle=0]{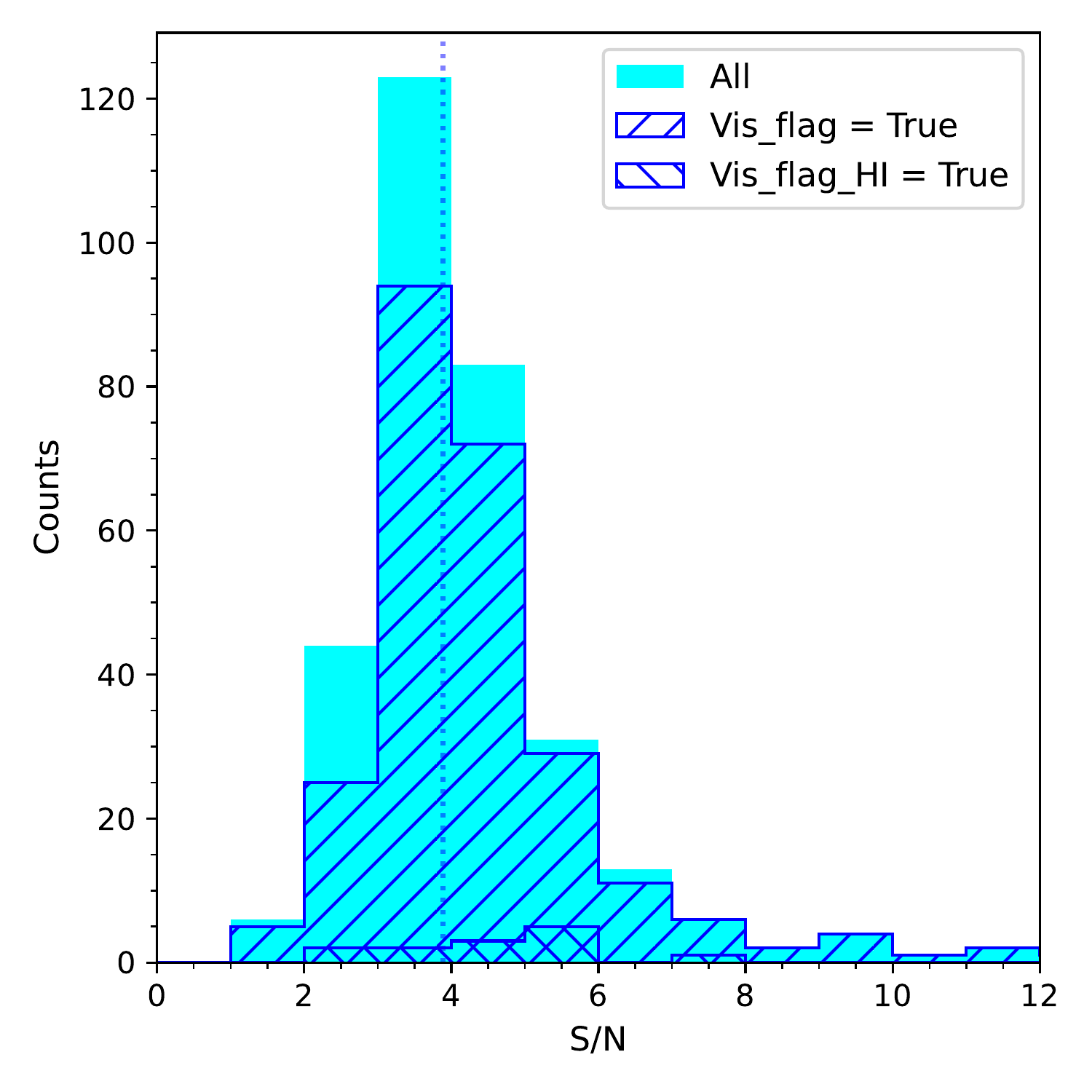}
    }
    \caption{
    Distributions of the positions of absorption peak i.e., {\tt pos\_max} ({\it left}) and peak S/N ({\it right}) for 1612\,MHz absorption candidates.  
    }
    \label{fig:ohsat1dist}
\end{figure*}
%%%%%%%%%%%%%

%%%%%%%%%%%%%%%
\begin{figure}
    \centering
    \vbox{
    \includegraphics[trim = {0cm 0cm 0cm 0cm}, width=0.38\textwidth,angle=0]{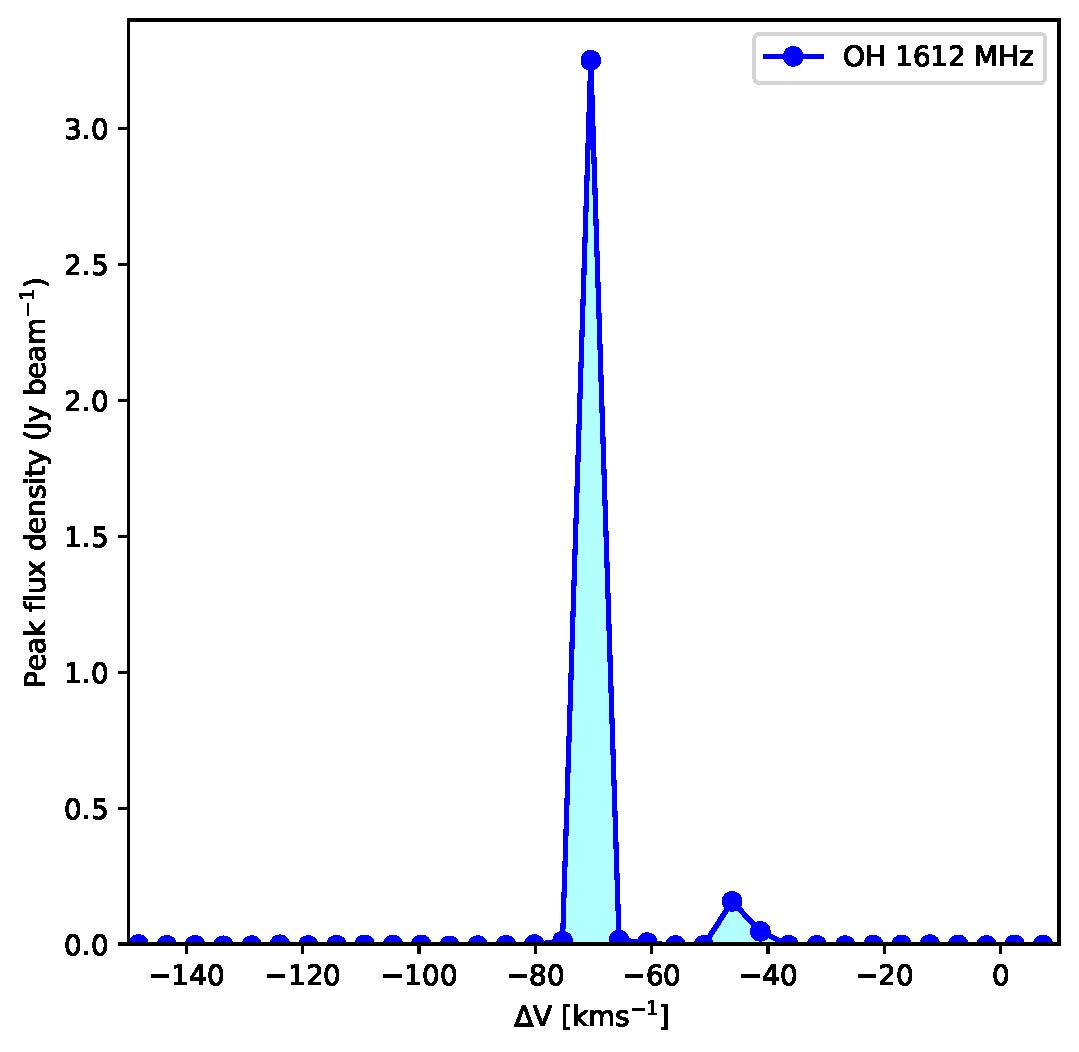}
    }
    \caption{
    OH 1612\,MHz emission line profile toward J171749.17$-$622903.4, associated with an AGB star.
    }
    \label{fig:ohsat-eml}
\end{figure}
%%%%%%%%%%%%%

%%%%%%%%%%%%%%%%%%
\begin{table}
\caption{Summary of OH 1667\,MHz and 1612\,MHz detections from the automated search.}
\vspace{-0.4cm}
\begin{center}
\begin{tabular}{lccccc}
\hline
\hline
                   &    OH 1667\,MHz$^\dag$    &   OH 1612\,MHz$^\dag$ \\
 \hline          
Total                            &   256  (30)         &  323 (35)     \\
{\it Central}                    &   14 (3)            &   11 (2)    \\
{\it Off-axis}                   &   242 (27)          &  312 (33)     \\
{\tt Vis\_flag} = {\tt True}     &   187  (28)         &  259 (33)     \\
{\tt Vis\_flag\_HI} = {\tt True} &   14   (6)          &  13 (1)      \\
\hline
%%%
\hline
\end{tabular}
%%%
\label{tab:ohstat}
\end{center}
$\dag$: The numbers in brackets are for absorbers with peak S/N $>$ 6.\\
\end{table}
%%%%%%%%%%%%%%%%%%%%%%%

%%%%%%%%%%%%%%%%%%%%%%%%%%%%%%%%%%%%%%%%%%
\renewcommand{\arraystretch}{1.45}
\setlength{\tabcolsep}{0.1cm}
\begin{table*}
\caption{Additional (discrepant) sight lines from initial automated search of MALS OH 1667\,MHz line discussed in Section~\ref{sec:sample}.  The column description is the same as in Table~\ref{tab:ohmain_det}.}
\vspace{-0.8cm}
\begin{center}
\begin{tabular}{cc|ccc|cc|ccc}
\hline
\hline
Sight line & $\log \frac{{\rm Pr}(D|M_{\rm abs})}{{\rm Pr}(D|M_{\rm cont})}$ & $v_{\rm 1667}$ & $\sigma_{1667}$ & $\tau_{\rm int}$ = $\int\tau_{1667}dv$& $\log N(\rm OH)$ & $\log N($\hi$)$ & S/N & {\tt Vis\_flag} & {\tt Vis\_flag\_HI} \\
& & (km\,s$^{-1}$) & (km\,s$^{-1}$) & (km\,s$^{-1}$) & (cm$^{-2}$) & (cm$^{-2}$) & & & \\
(1) & (2) & (3) & (4) & (5) & (6) & (7) & (8) & (9) & (10)\\
\hline
J200605.14$-$022343.5 &  6.4  &  $3.77^{+1.19}_{-0.27}$  &  $0.70^{+0.50}_{-0.70}$  & $0.0172^{+0.0032}_{-0.0031}$  & $13.49^{+0.19}_{-0.28}$ & $21.05^{+0.08}_{-0.10}$ & 7.5  & T  & T \\
J200609.17$-$022438.4 &  1.9  &  $3.70^{+1.20}_{-0.40}$  &  $0.90^{+0.60}_{-1.00}$  & $0.0309^{+0.0084}_{-0.0084}$  & $13.74^{+0.21}_{-0.29}$ & $21.14^{+0.08}_{-0.10}$ & 5.2  & T  & T \\
J040519.73$+$180610.9 &   0.1  &  $6.0^{+6.0}_{-4.0}$  &  $2.8^{+1.9}_{-1.9}$  & $0.56^{+0.36}_{-0.32}$  & $15.00^{+0.29}_{-0.41}$ & $21.38^{+0.08}_{-0.10}$ & 4.6 & T & F \\
J201857.74$+$110129.2 &   0.1  &  $15.0^{+3.6}_{-3.1}$  &  $1.5^{+1.0}_{-1.0}$  & $<1.9$  & $<15.6$ & $21.07^{+0.11}_{-0.13}$ & 2.2 & T & T \\
J151914.21$-$300048.2 &  <0  &  --  &  --  & --  & -- & -- & 2.4 & T & T \\
\hline
\hline
\end{tabular}
\label{tab:ohmain_add}
\end{center}
\end{table*}
%%%%%%%%%%%%%%%%%%%%%%%%%%%%%%%%%%%%%%%%%%

%%%%%%%%%%%%%%%%%%%%%%%%%%%%%%%%%%%%%%%%%%%
\renewcommand{\arraystretch}{1.45}
\setlength{\tabcolsep}{0.1cm}
\begin{table*}
\caption{Additional (discrepant) sight line from initial automated search of OH 1612\,MHz line discussed in Section~\ref{sec:oh_sat}.  The column description is the same as in Table~\ref{tab:ohsat_det}.}
\vspace{-0.4cm}
\begin{center}
\begin{tabular}{cc|ccc|c|ccc}
\hline
\hline
Sight line & $\log \frac{{\rm Pr}(D|M_{\rm abs})}{{\rm Pr}(D|M_{\rm cont})}$ & $v_{\rm 1612}$ & $\sigma_{1612}$ & $\tau_{\rm int}$ = $\int\tau_{1612}dv$  & OH main & S/N & {\tt Vis\_flag} & {\tt Vis\_flag\_HI} \\
& & (km\,s$^{-1}$) & (km\,s$^{-1}$) & (km\,s$^{-1}$) &  \\
(1) & (2) & (3) & (4) & (5) & (6) & (7) & (8) & (9) \\
\hline
J200325.73$-$023326.2 &  1.4  &  $1.8^{+1.0}_{-1.1}$  &  $1.20^{+0.80}_{-0.80}$  & $1.61^{+0.60}_{-0.52}$  & -- & 5.3 & T & T \\
\hline
\hline
\end{tabular}
\label{tab:ohsat_add}
\end{center}
\end{table*}
%%%%%%%%%%%%%%%%%%%%%%%%%%%%%%%%%%%%%%%%%%%

%%%%%%%%%%%%
The bottom panel of Fig.~\ref{fig:ohmaindist} shows peak S/N of all 1667\,MHz features.  Also shown are distributions for features confirmed through the visual inspection of only 1667\,MHz spectra, and both \hi\ and OH spectra.  The former with {\tt Vis\_flag} = {\tt True} in the catalog represents features that stand out in the spectrum and are within $\pm$5\,\kms\ of the cataloged value.  The latter, with {\tt Vis\_flag\_HI} = {\tt True}, also require a corresponding \hi\ absorption detection within 10\,\kms.  All of these are centered at $<v_{\rm LSR}>$[\kms] $\sim$ 0\,\kms\ and can have S/N as low as 2 (Fig.~\ref{fig:ohmaindist}; bottom panel).  

%%%%%
In Section~\ref{sec:sample}, we noted general consistency between the outcomes of initial search and the NS approach (Table~\ref{tab:ohmain_det}).  Here we provide details on five discrepant cases listed in Table~\ref{tab:ohmain_add}.  J151914.21$-$300048.2 and J201857.74$+$110129.2 have very low Bayesian evidence ratio but pass visual inspection test as the weak (S/N $\sim$2.2-2.4) feature identified by the initial search coincides with the \hi\ absorption feature.  The OH features are much broader than \hi\ absorption and are likely spurious.  This highlights the complementarity of the two approaches in reliably detecting weak absorption lines.

J200605.14$-$022343.5 and J200609.17$-$022438.4 are false detections due to the choice of a square aperture (60$\farcs$0) adopted for identifying the position of the peak pixel at which the spectrum is extracted.  Within this aperture there is only one unique detection i.e., J200608.43$-$022334.8, which is included as a credible detection in Table~\ref{tab:ohmain_det}.
Finally, in the case of J040519.73+180610.9 the Bayes evidence ratio is much lower than the expectation based on the peak S/N. This is because NS output corresponds to a weak feature at $\sim$10\,\kms\ coincident with \hi\ emission and absorption peaks, whereas the peak S/N and flags in the last three columns based on the initial search are for an unrelated feature at $\sim$45\,\kms.

%%%%%%%%%%
The search for OH 1612\,MHz line towards 19130 sight lines resulted in 323 features cataloged in Table~\ref{tab:ohsat1cat}.  The distributions of the positions of peak optical depth and peak S/N are presented in Fig.~\ref{fig:ohsat1dist} (see also Table~\ref{tab:ohstat}).  These look broadly similar to the distributions in Fig.~\ref{fig:ohmaindist}.  
As anticipated there is a peak at $\sim$0\,\kms\ for features coincident with \hi\ absorption ({\tt Vis\_flag\_HI} = {\tt True}).  However, there is a prominent peak at $-$71\,\kms.  The latter has no corresponding \hi\ absorption and belongs to 14 sight lines from the pointing J171844.67$-$625059.9.  We serendipitously detect a very strong OH 1612\,MHz emission (J171749.17$-$622903.4) associated with an AGB star at RA: 259.454901261 and  Dec: $-$62.484279534.  
The star is at $22\farcm8$ from the pointing center.  Since spectral line cubes considered here have not been deconvolved, the strong emission line introduces sidelobes across the entire field of view. The spectral profile of the deconvolved OH 1612\,MHz line is presented in Fig.~\ref{fig:ohsat-eml}.  We defer the further discussion of this detection to a future paper.
For completeness, Table~\ref{tab:ohsat_add} lists J200325.73$-$023326.2 as a false detection due to choice of aperture size described above.  Within this aperture there is only one true detection (J200322.53$-$023347.9) that is included in Table~\ref{tab:ohsat_det}.
Finally, we note that the following entries in the OH 1612\,MHz line catalog (Table~\ref{tab:ohsat1cat}) are spurious detections due to a ripple in bandpass: J223737.36$-$171240.2, J231635.20+040517.6 and J231621.12$-$433746.8.

\clearpage

%%%%%%%%%%%%%%%%%%%%%%%%%%%%%%%%%%%%%
\section{Validation of the Bayesian detection threshold}
\label{sec:oh-check}
%%%%%%%%%%%%%%%%%%%%%%%%%%%%%%%%%%%%%

\begin{figure}
    \centering
    \vbox{
    \includegraphics[trim = {0cm 0cm 0cm 0cm}, width=0.45\textwidth,angle=0]{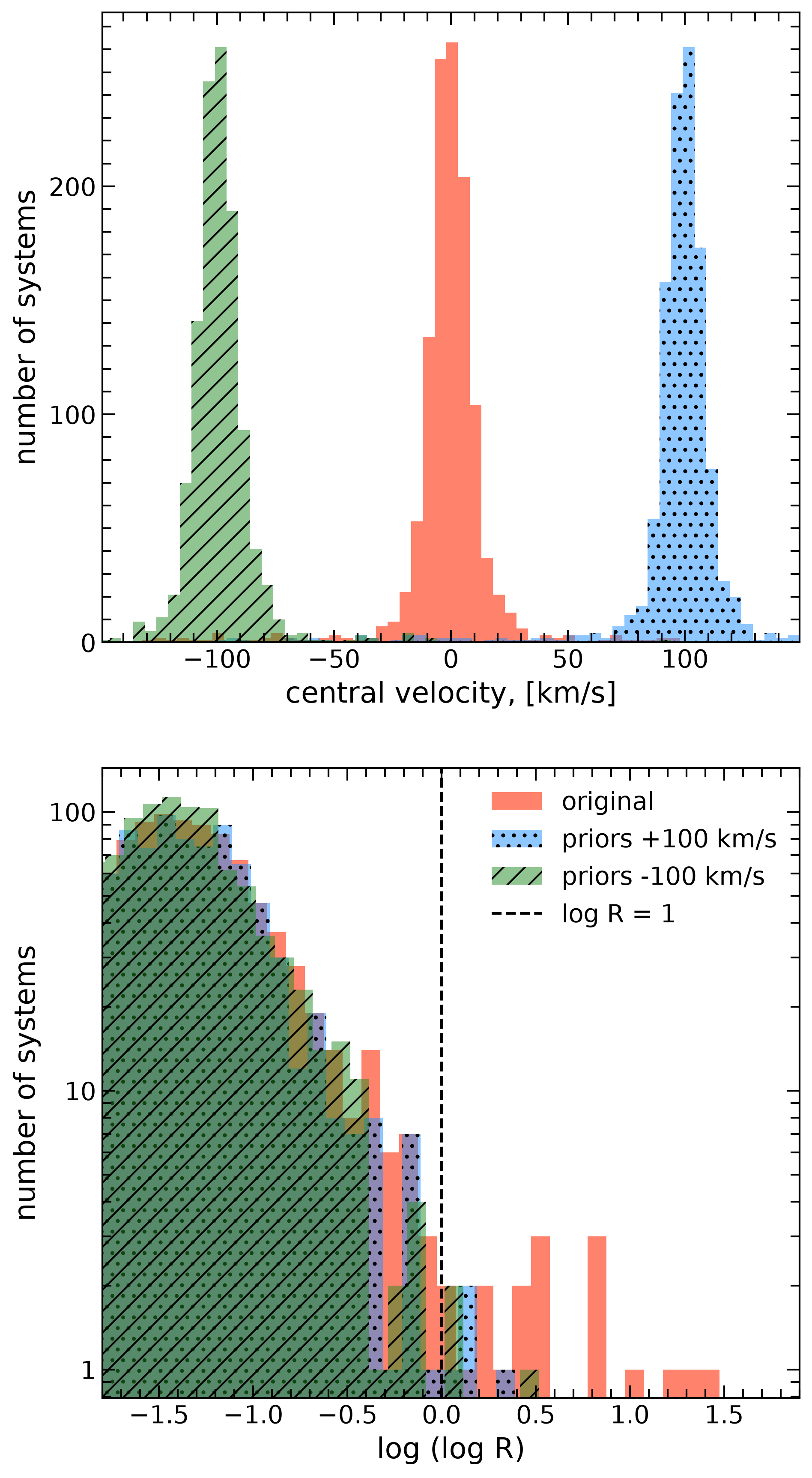}
    }
    \caption{Distributions of the most probable OH absorption velocities (\textit{top}) and Bayesian evidence ratio $\log\,\mathcal{R}$ (\textit{bottom}) for non-shifted priors (red) and priors shifted by $+100$ (blue) and $-100$ (green)\,\kms. The shifted realizations serve as a false-detection reference sample. The dashed line shows $\log\,\mathcal{R}=1$, the confident detection threshold adopted in Table~\ref{tab:ohmain_det}.  
    }
    \label{fig:sanity}
\end{figure}

To test whether the chosen threshold for confident detection ($\log\,\mathcal{R} > $1) presented in Section~\ref{sec:sample} is appropriate,  we perform a simple sanity check. We applied the same Bayesian nested sampling approach with one modification -- we artificially shifted the velocity prior for the position of the OH line by $+100$ and $-100$\,\kms. This ensures that we are probing false-positive detections, since no OH absorption is expected at these shifted positions, while the noise conditions remain similar. The resulting distributions for these shifted and non-shifted realizations are compared in Fig.~\ref{fig:sanity}. One can see that the false-detection samples show a few sight lines with $1 < \log\,\mathcal{R} \lesssim 3$. We verified that these spurious `detections' do not coincide in velocity with the OH detections in the original sample. Since $\log\,\mathcal{R} = 1$ was adopted in Section~\ref{sec:sample} as the confident detection threshold, this implies that a similar number of false detections may be present among the 15 sight lines with $\log\,\mathcal{R} > 1$ presented in Table~\ref{tab:ohmain_det}. However, with only two velocity shift realizations,  quantitative conclusions are limited. In principle, this approach could be extended to a bootstrap analysis, to accurately calculate the significance of the chosen threshold level, though in practice, this is computationally expensive given the large sample size, as it would require many realizations further limited by the $\sim$500\,\kms\ search window and the proximity of the 1665\,MHz line.

%%%%%%%%%%%%%%%
%%%%%%%%%%%%%

\clearpage
\onecolumn

%%%%%%%%%%%%%%%%%%%%%%%%%%%%%%%%%%%%%
\section{OH main line detections for the confident sample}
\label{sec:ohmain-confid-spec}
%%%%%%%%%%%%%%%%%%%%%%%%%%%%%%%%%%%%%
\suppressfloats[t]
We present the OH 1667 and 1665\,MHz absorption spectra, along with the 1612\,MHz profiles and the corresponding \hi\ emission and absorption spectra, for the confident detections listed in Table~\ref{tab:ohmain_det}, except J152149.13$-$294244.8, which is shown in Fig.~\ref{fig:ohmain-hi}.

\begin{figure*}
    \centering
    \vbox{
    \includegraphics[trim = {0cm -0.5cm 0cm 0cm}, width=1.0\textwidth,angle=0]{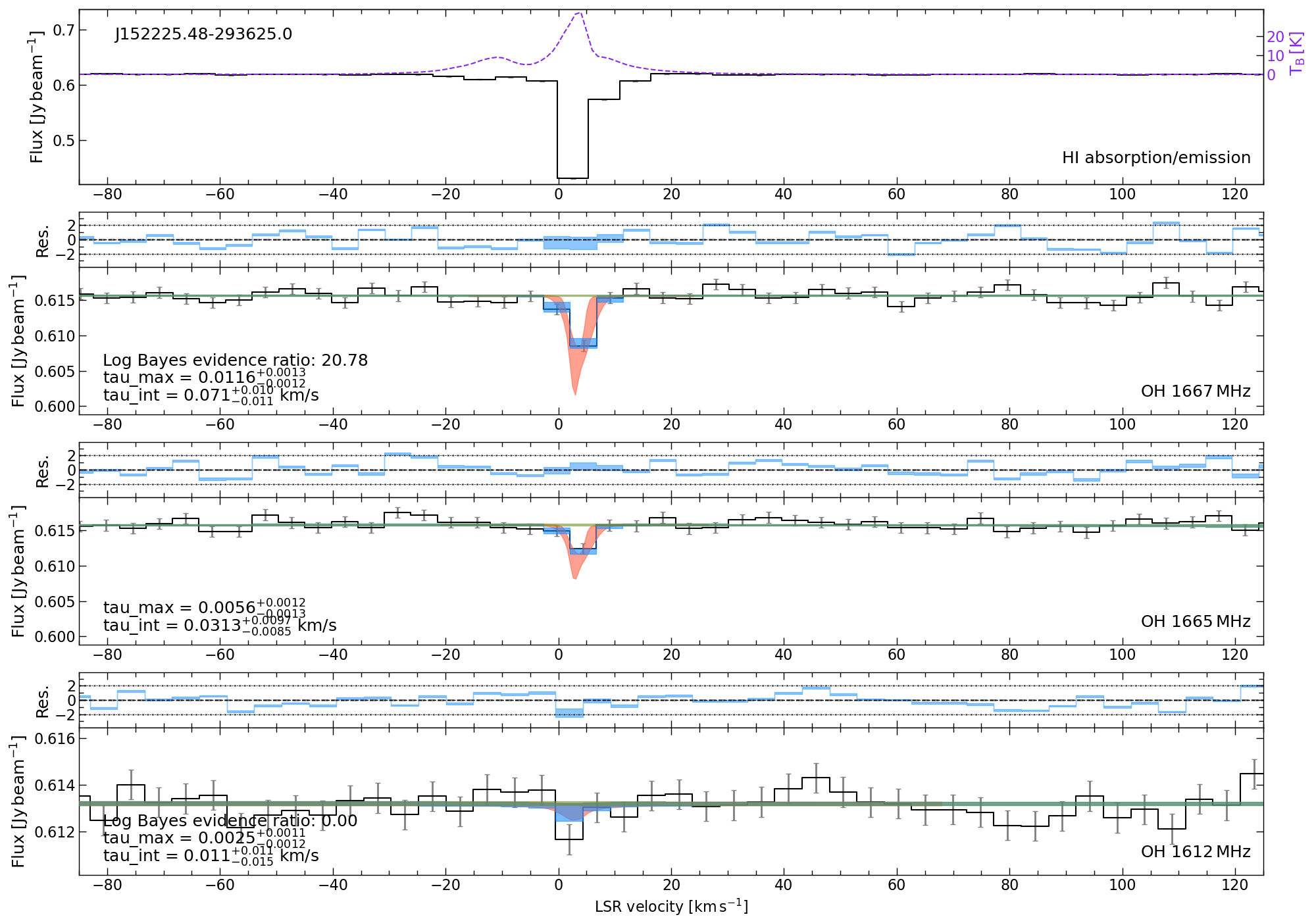} \\
    }
    \caption{The spectra of the sight line toward J152225.48$-$293625.0 showing detections of OH absorptions. The graphical elements are the same as in Fig.~\ref{fig:ohmain-hi}.}
    \label{fig:J152225.48-293625.0}
\end{figure*}

\begin{figure*}
    \centering
    \vbox{
    \includegraphics[trim = {0cm -0.5cm 0cm 0cm}, width=1.0\textwidth,angle=0]{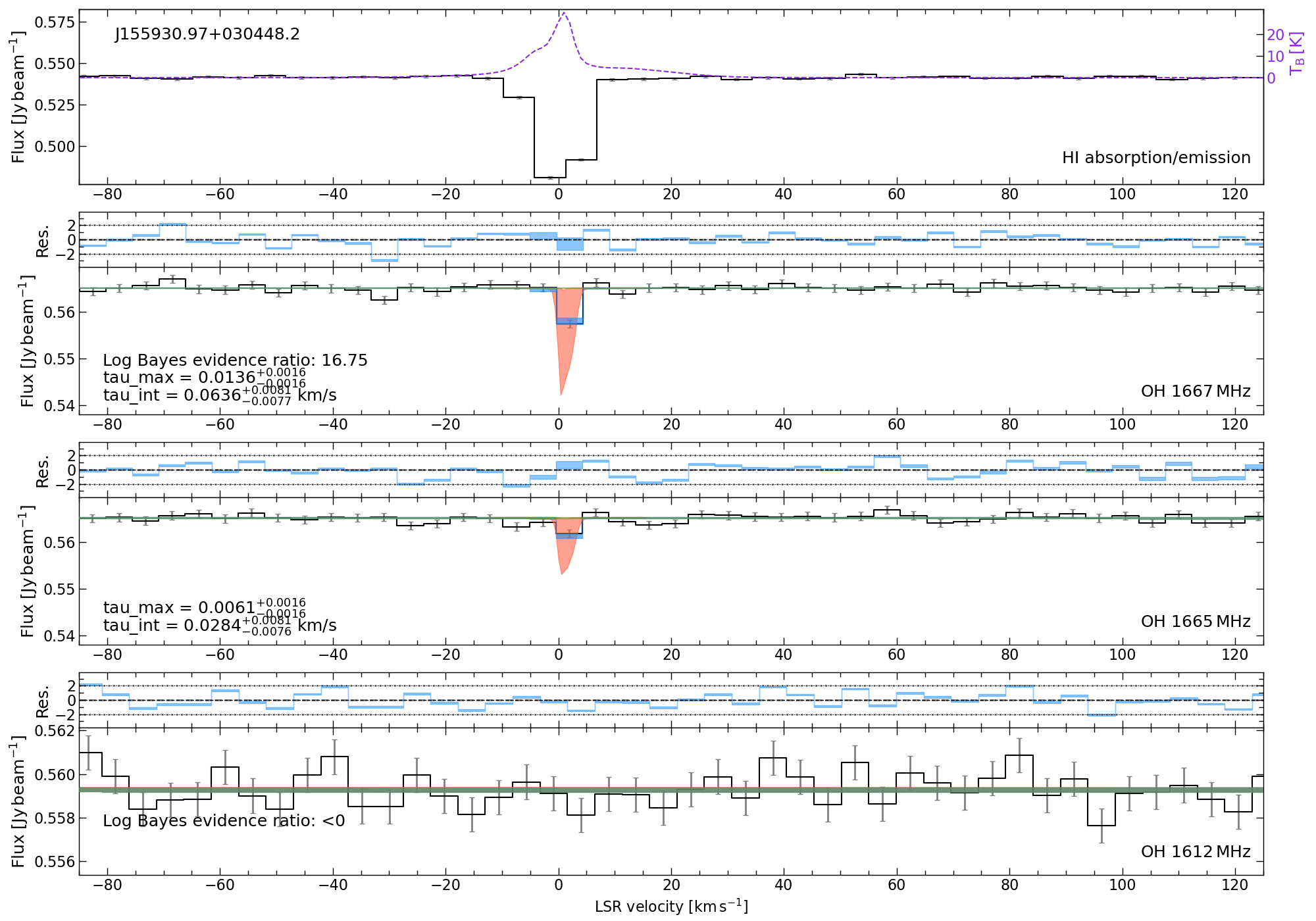} \\
    }
    \caption{The spectra of the sight line toward J155930.97$+$030448.2 showing detections of OH absorptions. The graphical elements are the same as in Fig.~\ref{fig:ohmain-hi}.}
    \label{fig:J155930.97+030448.2}
\end{figure*}

\begin{figure*}
    \centering
    \vbox{
    \includegraphics[trim = {0cm -0.5cm 0cm 0cm}, width=1.0\textwidth,angle=0]{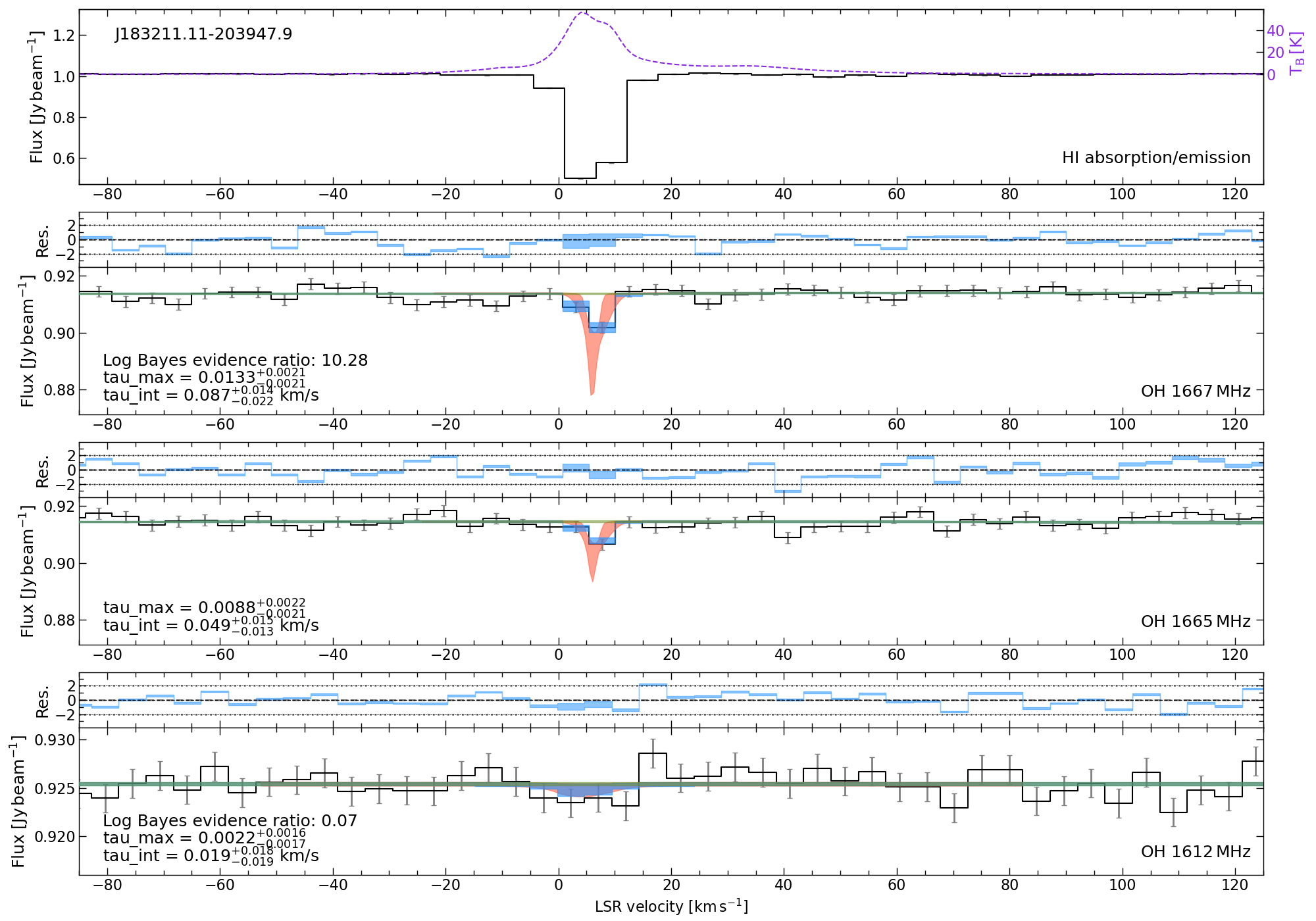} \\
    }
    \caption{The spectra of the sight line toward J183211.11$-$203947.9 showing detections of OH absorptions. The graphical elements are the same as in Fig.~\ref{fig:ohmain-hi}.}
    \label{fig:J183211.11-203947.9}
\end{figure*}

\begin{figure*}
    \centering
    \vbox{
    \includegraphics[trim = {0cm -0.5cm 0cm 0cm}, width=1.0\textwidth,angle=0]{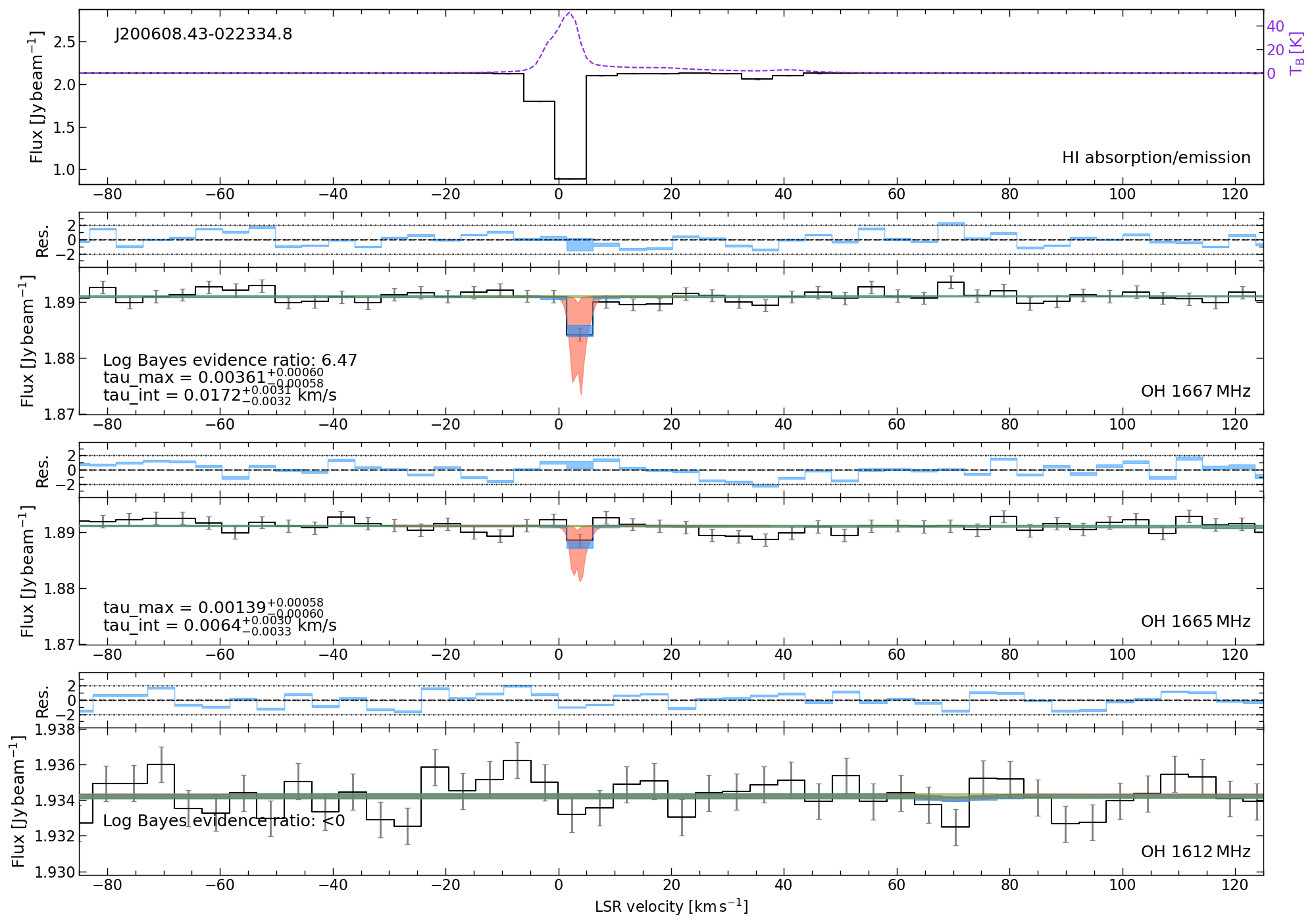} \\
    }
    \caption{The spectra of the sight line toward J200608.43$-$022334.8 showing detections of OH absorptions. The graphical elements are the same as in Fig.~\ref{fig:ohmain-hi}.}
    \label{fig:J200608.43-022334.8}
\end{figure*}

\begin{figure*}
    \centering
    \vbox{
    \includegraphics[trim = {0cm -0.5cm 0cm 0cm}, width=1.0\textwidth,angle=0]{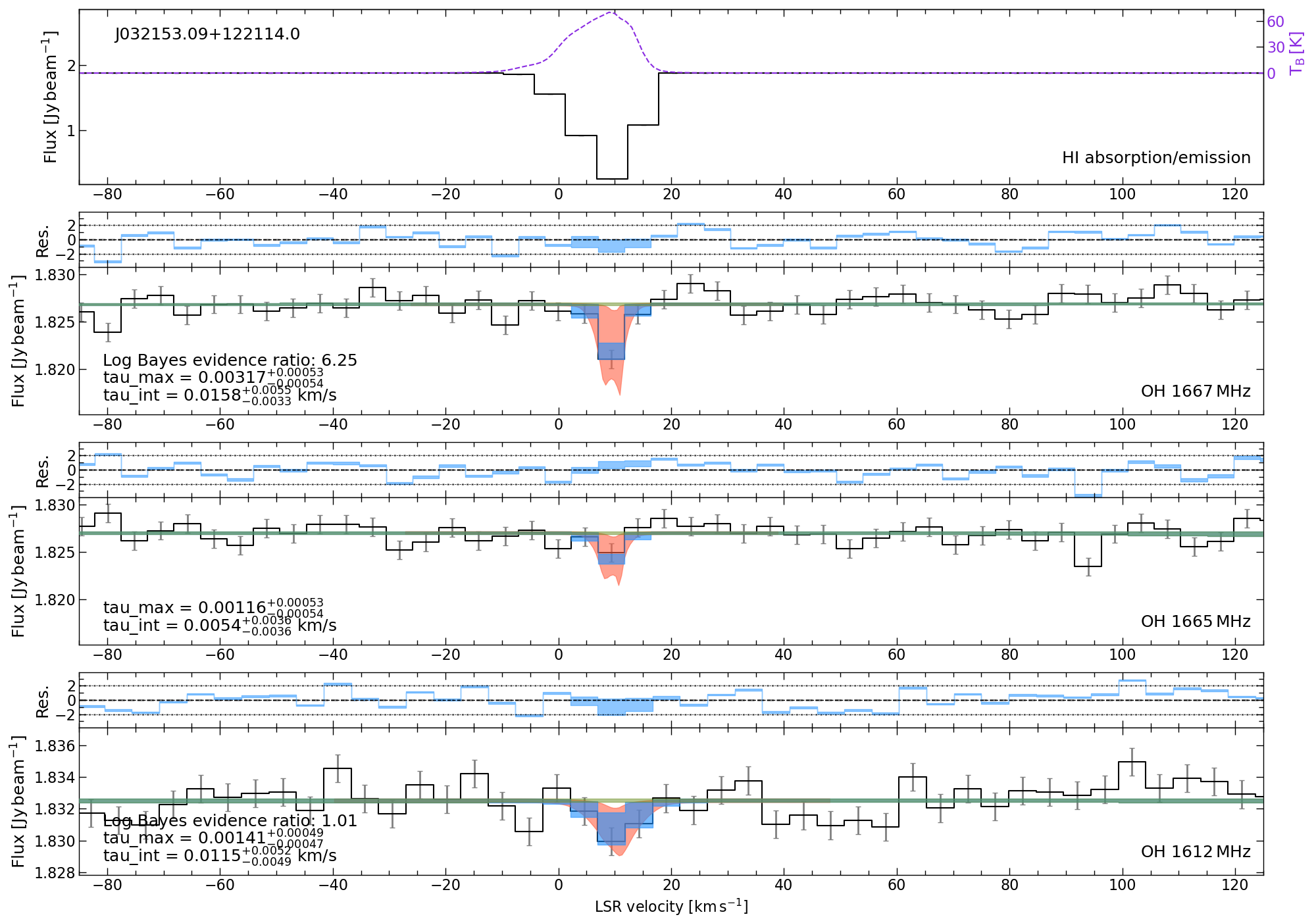} \\
    }
    \caption{The spectra of the sight line toward J032153.09$+$122114.0 showing detections of OH absorptions. The graphical elements are the same as in Fig.~\ref{fig:ohmain-hi}.}
    \label{fig:J032153.09+122114.0}
\end{figure*}

\begin{figure*}
    \centering
    \vbox{
    \includegraphics[trim = {0cm -0.5cm 0cm 0cm}, width=1.0\textwidth,angle=0]{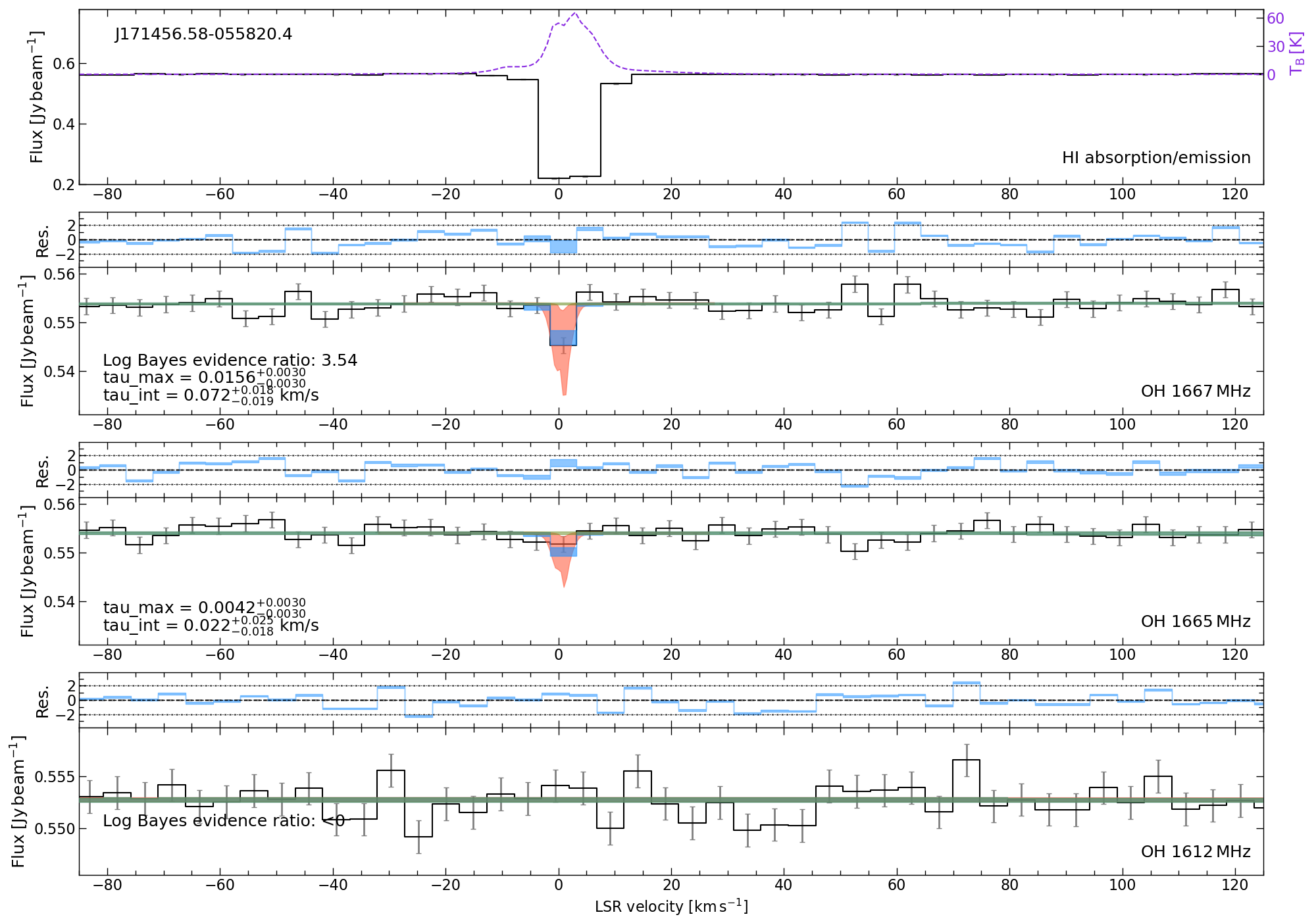} \\
    }
    \caption{The spectra of the sight line toward J171456.58$-$055820.4 showing detections of OH absorptions. The graphical elements are the same as in Fig.~\ref{fig:ohmain-hi}.}
    \label{fig:J171456.58-055820.4}
\end{figure*}

\begin{figure*}
    \centering
    \vbox{
    \includegraphics[trim = {0cm -0.5cm 0cm 0cm}, width=1.0\textwidth,angle=0]{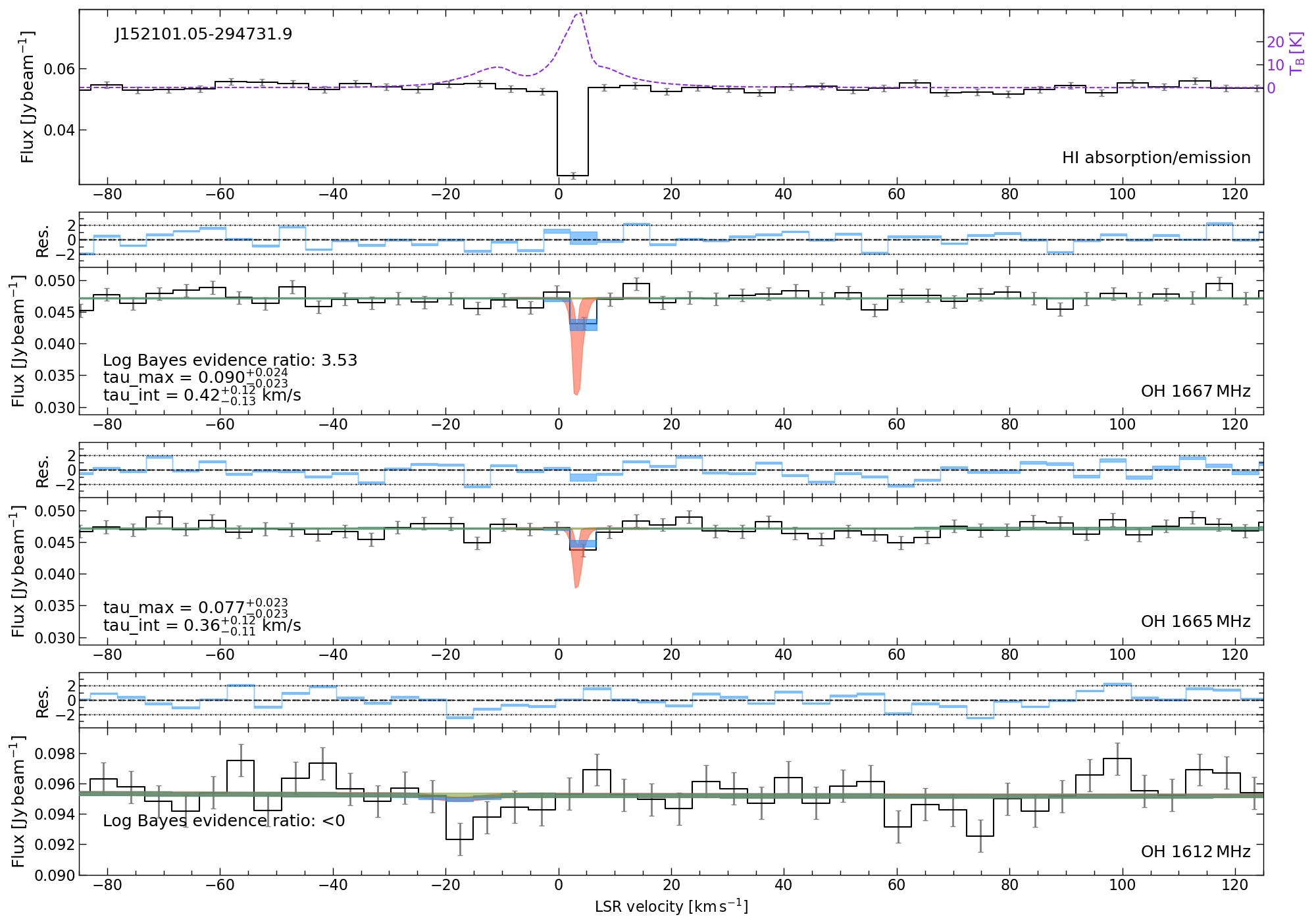} \\
    }
    \caption{The spectra of the sight line toward J152101.05$-$294731.9 showing detections of OH absorptions. The graphical elements are the same as in Fig.~\ref{fig:ohmain-hi}.}
    \label{fig:J152101.05-294731.9}
\end{figure*}

\begin{figure*}
    \centering
    \vbox{
    \includegraphics[trim = {0cm -0.5cm 0cm 0cm}, width=1.0\textwidth,angle=0]{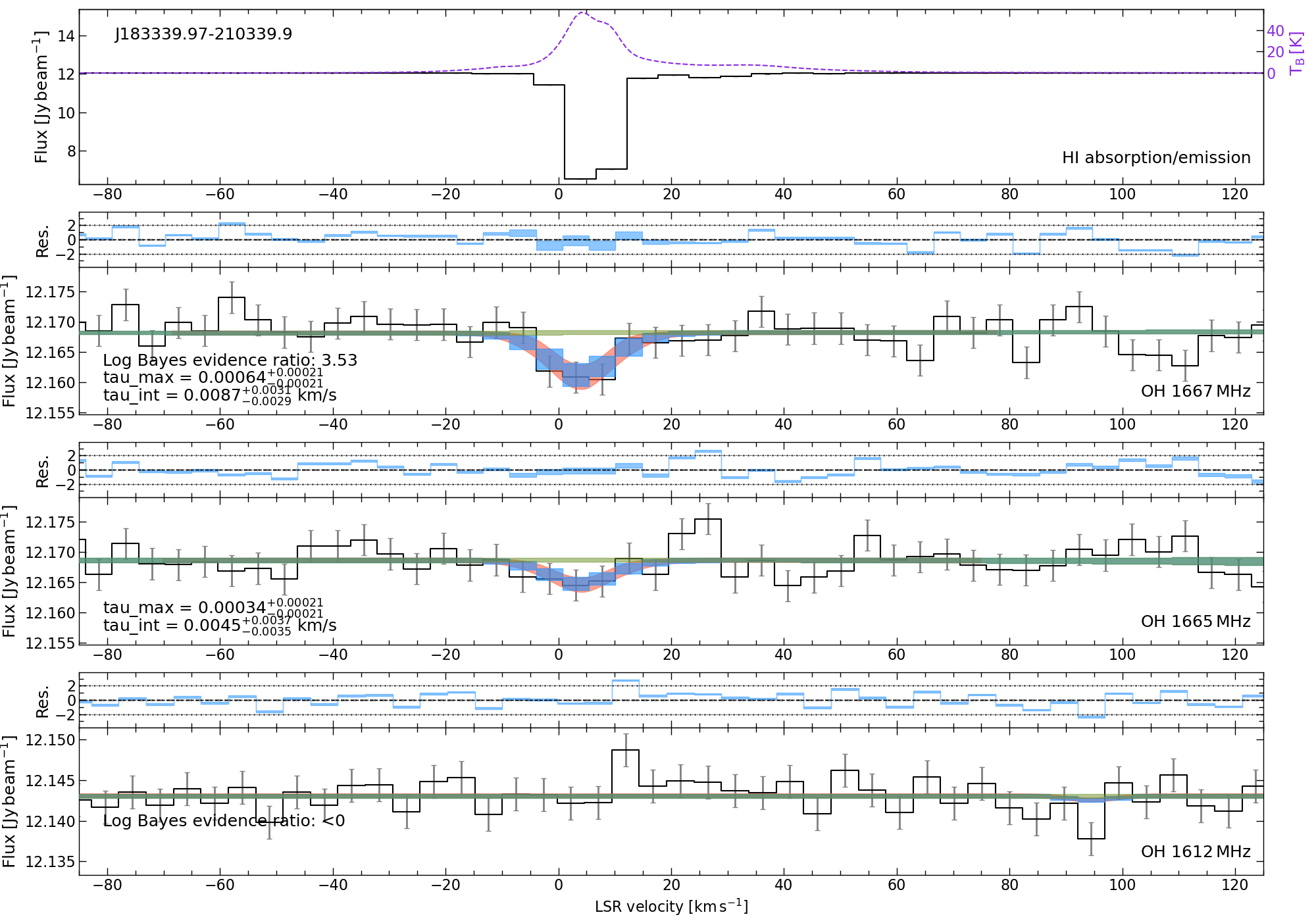} \\
    }
    \caption{The spectra of the sight line toward J183339.97$-$210339.9 showing detections of OH absorptions. The graphical elements are the same as in Fig.~\ref{fig:ohmain-hi}.}
    \label{fig:J183339.97-210339.9}
\end{figure*}

\begin{figure*}
    \centering
    \vbox{
    \includegraphics[trim = {0cm -0.5cm 0cm 0cm}, width=1.0\textwidth,angle=0]{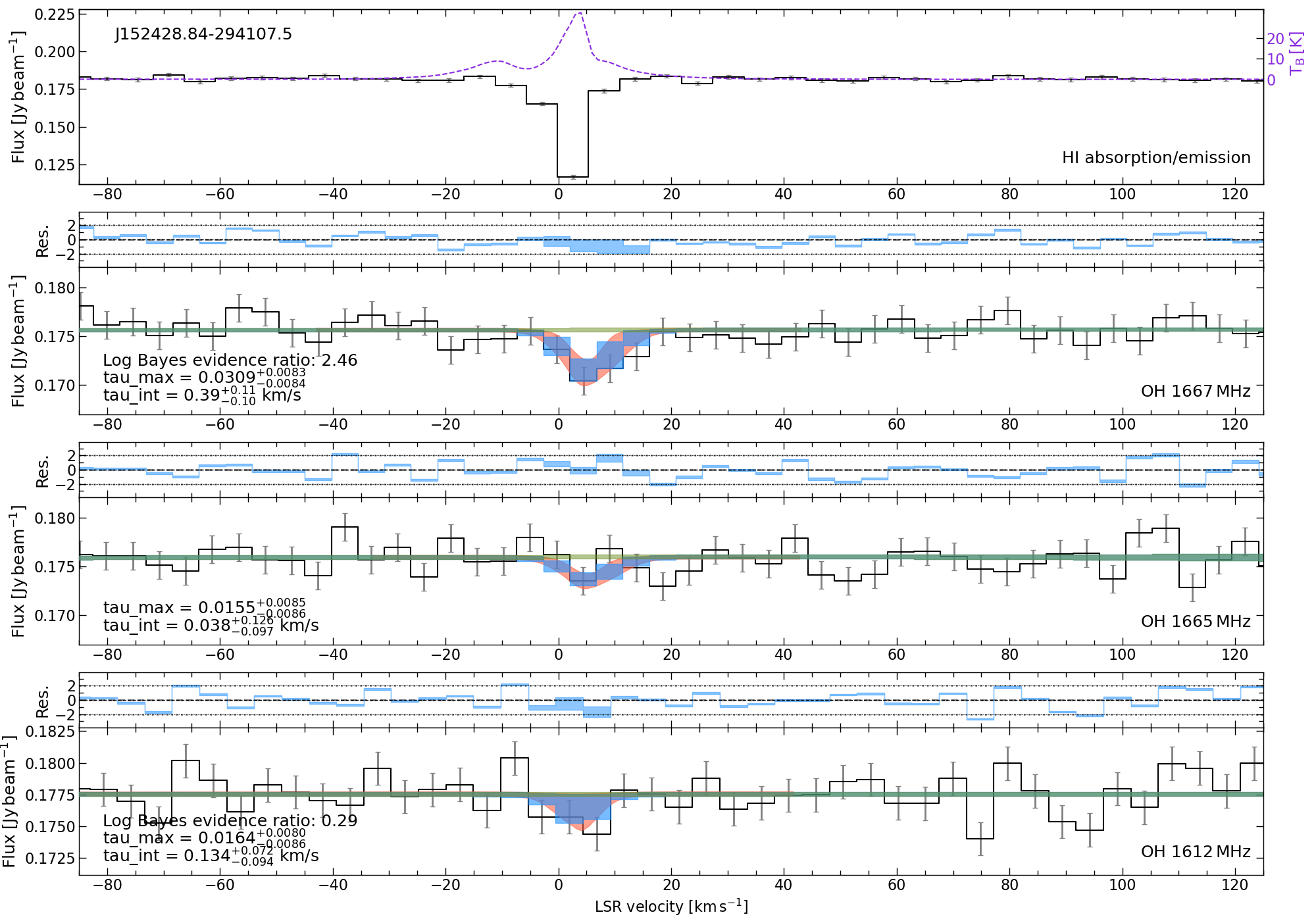} \\
    }
    \caption{The spectra of the sight line toward J152428.84$-$294107.5 showing detections of OH absorptions. The graphical elements are the same as in Fig.~\ref{fig:ohmain-hi}.}
    \label{fig:J152428.84-294107.5}
\end{figure*}

\begin{figure*}
    \centering
    \vbox{
    \includegraphics[trim = {0cm -0.5cm 0cm 0cm}, width=1.0\textwidth,angle=0]{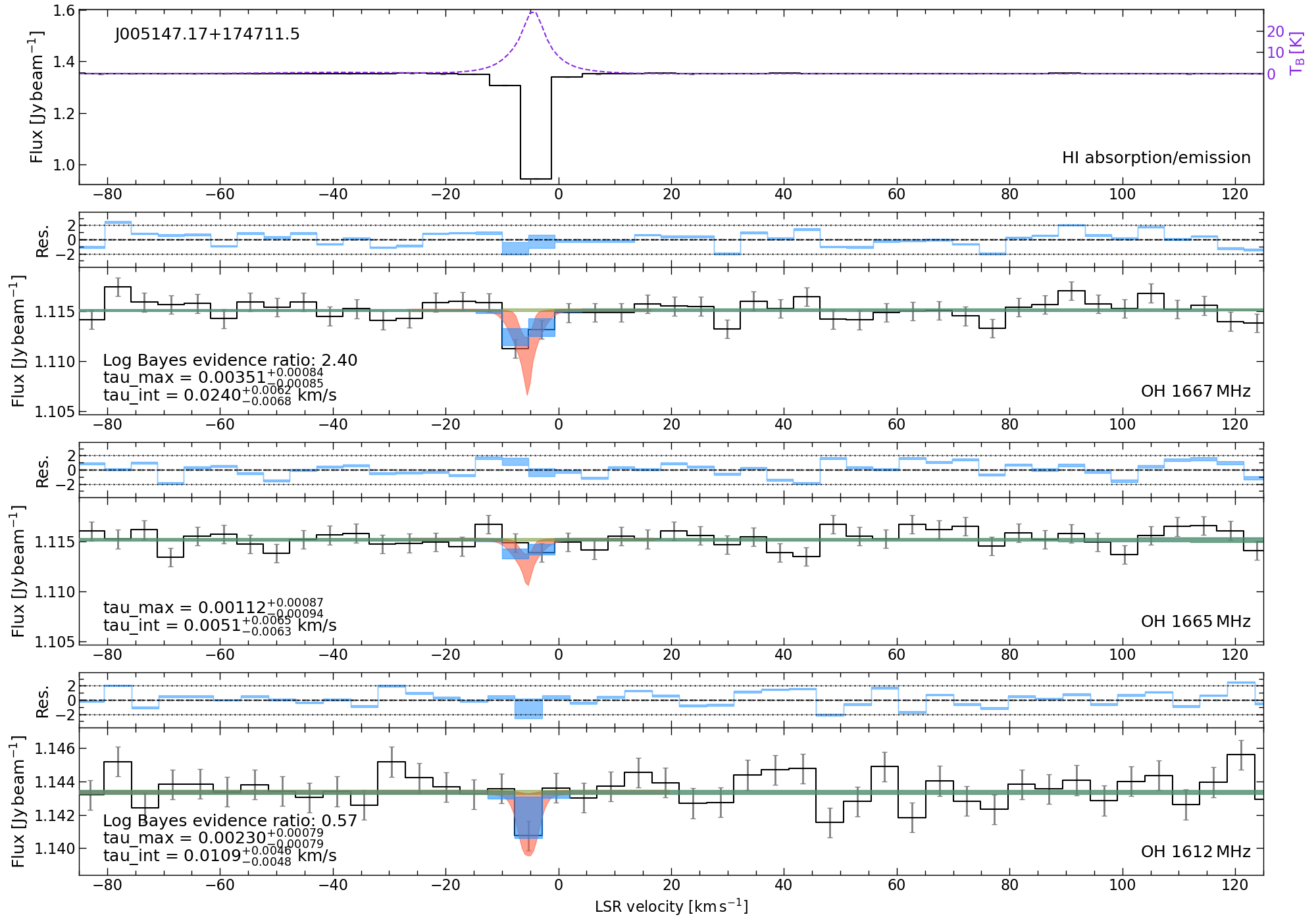} \\
    }
    \caption{The spectra of the sight line toward J005147.17$+$174711.5 showing detections of OH absorptions. The graphical elements are the same as in Fig.~\ref{fig:ohmain-hi}.}
    \label{fig:J005147.17+174711.5}
\end{figure*}

\begin{figure*}
    \centering
    \vbox{
    \includegraphics[trim = {0cm -0.5cm 0cm 0cm}, width=1.0\textwidth,angle=0]{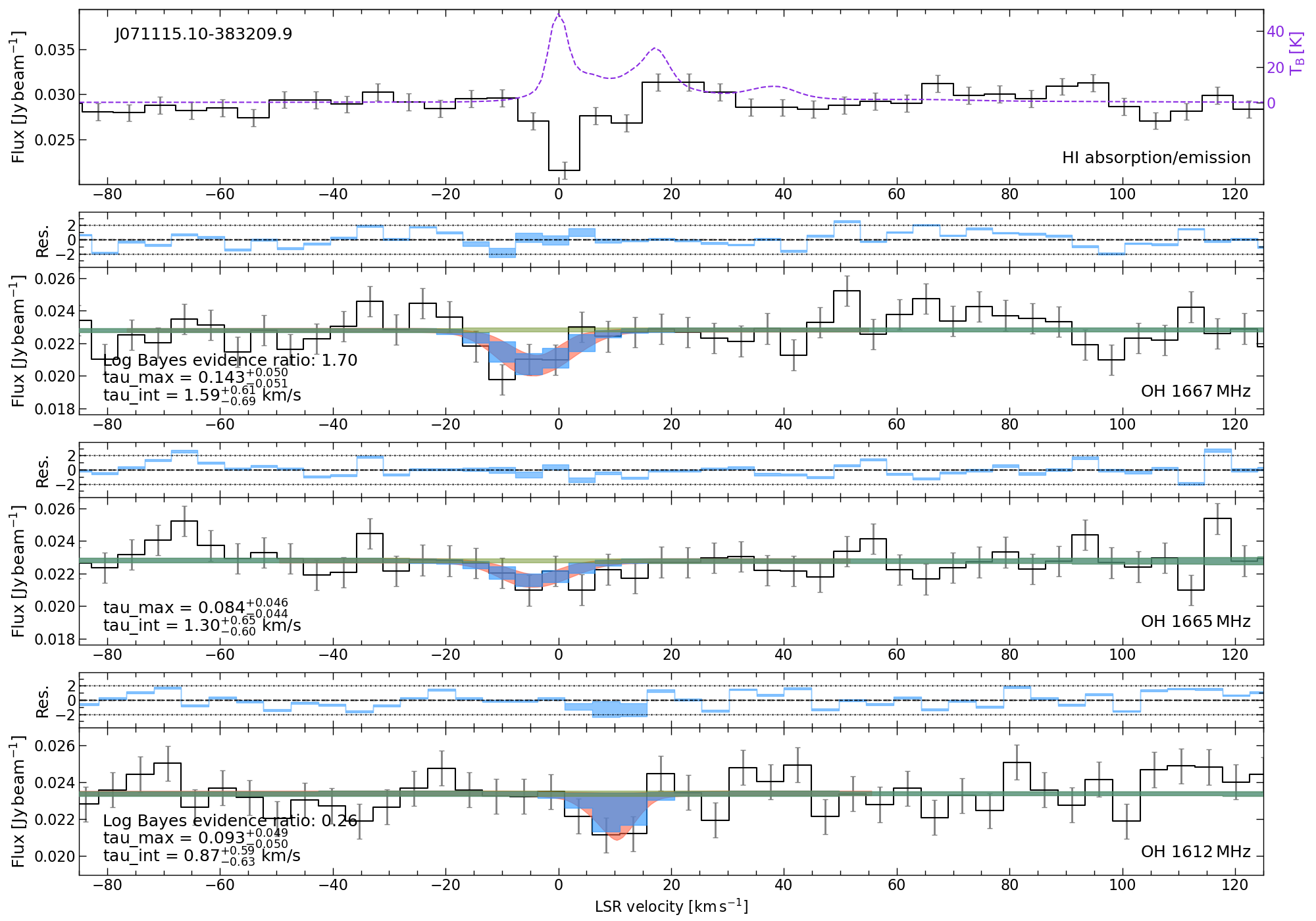} \\
    }
    \caption{The spectra of the sight line toward J071115.10$-$383209.9 showing detections of OH absorptions. The graphical elements are the same as in Fig.~\ref{fig:ohmain-hi}.}
    \label{fig:J071115.10-383209.9}
\end{figure*}

\begin{figure*}
    \centering
    \vbox{
    \includegraphics[trim = {0cm -0.5cm 0cm 0cm}, width=1.0\textwidth,angle=0]{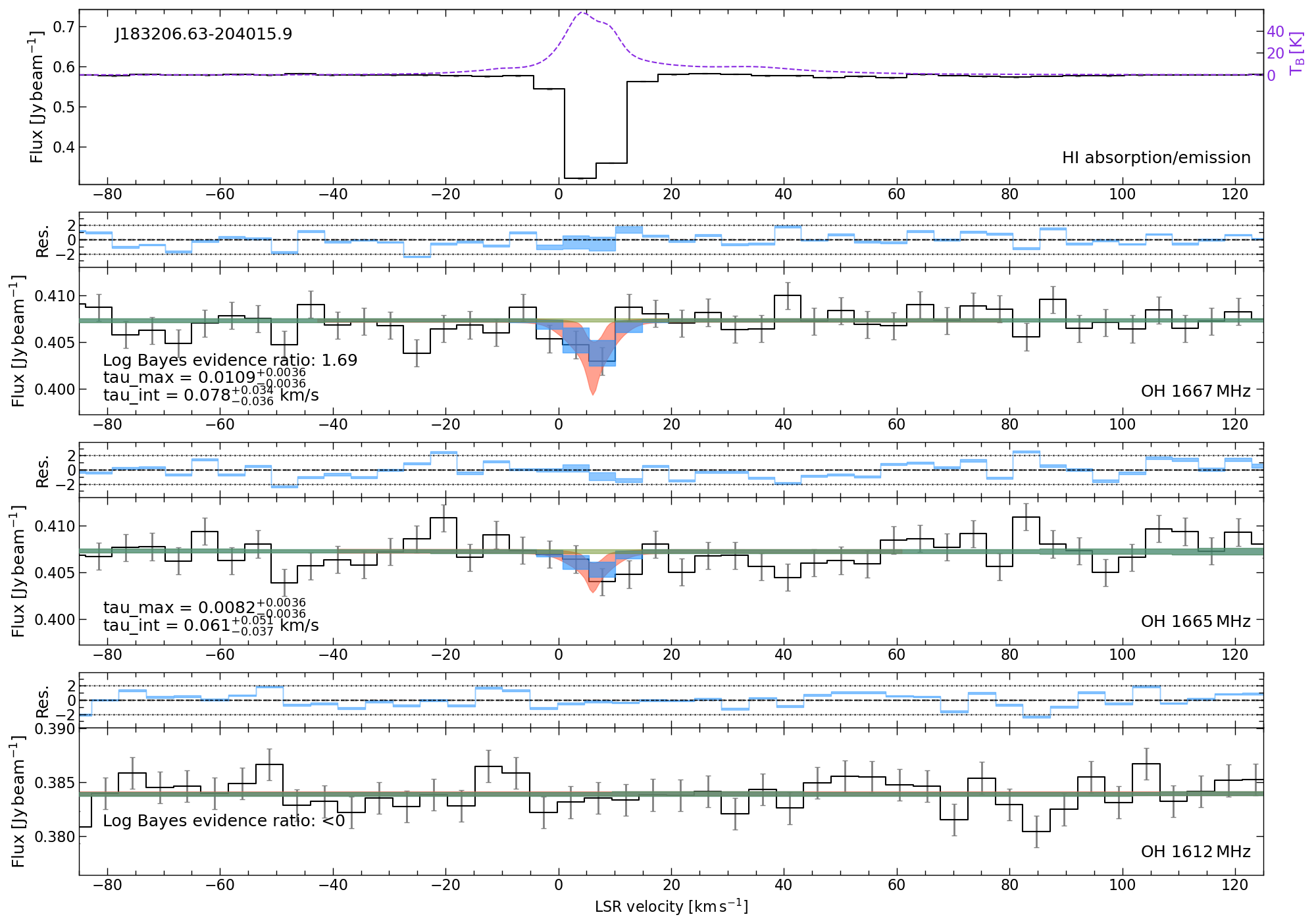} \\
    }
    \caption{The spectra of the sight line toward J183206.63$-$204015.9 showing detections of OH absorptions. The graphical elements are the same as in Fig.~\ref{fig:ohmain-hi}.}
    \label{fig:J183206.63-204015.9}
\end{figure*}

\begin{figure*}
    \centering
    \vbox{
    \includegraphics[trim = {0cm -0.5cm 0cm 0cm}, width=1.0\textwidth,angle=0]{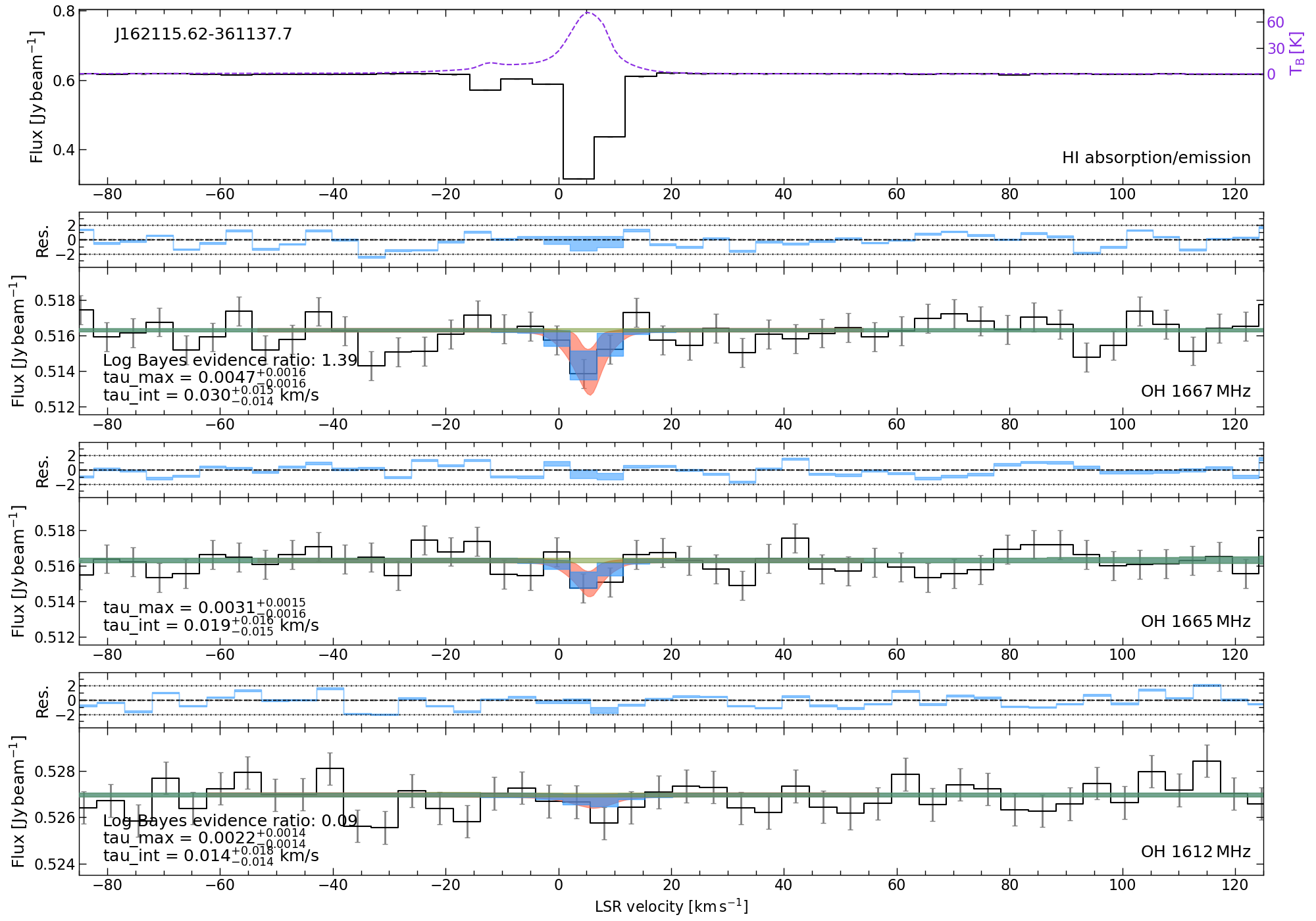} \\
    }
    \caption{The spectra of the sight line toward J162115.62$-$361137.7 showing detections of OH absorptions. The graphical elements are the same as in Fig.~\ref{fig:ohmain-hi}.}
    \label{fig:J162115.62-361137.7}
\end{figure*}

\begin{figure*}
    \centering
    \vbox{
    \includegraphics[trim = {0cm -0.5cm 0cm 0cm}, width=1.0\textwidth,angle=0]{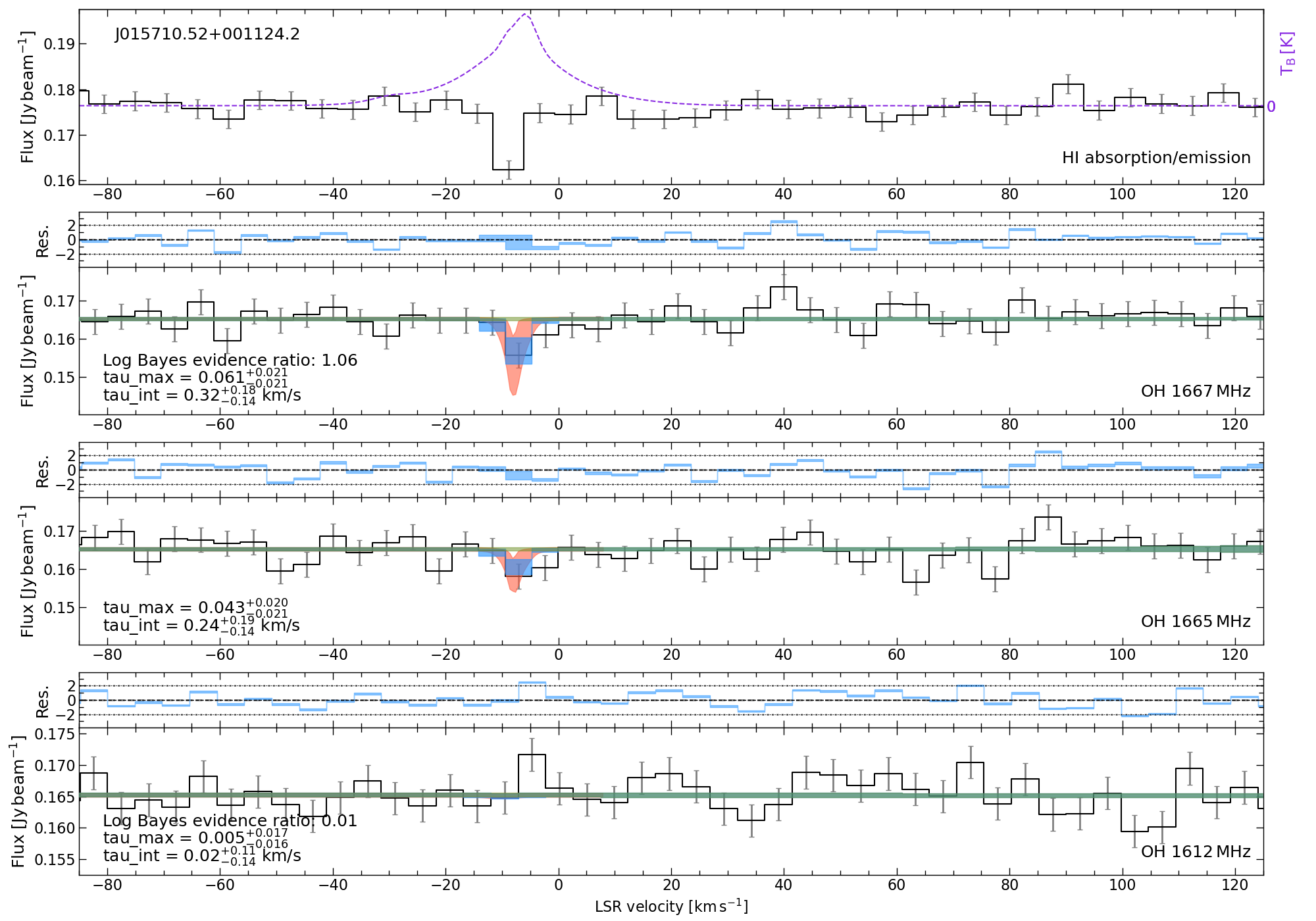} \\
    }
    \caption{The spectra of the sight line toward J015710.52$+$001124.2 showing detections of OH absorptions. The graphical elements are the same as in Fig.~\ref{fig:ohmain-hi}.}
    \label{fig:J015710.52+001124.2}
\end{figure*}
%\end{comment}
%%%%%%%%%%%%%%%%%%%%%%%%%%%%%%%

\clearpage

%%%%%%%%%%%%%%%%%%%%%%%%%%%%%%%%%%%%%%%%%%%%%%%
\section{CO overlays for pointings with OH detections}
\label{sec:co-overlays}
%%%%%%%%%%%%%%%%%%%%%%%%%%%%%%%%%%%%%%%%%%%%%%%
\suppressfloats[t]

We present the MALS pointings containing the confident OH detections in Table~\ref{tab:ohmain_det}, overlaid with CO $J=1\rightarrow0$ emission \citep{Ghosh2024}. The J152225.50$-$293624.8 field is shown in Fig.~\ref{fig:poi}. We also include the pointing of J045541.91$+$185010.9 discussed in Section~\ref{sec:global}.

%%%%%%%%%%%%%%%%%%%%
\begin{figure*}
    \centering
    \includegraphics[trim = {0cm 0cm 0cm 0cm}, width=1.0\textwidth,angle=0]{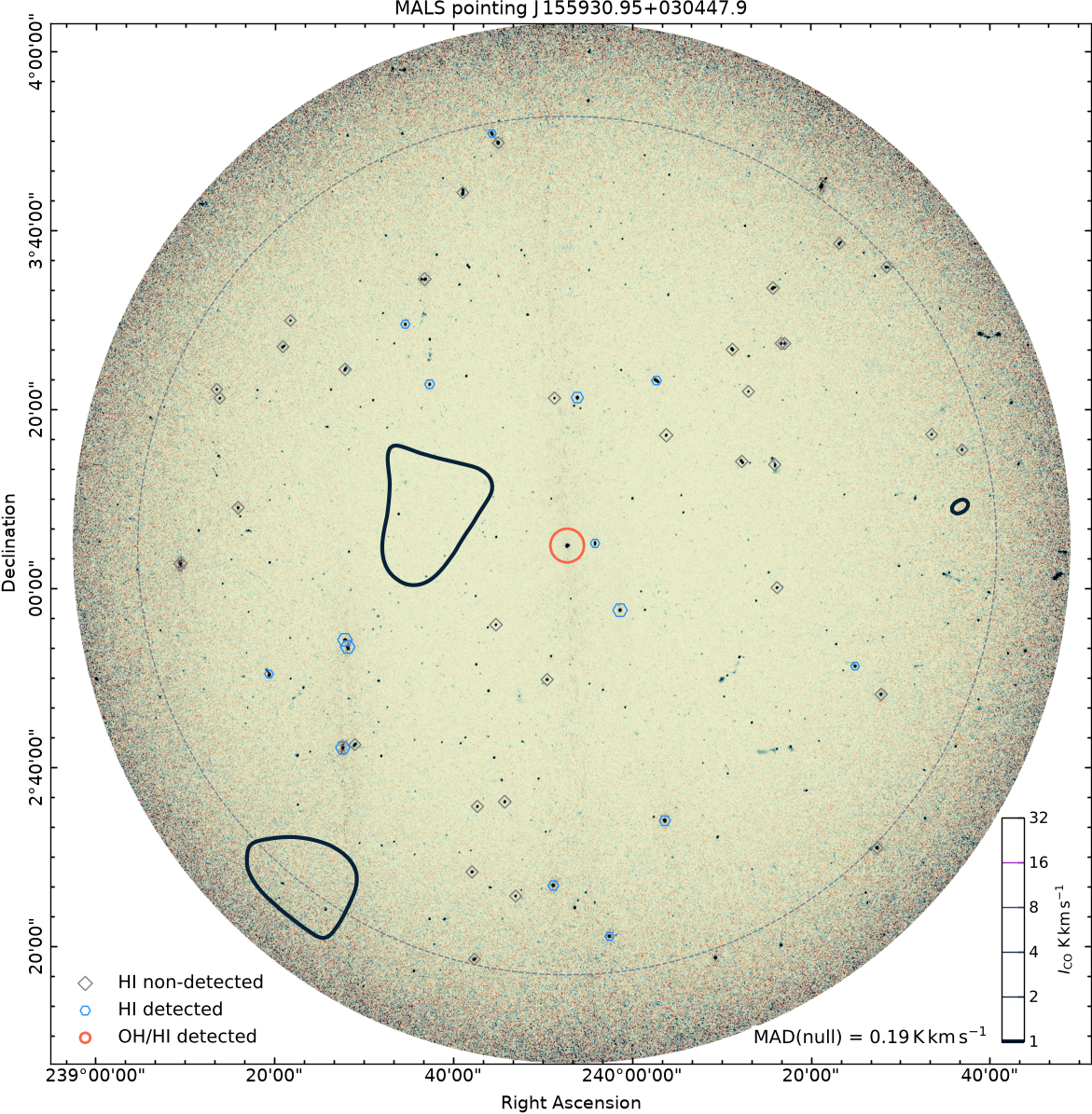}
    \caption{MALS field around J155930.95$+$030447.9. The symbols are the same as in Fig.~\ref{fig:poi}. 
    }
\end{figure*}

\begin{figure*}
    \centering
    \includegraphics[trim = {0cm 0cm 0cm 0cm}, width=1.0\textwidth,angle=0]{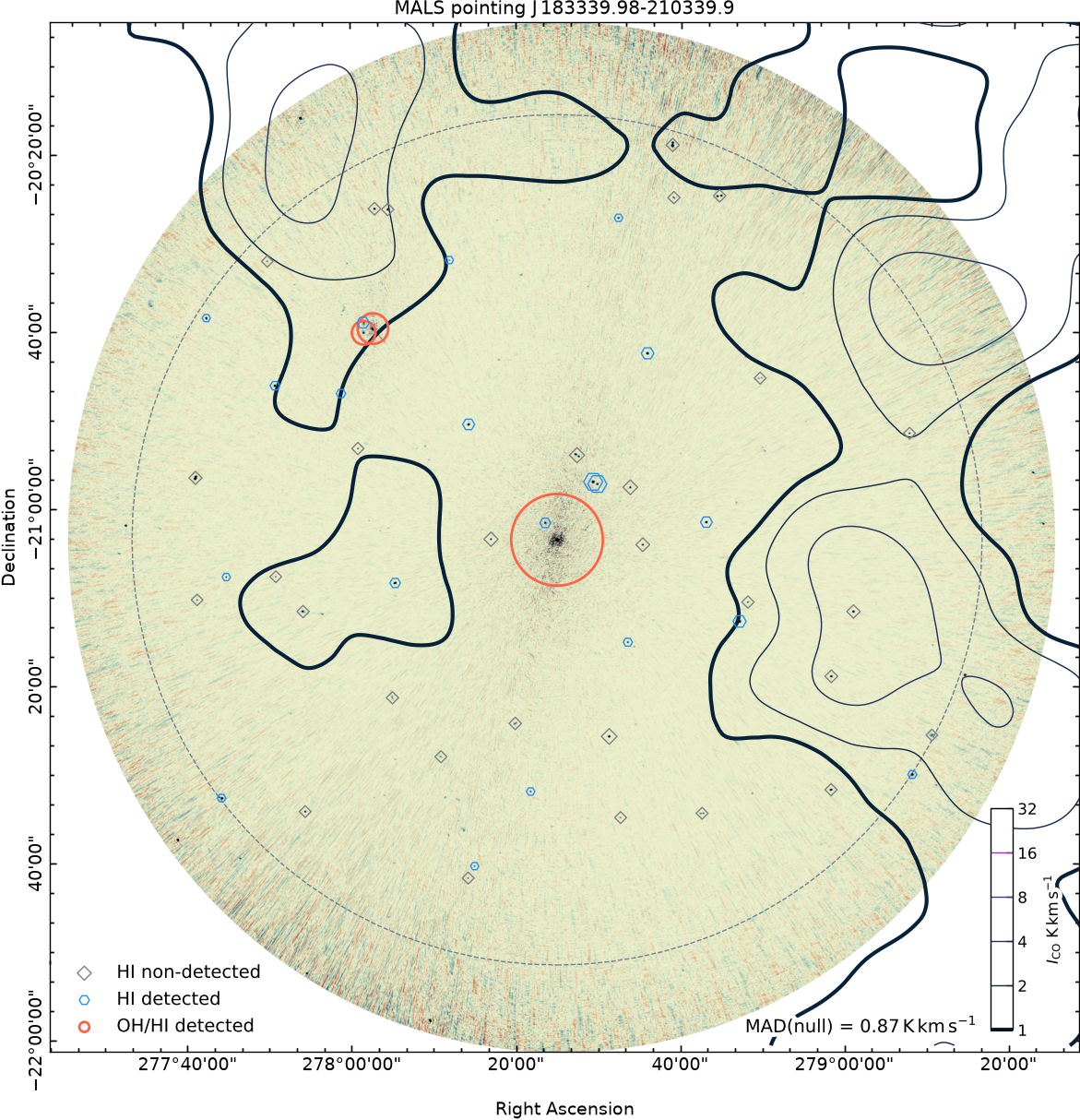}
    \caption{MALS field around J183339.98$-$210339.9. The symbols are the same as in Fig.~\ref{fig:poi}. 
    }
\end{figure*}

\begin{figure*}
    \centering
    \includegraphics[trim = {0cm 0cm 0cm 0cm}, width=1.0\textwidth,angle=0]{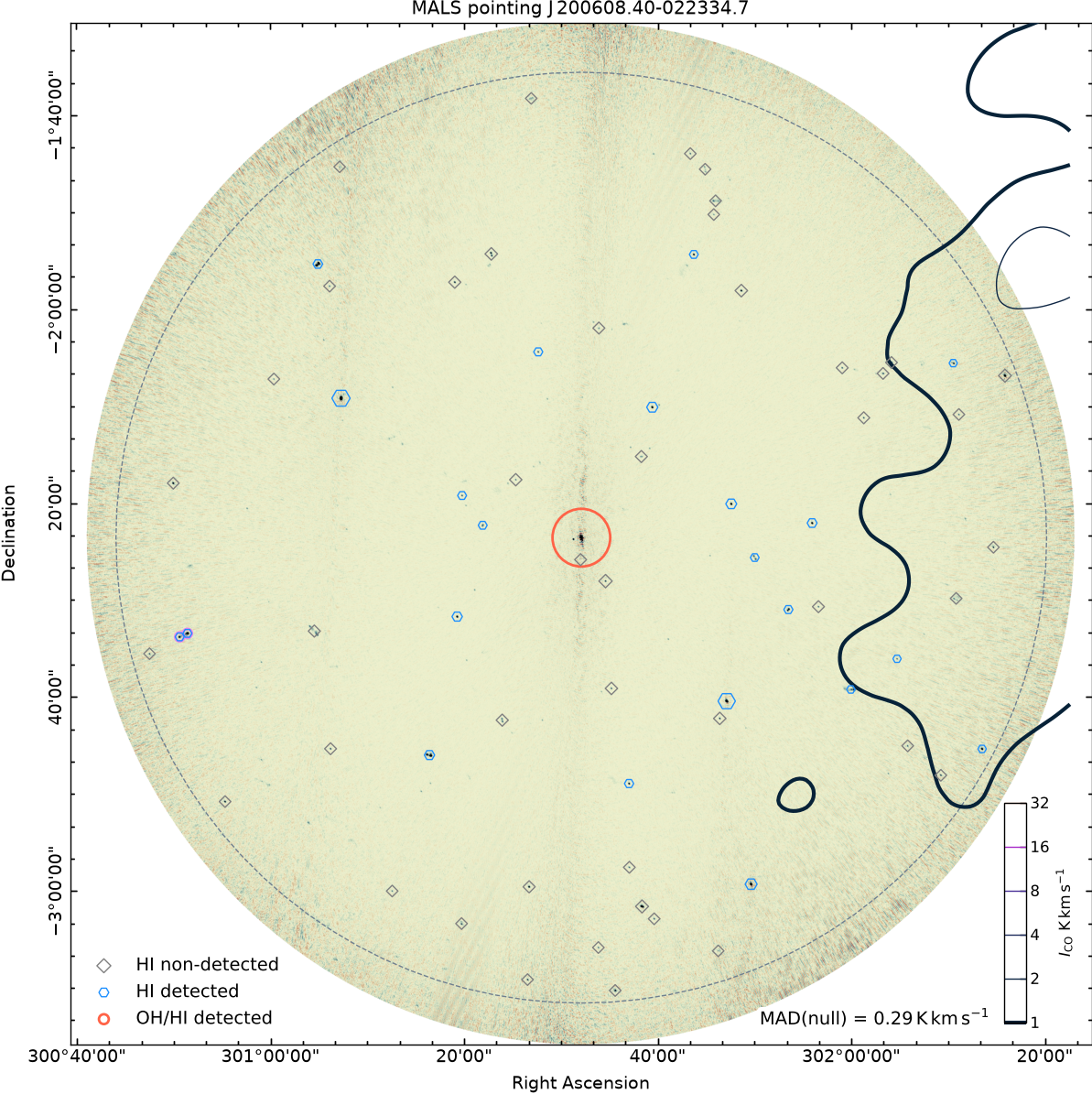}
    \caption{MALS field around J200608.40$-$022334.7. The symbols are the same as in Fig.~\ref{fig:poi}. 
    }
\end{figure*}

\begin{figure*}
    \centering
    \includegraphics[trim = {0cm 0cm 0cm 0cm}, width=1.0\textwidth,angle=0]{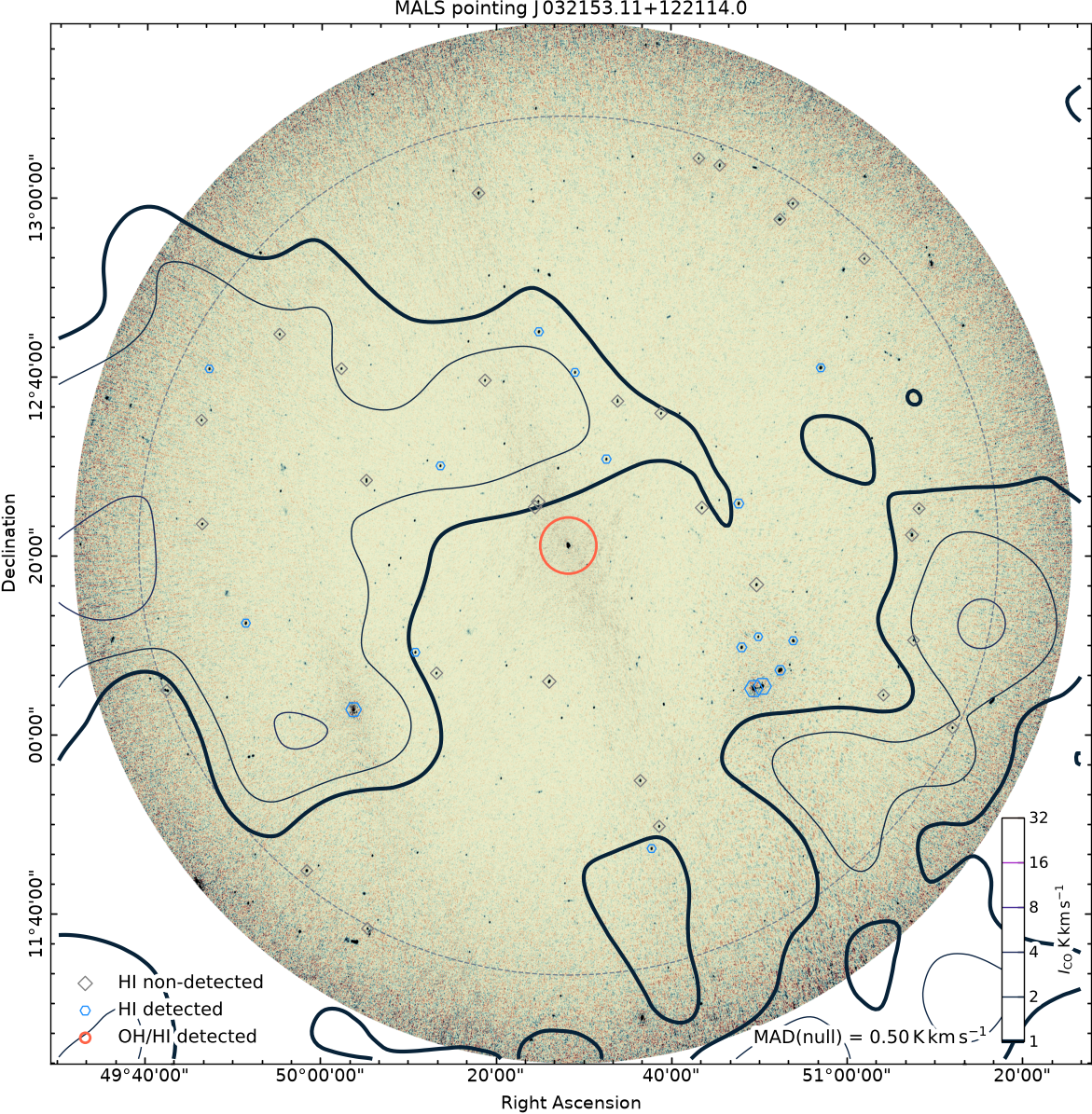}
    \caption{MALS field around J032153.11$+$122114.0. The symbols are the same as in Fig.~\ref{fig:poi}. 
    }
\end{figure*}

\begin{figure*}
    \centering
    \includegraphics[trim = {0cm 0cm 0cm 0cm}, width=1.0\textwidth,angle=0]{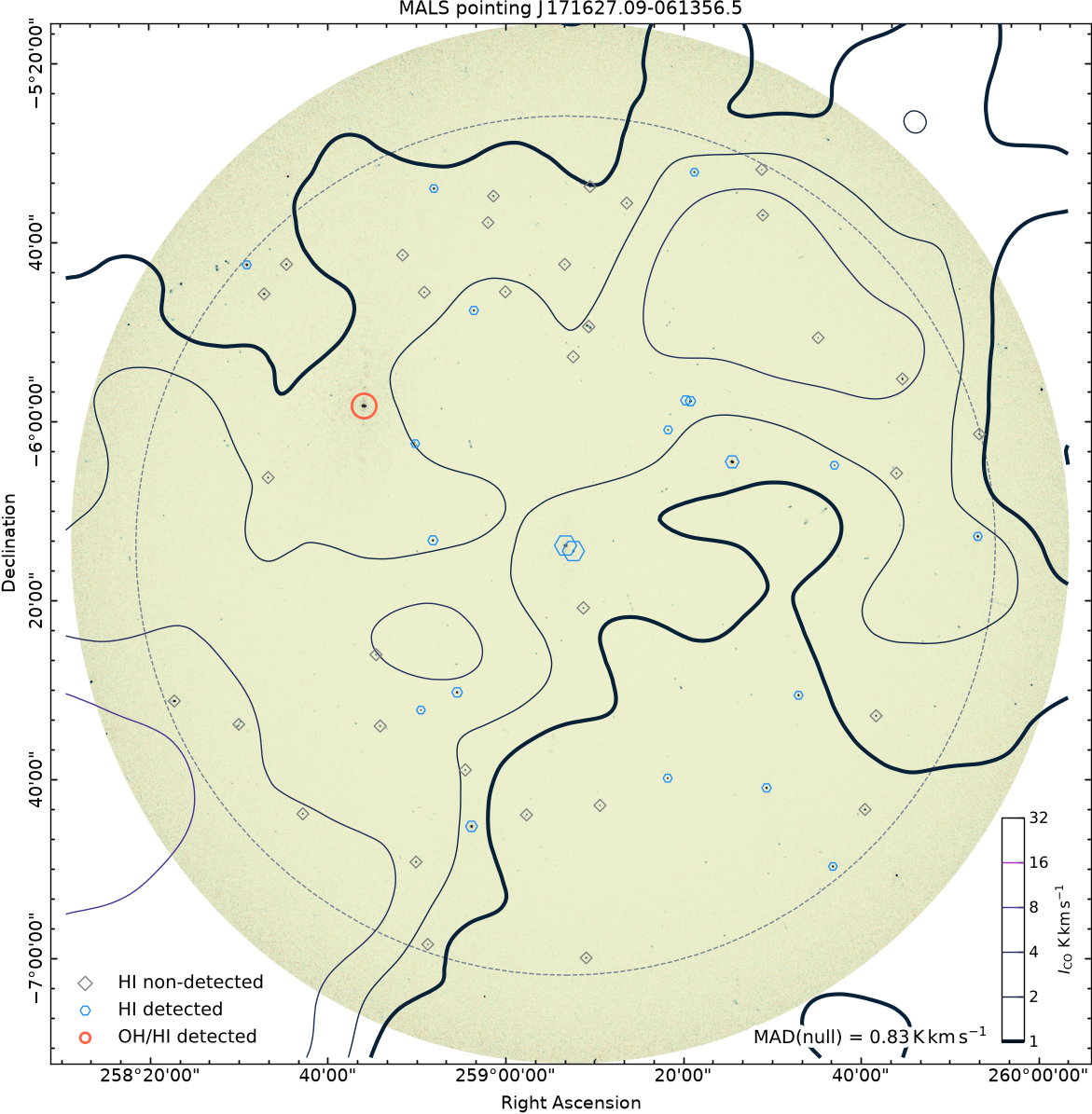}
    \caption{MALS field around J171627.09$-$061356.5. The symbols are the same as in Fig.~\ref{fig:poi}. 
    }
\end{figure*}

\begin{figure*}
    \centering
    \includegraphics[trim = {0cm 0cm 0cm 0cm}, width=1.0\textwidth,angle=0]{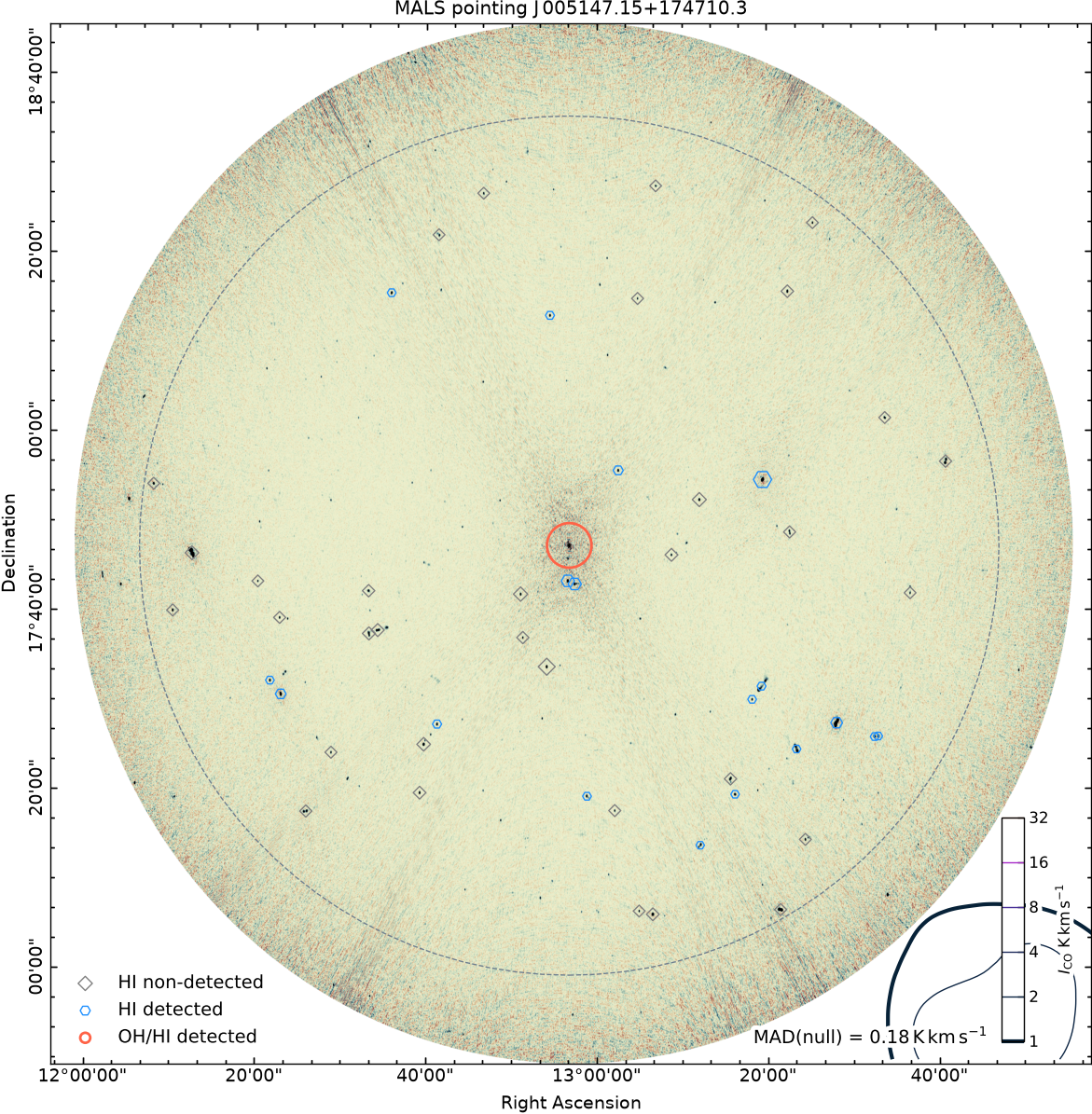}
    \caption{MALS field around J005147.15$+$174710.3. The symbols are the same as in Fig.~\ref{fig:poi}. 
    }
\end{figure*}

\begin{figure*}
    \centering
    \includegraphics[trim = {0cm 0cm 0cm 0cm}, width=1.0\textwidth,angle=0]{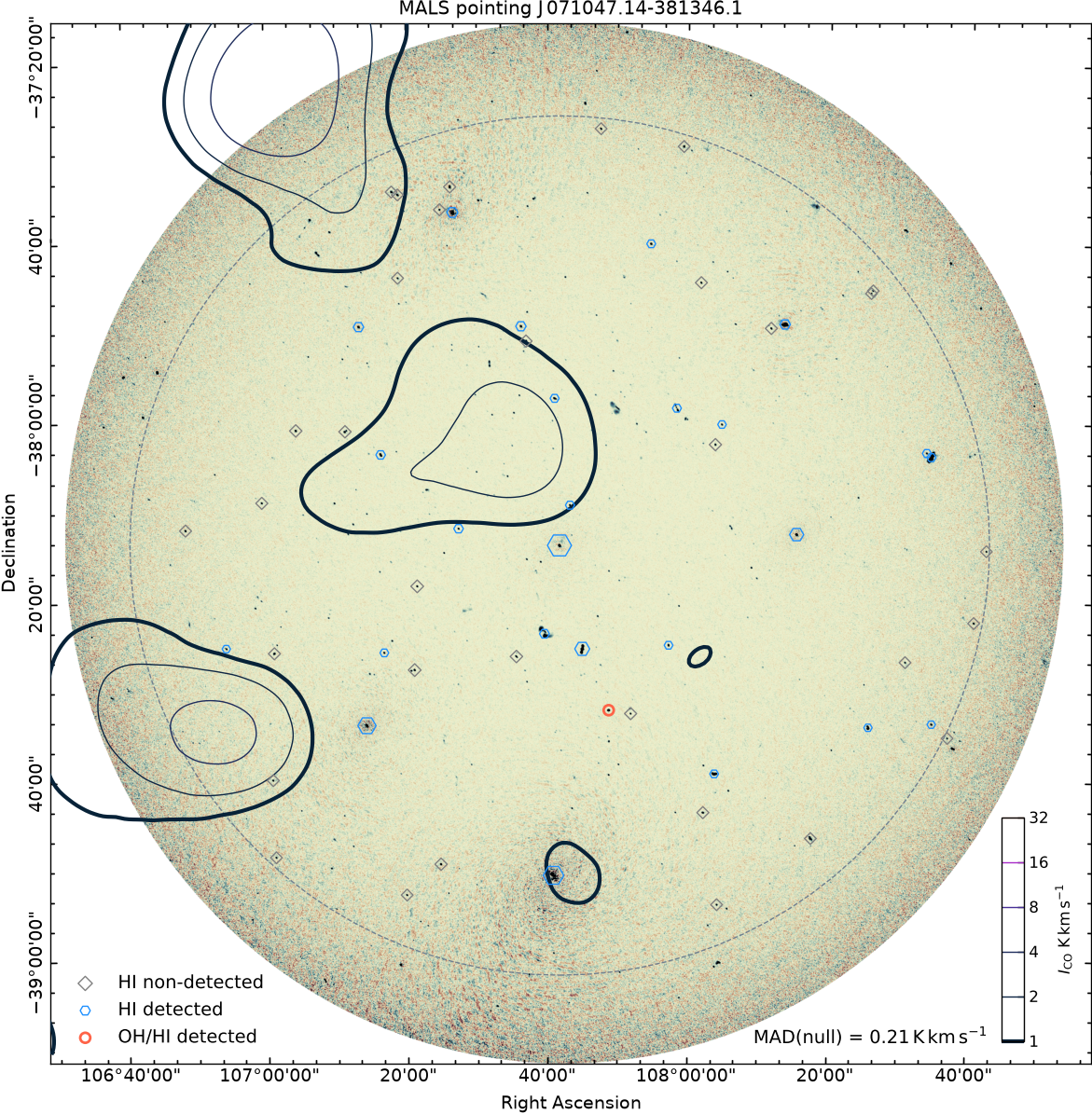}
    \caption{MALS field around J071047.14$-$381346.1. The symbols are the same as in Fig.~\ref{fig:poi}. 
    }
\end{figure*}

\begin{figure*}
    \centering
    \includegraphics[trim = {0cm 0cm 0cm 0cm}, width=1.0\textwidth,angle=0]{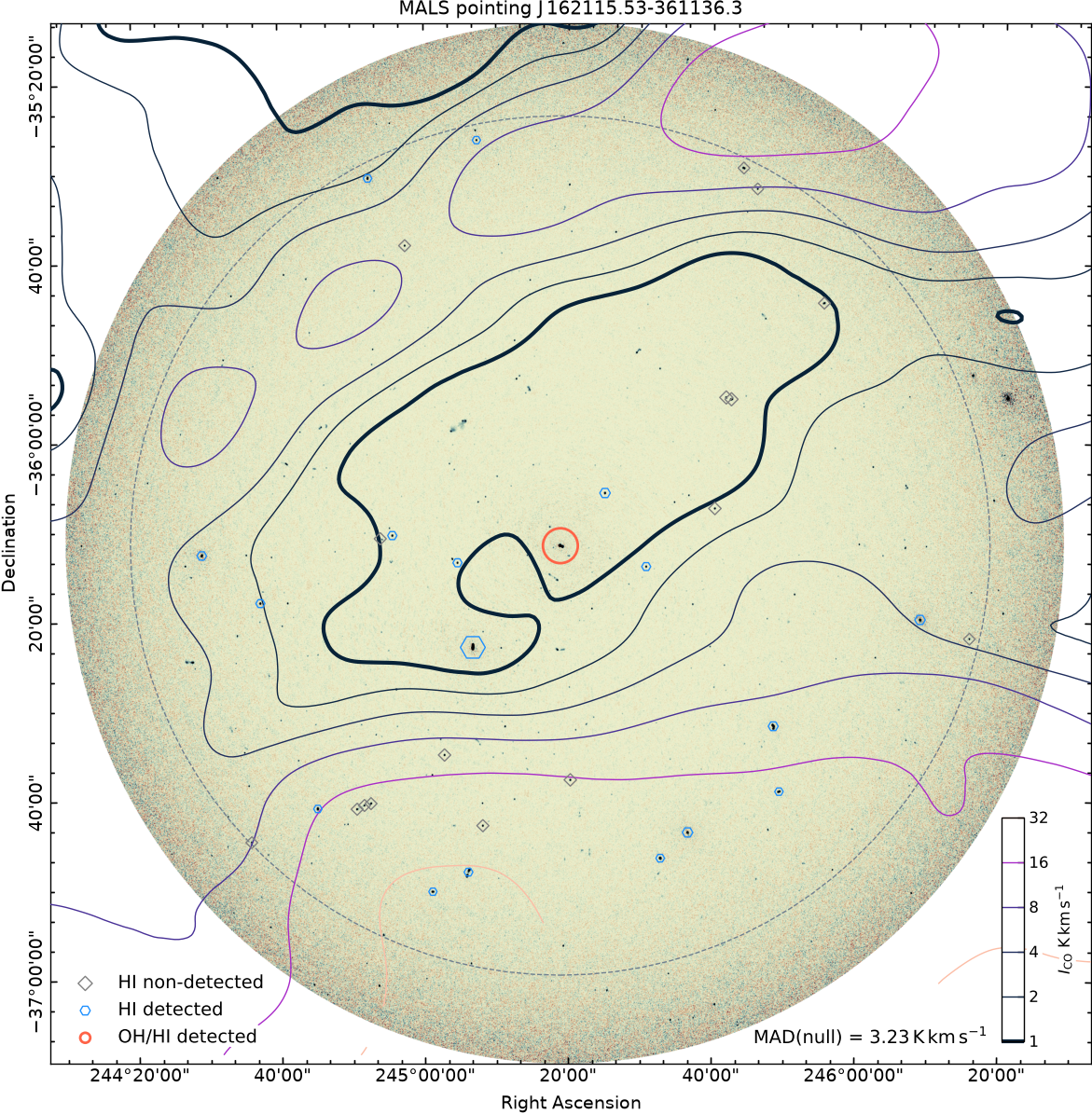}
    \caption{MALS field around J162115.53$-$361136.3. The symbols are the same as in Fig.~\ref{fig:poi}. 
    }
\end{figure*}

\begin{figure*}
    \centering
    \includegraphics[trim = {0cm 0cm 0cm 0cm}, width=1.0\textwidth,angle=0]{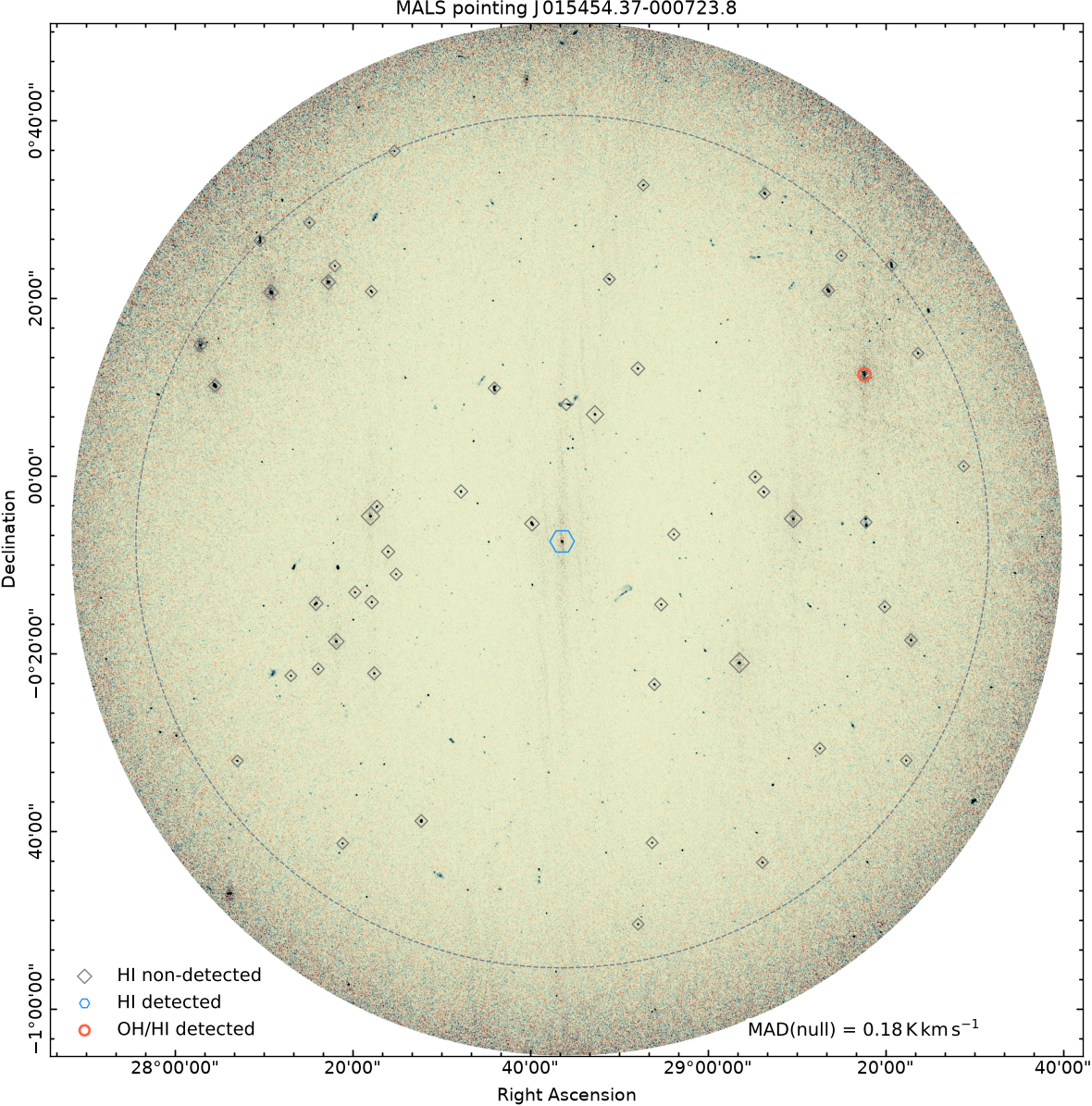}
    \caption{MALS field around J015454.37$-$000723.8. The symbols are the same as in Fig.~\ref{fig:poi}. 
    }
\end{figure*}

\begin{figure*}
    \centering
    \includegraphics[trim = {0cm 0cm 0cm 0cm}, width=1.0\textwidth,angle=0]{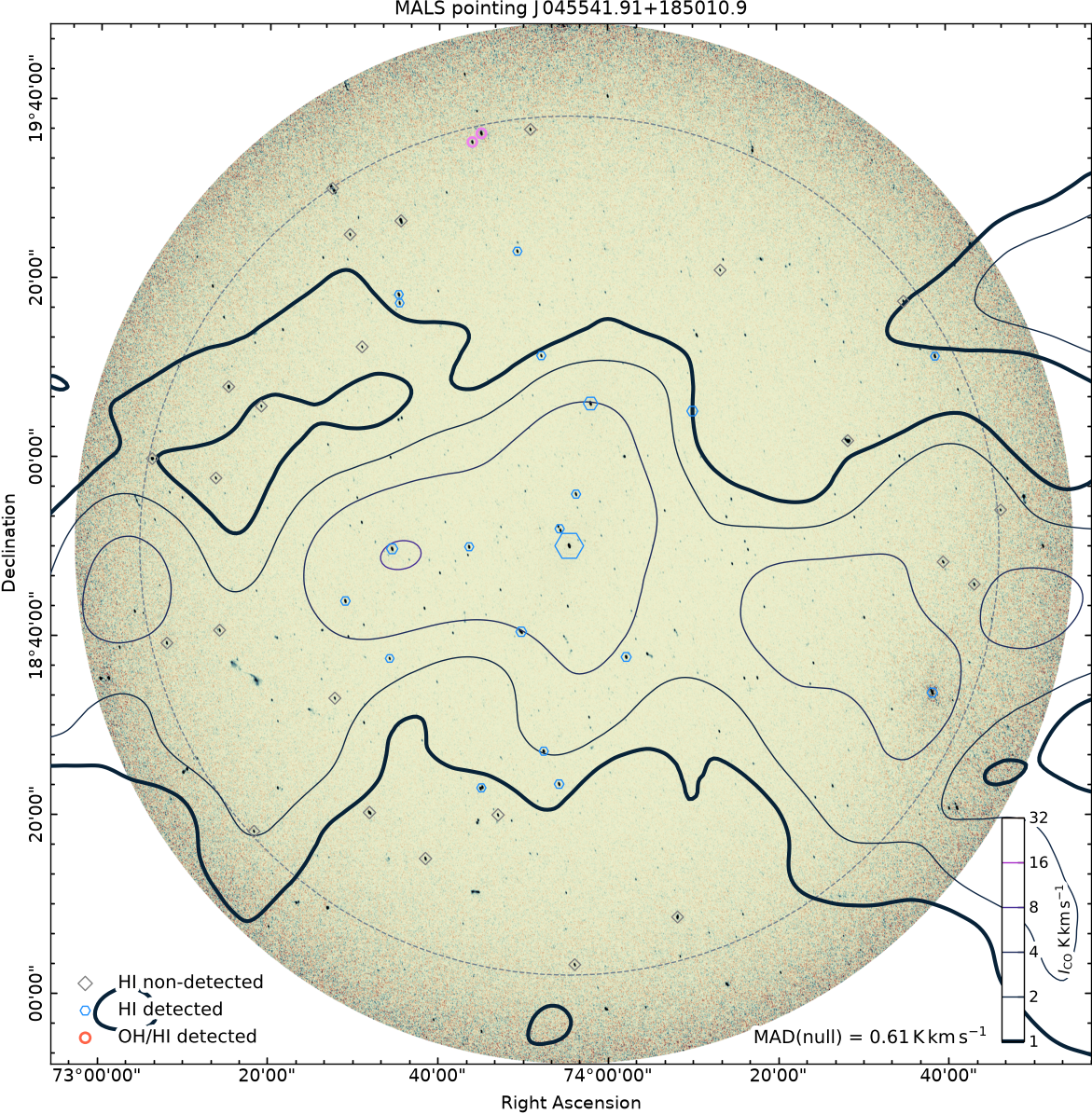}
    \caption{MALS field around J045541.91$+$185010.9. The symbols are the same as in Fig.~\ref{fig:poi} except that pink circles indicate the sight lines, with $0.3 < \log\mathcal{R} < 1$, lower than was used for the confident detections. 
    \label{fig:poi_J0445}
    }
\end{figure*}
%%%%%%%%%%%%%%%%%%%%

\clearpage

%%%%%%%%%%%%%%%%%%%%%%%%%%%%%%%%%%%%%
\section{OH 1612 MHz satellite line detections without main line counterparts}
\label{sec:ohsat-nomain-spec}
%%%%%%%%%%%%%%%%%%%%%%%%%%%%%%%%%%%%%
\suppressfloats[t]

We present the OH 1612\,MHz absorption spectra for J114335.39$-$261102.0 and J200322.53$-$023347.9, the two satellite line candidates without detected OH main line counterparts (see Table~\ref{tab:ohsat_det}).

%%%%%%%%%%%%%%%%%%%%%%%%%%%%%%%
%%%% >>>>>>> OH sattelite detections 
\begin{figure*}
    \centering
    \vbox{
    \includegraphics[trim = {0cm -0.5cm 0cm 0cm}, width=1.0\textwidth,angle=0]{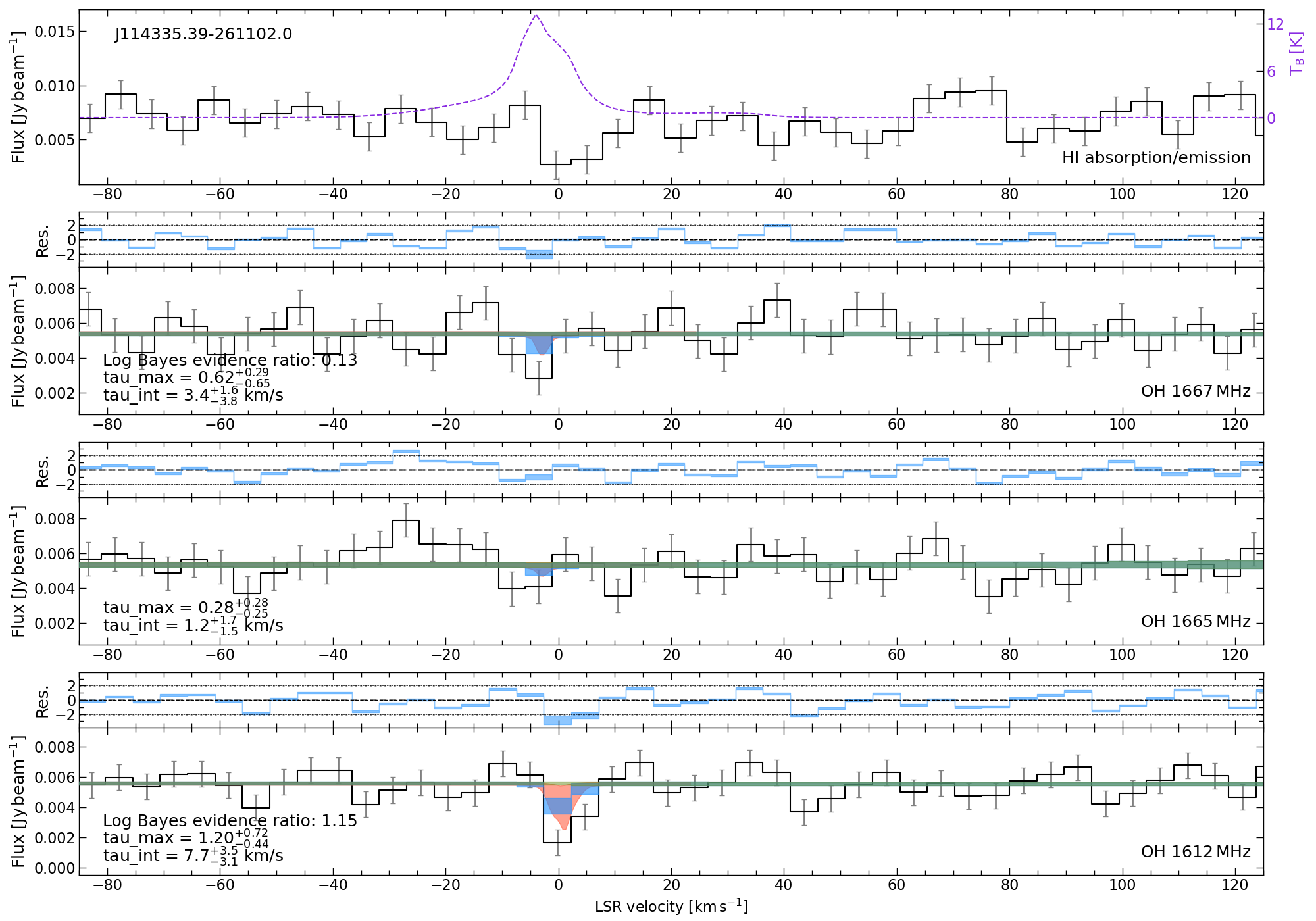} \\
    }
    \caption{The spectra of the sight line toward J\,114335.39$-$261102.0 showing no detection of OH main absorptions, while the satellite line is likely present. The graphical elements are the same as in Fig.~\ref{fig:ohmain-hi}.}
    \label{fig:J114335.39-261102.0}
\end{figure*}
\begin{figure*}
    \centering
    \vbox{
    \includegraphics[trim = {0cm -0.5cm 0cm 0cm}, width=1.0\textwidth,angle=0]{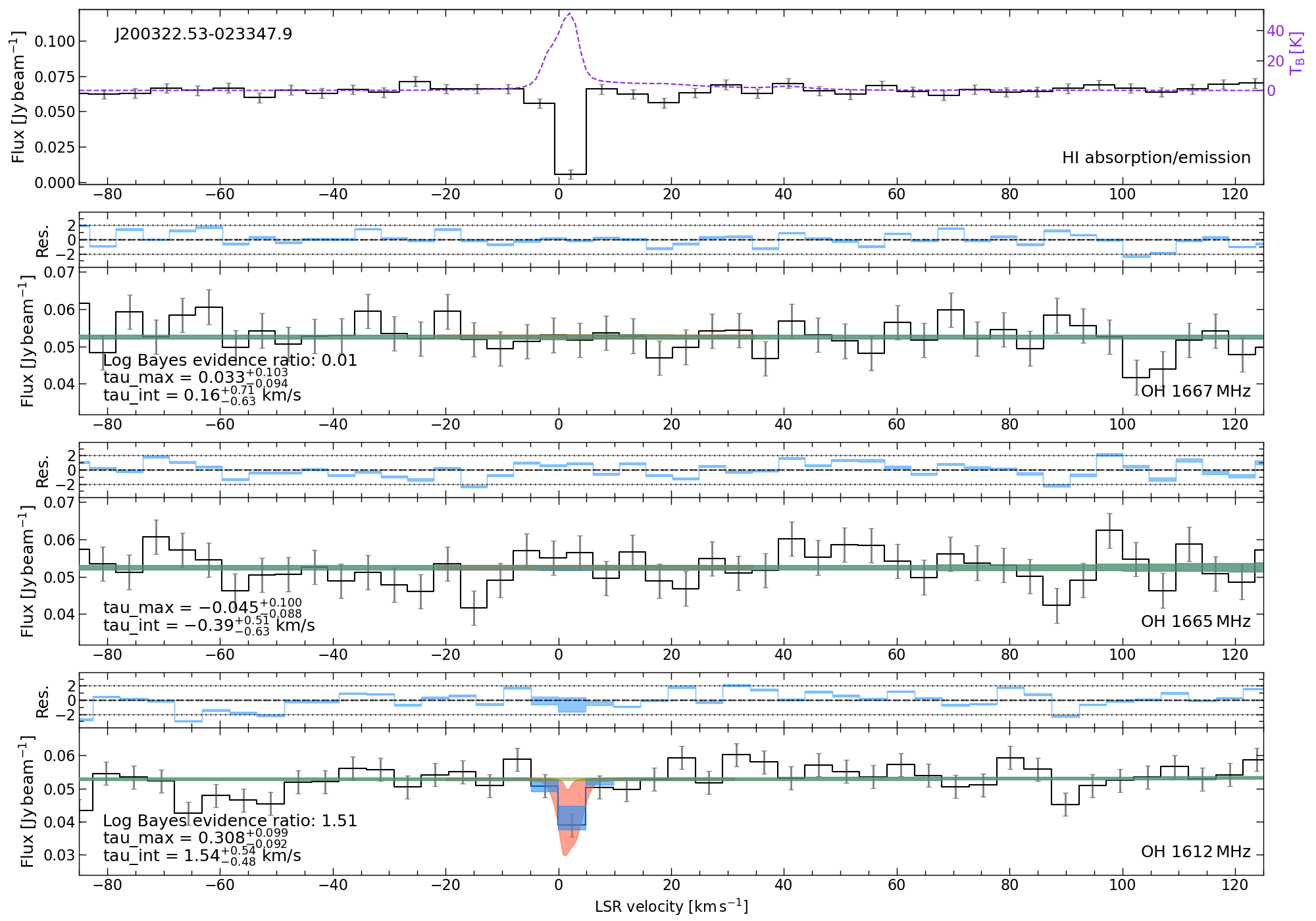} \\
    }
    \caption{The spectra of the sight line toward J\,200322.53$-$023347.9 showing no detection of OH main absorptions, while the satellite line is likely present. The graphical elements are the same as in Fig.~\ref{fig:ohmain-hi}.}
    \label{fig:J200322.53-023347.9}
\end{figure*}
%%%%%%%%%%%%%%%%%%%%%%%%%%%%%%%

\clearpage

%%%%%%%%%%%%%%%%%%%%%%%%%%%%%%%%%%%%%
\section{Thermal balance}
%%%%%%%%%%%%%%%%%%%%%%%%%%%%%%%%%%%%%

We present here the comparison of model calculations with and without the thermal balance assumption, as described in Section~\ref{sec:phys}.

%%%%%%%%%%%%%%%%%%%%%%%%%%%%%%
\begin{figure*}
    \centering
    \includegraphics[trim = {0cm 0cm 0cm 0cm}, width=1.0\textwidth,angle=0]{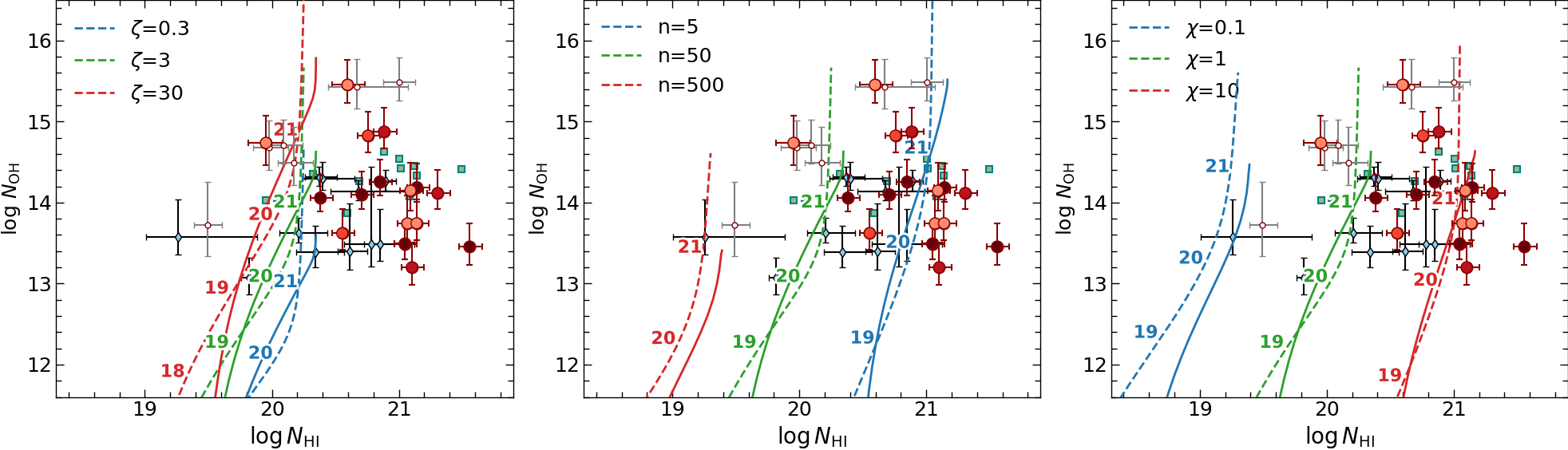}
    \caption{
    Comparison of model calculations, described in Section~\ref{sec:phys}, for fixed $T=100$\,K \citep[as was done in][dashed lines]{Balashev2021} and with the thermal balance \citep[see detailed description in][solid lines]{Balashev2024}. The colors of the  curves on each panel correspond to the variation of the cosmic ray ionization rate (left), number density (middle) and UV field (right) from the base model \citep[similar to Fig.~6 in][]{Balashev2021}. The symbols denote the observational data with the color coding same as in Fig.~\ref{fig:ohhi}. Briefly, the assumed temperature does not significantly affect the derived OH column densities.}
    \label{fig:thermal}
\end{figure*}
%%%%%%%%%%%%%%%%%%%%%%%%%%%%%%

\clearpage

%%%%%%%%%%%%%%%%%%%%%%%%%%%%%%%%%%%%%
\section{Physical conditions posteriors}
%%%%%%%%%%%%%%%%%%%%%%%%%%%%%%%%%%%%%
\suppressfloats[t]

We present the posterior distributions of the physical conditions derived from the semi-analytical ISM model for the OH detection sample, as described in Section~\ref{sec:phys}.

%%%%%%%%%%%%%%%%%%%%%%%%%%%%%%
\begin{figure*}
    \centering
    \includegraphics[trim = {0cm 0cm 0cm 0cm}, width=1.0\textwidth,angle=0]{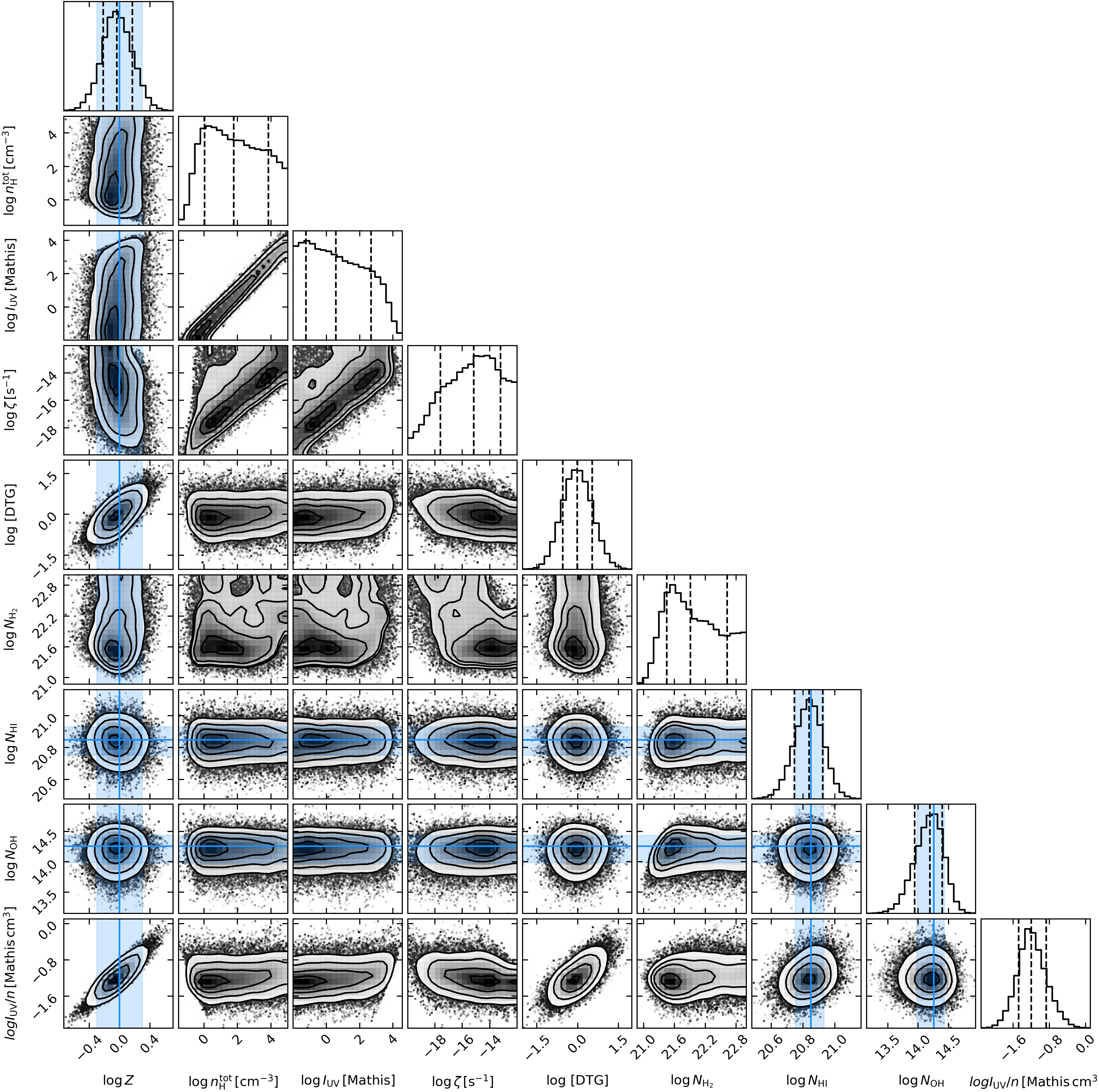}
    \caption{
    Corner plot of the posterior distribution functions for the modeling of OH/\hi\ abundances towards J152149.13$-$294244.8. The diagonal and non-diagonal panels show 1d and 2d marginal posterior distribution functions sampled using the model described in Section~\ref{sec:phys}. The values of $N(\rm \hi)$ and $N({\rm OH})$ are used as observational constraints, while the combination $I_{\rm UV}/n$ (last row) is derived from the sampled parameters. The blue lines and shaded stripes show the point estimates and credibility intervals that were used for $N(\rm \hi)$ and $N({\rm OH})$, as the data, and metallicity, as a prior.
    }
    \label{fig:J1521_phys}
\end{figure*}
%%%%%%%%%%%%%%%%%%%%%%%%%%%%%%

%%%%%%%%%%%%%%%%%%%%%%%%%%%%%%
\begin{figure*}
    \centering
    \includegraphics[trim = {0cm 0cm 0cm 0cm}, width=1.0\textwidth,angle=0]{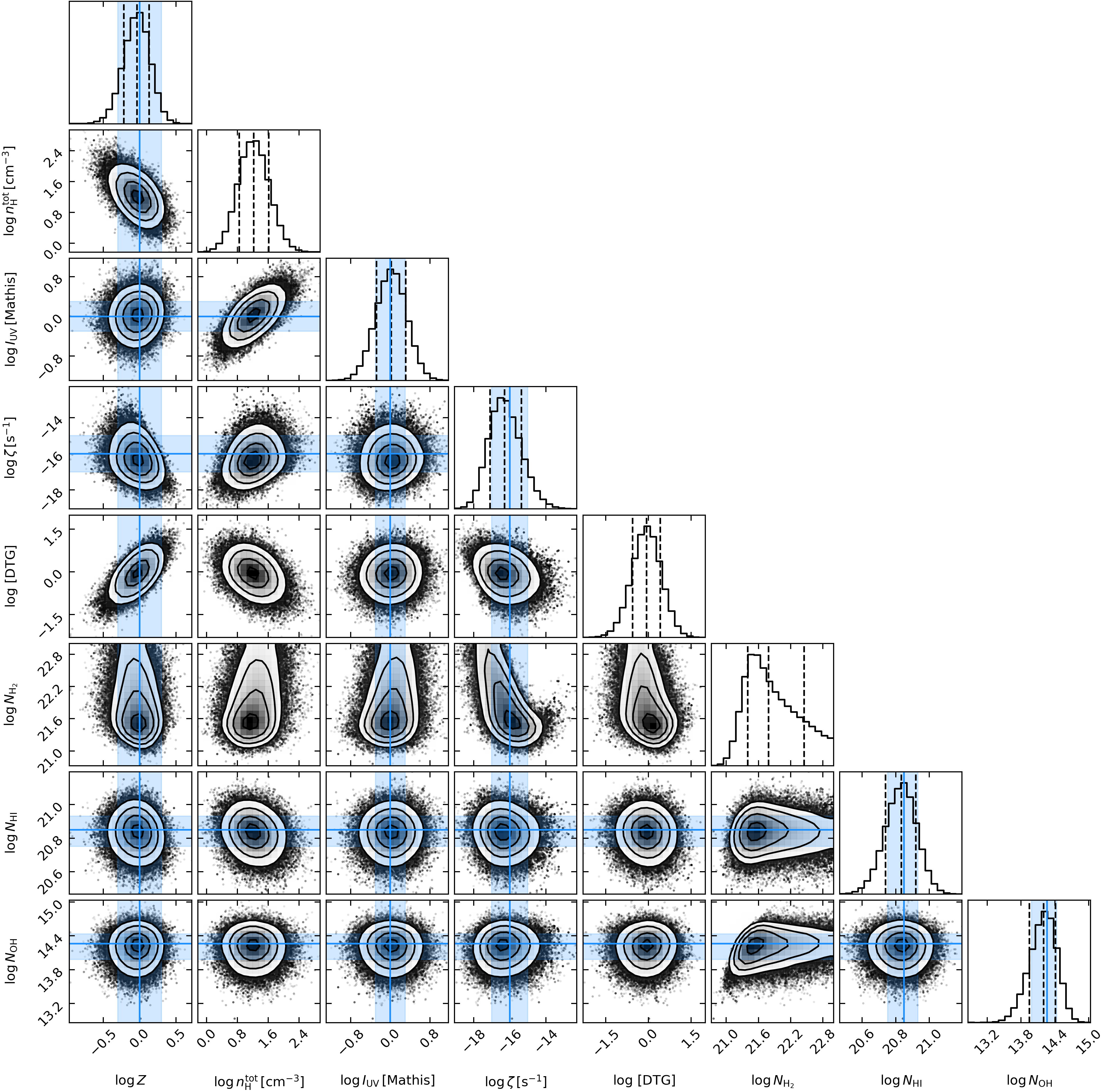}
    \caption{
    Same as Fig.~\ref{fig:J1521_phys} but assuming an informative Gaussian prior, with 0.0 mean and $0.3$ standard deviation, on the UV flux, $\log I_{\rm UV}$. 
    }
    \label{fig:1521_phys_uv}
\end{figure*}
%%%%%%%%%%%%%%%%%%%%%%%%%%%%%%

\twocolumn

%%%%%%%%%%%%%%%%%%%%%%%%%%%%%%%%%%%%%%%%%%%%%%%%%%%%%%%%%%%%
\bsp
\label{lastpage}

\end{document}